\documentclass[dvips,aoas,preprint]{imsart}

\RequirePackage[OT1]{fontenc}
\RequirePackage{amsthm,amsmath}
\RequirePackage[numbers]{natbib}
\RequirePackage[colorlinks,citecolor=blue,urlcolor=blue]{hyperref}
\RequirePackage{hypernat}

\usepackage{amssymb}
\usepackage{mathrsfs}
\usepackage{amsmath}
\usepackage{graphicx}

\startlocaldefs
\numberwithin{equation}{section}
\theoremstyle{plain}

\endlocaldefs

\begin{document}

\begin{frontmatter}

\title{Penalized Likelihood Regression in Reproducing Kernel Hilbert Spaces with
Randomized Covariate Data}
\runtitle{Penalized Likelihood with Randomized Covariate}

\begin{aug}
\author{\fnms{Xiwen} \snm{Ma}\thanksref{t1}\ead[label=e1]{xiwenma@stat.wisc.edu}},
\author{\fnms{Bin} \snm{Dai}\thanksref{t1}\ead[label=e2]{dai@stat.wisc.edu}},
\author{\fnms{Ronald} \snm{Klein}\thanksref{t2}\ead[label=e3]{kleinr@epi.ophth.wisc.edu}},
\author{\fnms{Barbara E.K.} \snm{Klein}\thanksref{t2}\ead[label=e4]{kleinb@epi.ophth.wisc.edu}},
\author{\fnms{Kristine E.} \snm{Lee}\thanksref{t3}\ead[label=e5]{klee@epi.ophth.wisc.edu}}
and
\author{\fnms{Grace} \snm{Wahba}\thanksref{t1}\ead[label=e6]{wahba@stat.wisc.edu}}

\thankstext{t1}{Research supported in part by NIH Grant EY09946,
NSF Grant DMS-0604572, NSF Grant DMS-0906818 and
ONR Grant N0014-09-1-0655.}
\thankstext{t2}{Supported in part by NIH Grant EY06594, and by the Research to Prevent Blindness Senior Scientific Investigator Awards, New York, NY.}
\thankstext{t3}{Supported in part by NIH Grant EY06594.}

\runauthor{Ma, Dai, Klein, Klein, Lee and Wahba}

\affiliation{University of Wisconsin} 

\address{Xiwen Ma\\
Grace Wahba\\
Bin Dai\\
Department of Statistics\\
University of Wisconsin\\
1300 University Avenue\\
Madison, WI 53706\\
\printead{e1}\\
\phantom{E-mail:\ }\printead*{e2}\\
\phantom{E-mail:\ }\printead*{e6} }

\address{Ronald Klein\\
Barbara E.K. Klein\\
Kristine E. Lee\\
Department of Epidemiology and Visual Science\\
University of Wisconsin\\
610 N. Walnut Street\\
Madison, WI 53726\\
\printead{e3}\\
\phantom{E-mail:\ }\printead*{e4}\\
\phantom{E-mail:\ }\printead*{e5} }
\end{aug}

\begin{abstract}
Classical penalized likelihood regression problems deal with the case that the independent variables data are known exactly. In practice, however, it is common to observe data with incomplete covariate information. We are concerned with a fundamentally important case where some of the observations do not represent the exact covariate information, but only a probability distribution. In this case, the maximum penalized likelihood method can be still applied to estimating the regression function.  We first show that the maximum
penalized likelihood estimate exists under a mild condition. In the computation, we propose a dimension reduction technique to minimize the penalized likelihood and derive a GACV (Generalized Approximate Cross Validation) to choose the smoothing parameter.  Our methods are extended to handle more complicated incomplete data problems, such as, covariate measurement error and partially missing covariates.
\end{abstract}

\begin{keyword}
\kwd{Penalized likelihood regression}
\kwd{reproducing kernel Hilbert spaces}
\kwd{randomized covariate data}
\kwd{generalized approximate cross validation}
\kwd{errors in variables}
\kwd{covariate measurement error}
\kwd{partially missing covariates.}
\end{keyword}

\end{frontmatter}

\tableofcontents

\section{ Introduction}

\subsection{Penalized likelihood regression in reproducing kernel Hilbert spaces}
We are concerned with non or semi parametric
regression for data from a non-Gaussian exponential
family.  Suppose that we have $n$ independent observations $(y_i,x_i), i=1,...,n$, where each $y_i$ denotes the response and each $x_i$ denotes the
covariate information.  The goal
is to fit a probability mechanism, assuming that the conditional distribution of $y_i$ given $x_i$ has a density in the exponential family with the form
\begin{eqnarray}\label{den0}
p(y_i|x_i, f)  = \exp\{(y_i \cdot f(x_i)- b(f(x_i)))/a(\phi)
+c(y_i, \phi)\}
\end{eqnarray}
where $b(\cdot)$ and $c(\cdot)$ are given functions with $b(\cdot)$
strictly convex, $\phi$ is the scale parameter and $f$ is the regression function to be estimated.
We assume throughout this paper that $\phi$ is known, as, for example, Binomial data and Poisson data.  In this case, (\ref{den0}) can be simplified by
\begin{eqnarray}
\label{den1}p(y_i|x_i, f)  = \exp\{y_i \cdot f(x_i)- b(f(x_i))
+c(y_i)\}.
\end{eqnarray}
Note that the methods of this paper can also be extended to the situation when $\phi$ is unknown, but may be more computationally complicated.

The regression function $f$ will be estimated non or semi parametrically in some reproducing kernel Hilbert
space (RKHS) $\cal{H}$ by
minimizing the penalized likelihood
\begin{eqnarray}
\label{plk}I_\lambda(f) = -\frac{1}{n}\sum_{i=1}^n\log p(y_i|x_i, f)
+ \frac{\lambda}{2}{J}(f)
\end{eqnarray}
where the penalty ${J}(\cdot)$ is a norm or semi-norm in $\cal{H}$ with finite
dimensional null space $\mathcal {H}_0 = \{f \in \mathcal
{H}~|~{J}(f) = 0\}$ and $\lambda$ is the smoothing parameter which
balances the tradeoff between model fitting and smoothness. In this case, if the null space $\mathcal{H}_0$ satisfies some condition, saying that $I_\lambda(f)$
has a unique minimizer in $\mathcal{H}_0$, then the minimizer of $I_\lambda(f)$ in $\mathcal {H}$ exists in a known $n$-dimensional subspace spanned by $\mathcal{H}_0$
and functions of the reproducing kernel. See, for example, Kimeldorf and Wahba (1971)\cite{Kimeldorf1971}, O'Sullivan, Yandell and Raynor (1983)\cite{Sullivan1983}, Wahba (1990)\cite{Wahba1990} and Xiang and Wahba (1996)\cite{Xiang1996}. This model building technique, known as penalized likelihood regression with RKHS penalty, allows for more flexibility than
parametric regression models.
We will not review the general literature, other
than to note two books and references therein.
Wahba (1990)\cite{Wahba1990} offers a general introduction of spline
models.  Gu (2002)\cite{Gu2002} comprehensively reviews the
smoothing spline analysis of variance (SS-ANOVA), an important
implementation of penalized likelihood regression in multivariate function
estimation.

\subsection{Randomized covariate data and related problems}
In this paper, the issue we are concerned about is the situation where components of $x_i$ are not observable but only known to have come from a particular probability distribution. This concept of randomized covariate, without the requirement of any actual measure of $x_i$, is more flexible than the common sense of covariate measurement error.  In this case, a natural likelihood-based approach is to treat $x_i$'s as latent variables and minimize a randomized version of penalized likelihood that integrates $x_i$'s out of the likelihood.  This approach, however, typically leads to a non-convex and infinite dimensional optimization problem in RKHS.   Therefore we shall first prove that the randomized penalized likelihood is minimizable.
This is the subject of Section 2.  Afterwards, two computational issues will be addressed in Section 3:  (1) how to numerically compute an estimator and (2) how to select the smoothing parameter.

Randomized covariate data is a basic version of incomplete data. Our methods can be extended to other incomplete data problems. For example, in the survey or medical research, it is common to obtain data where the covariates are measured with error. More specifically, $x_i$ is not directly observed but instead
$
x^{err}_i = x_i + {u}_i
$
is observed, where ${u}_i,i=1,...,n$ are iid random perturbations.
Fan and Truong (1993)\cite{Fan1993} regarded this measurement error problem in the context of nonparametric regression, using the methods based on kernel deconvolution.
Their technique was later studied and extended by, for example, Ioannides and Alevizo (1997)\cite{Ioannides1997}, Schennach (2004)\cite{Schennach2004}, Carroll, Ruppert and Stefanski (2006)\cite{Carroll2006} and Delaigle, Fan and Carroll (2009)\cite{Delaigle2009}.
More recently, penalized likelihood regression have been considered in the measurement error literature.
Carroll, Maca and Ruppert (1999)\cite{Carroll1999} suggested to use the SIMEX method (Cook and Stefanski, 1994\cite{Cook1994}) to build nonparametric regression models including both kernel regression and penalized likelihood regression. Berry, Carroll and Ruppert (2001)\cite{Berry2001} described Bayesian approaches for smoothing splines and regression P-splines.  Cardot, Crambes, Kneip and Sarda (2007)\cite{Cardot2007} used the total least square method (Van Huffel and Vandewalle, 1991\cite{VanHuffel1991}) to compute a smoothing spline estimator from noisy covariates. These papers mainly discussed the situation of Gaussian responses but very little literature concerns other responses. As a sequel to these works, in this paper, we treat measurement error as a special case of randomized covariates, because each $x_i$ can be viewed as a random variable (vector) distributed as $x^{err}_i-{u}_i$.  Therefore the methodology of randomized penalized likelihood estimate can be employed.

We will as well be able to make another modest extension to treat the important situation where some components of some $x_i$'s are completely missing.  In this case, we may write $x_i =( x_i^{obs}, x_i^{mis})$, where $x_i^{obs}$ and $x_i^{mis}$ denote the observed and the missing components.  It is well-known (Little and Rubin,
2002\cite{Little2002}) that a complete case analysis that deletes the
cases with missing information often leads to bias or inefficient estimates. Various methods for missing covariate data have been developed
in the context of parametric regression models, but to date few methods have been proposed for nonparametric penalized likelihood regression in RKHS.  For parametric regression, one popular approach is the method of weights initially proposed by Ibrahim (1990)\cite{Ibrahim1990}. His suggestion is to
assume the $x_i$'s to be independent observations from a marginal distribution depending on some parameters and to maximize the joint distribution of $(y_i,x_i)$ by the expectation-maximization (EM) algorithm. Discussions and extensions of this method appear in Ibrahim, Lipsitz and Chen (1999)\cite{Ibrahim1999}, Horton and Laird (1999)\cite{Horton1999}, Huang, Chen and Ibrahim
(2005)\cite{Huang2005}, Ibrahim, Chen, Lipsitz and Herring
(2005)\cite{Ibrahim2005}, Horton and Kleinman
(2007)\cite{Horton2007}, Chen and Ibrahim (2006)\cite{Chen2006}, Chen, Zeng and Ibrahim (2007)\cite{Chen2007} and elsewhere.
Ibrahim's method can also be employed to build nonparametric regression models.  Actually, in the framework of Ibrahim's method, the missing components $x_i^{mis}$ can be viewed as a random vector depending on the observed components $x_i^{obs}$ and the covariate marginal distribution.
Therefore in this paper, missing covariate data is treated as a special case of randomized covariate data, and thus our methods can be extended.

\subsection{Outline of paper}
The rest of the paper is organized as follows.  In Section 2, we prove the existence of the randomized covariate penalized likelihood estimation in the general smoothing spline set-up.  Computational techniques are presented in Section 3.  Sections 4 and 5 extend our methods to the problem of covariate measurement error.  Sections 6 and 7 describe penalized likelihood regression with missing covariate data.  Section 8 provides some numerical results.  We conclude our paper in Section 9.  

\section{Randomized covariate penalized likelihood estimation (theory)}

Consider the general smoothing spline set-up, where $x$ is
allowed to be from some arbitrary index set $\mathscr{T}$ on which an
RHKS can be defined.  Randomized covariate data is defined in the
way that we ``observe" for each subject $i$ a \emph{probability space}
$(\mathcal {X}_i, \mathcal{F}_i, P_{i})$, rather than a
realization of $x_i$, where $\mathcal {X}_i \subseteq \mathscr{T}$ denotes the domain of $x_i$,
$\mathcal{F}_i$ is a $\sigma-$algebra and $P_{i}$ is a
probability measure over $(\mathcal {X}_i,\mathcal{F}_i)$.

In this case, each $x_i$ can be treated as a latent random variable.  Thus, given a regression function $f$, the distribution of $[y_i|f]$ has a density
\begin{equation}\label{ymargin}
 p(y_i|f) = \int_{\mathcal {X}_i} p(y_i|x_i, f)d P_{i}{}{.}
\end{equation}
Note that, throughout this paper,  we use the labels $[A|B]$ and $p(A|B)$ to denote the conditional distribution of $A$ given $B$ and the density function for this distribution.
According to (\ref{ymargin}), the penalized likelihood estimate of $f$ is the minimizer of
\begin{equation}
\label{plkr} I^R_\lambda(f) = -\frac{1}{n}\sum_{i=1}^n \log
\int_{\mathcal {X}_i} p(y_i|x_i, f) d P_{i}{}+
\frac{\lambda}{2}{J}(f)
\end{equation}
where $R$ denotes the ``randomness" of the covariates and $f$ is restricted on the Borel measurable subset
\begin{equation}
\label{HB} \mathcal {H}_B = \{f \in \mathcal {H}~:~f \text{ is Borel
measurable on } (\mathcal {X}_i, \mathcal{F}_i) , i=1,...,n\}
\end{equation}
in which the Lebesgue integrals in (\ref{plkr}) can be defined.
It can be shown that $\mathcal{H}_B$ is a subspace of
$\mathcal {H}$. \\

PROPOSITION 2.1. \emph{$\mathcal{H}_B$ is a subspace of
$\mathcal {H}$.}\\

\textbf{Proof} See Appendix A. $~\Box$\\

This methodology can be referred to as \textbf{randomized covariate penalized likelihood estimation} or \textbf{RC-PLE}. Note that RC-PLE includes the classical penalized likelihood regression where $x_i$'s are observed exactly.
Actually, $I^R_\lambda(f)$ equals $I_\lambda(f)$ if every $(\mathcal {X}_i, \mathcal{F}_i, P_{i})$ stands for a single point probability.

However, computation of RC-PLE is extremely difficult.
Firstly, since each $p(y_i|x_i, f)$ is log-concave as a function of $f$,
$I^R_\lambda(f)$ is in general not convex due to the integrals.  Secondly, if at least one
$(\mathcal{X}_i, \mathcal {F}_i, P_{i})$ has infinite
support, then there is no finite dimensional subspace in which $f_\lambda$ is known
a {\it priori} to lie,  as
can be concluded from the arguments in Kimeldorf and Wahba
(1971)\cite{Kimeldorf1971}.  Therefore, we shall first prove that
$I^R_\lambda(f)$ is minimizable and hence the phrase ``penalized likelihood
estimate" is meaningful.  Computational techniques will be described
in Section 3.

Recall that for the classical penalized likelihood regression, the unique solution in the
null space is sufficient to ensure the existence of the penalized
likelihood estimate.  In the case of randomized covariate data, we
extend this condition as follows:\\

ASSUMPTION A.1 (Null space condition). There exist exactly observed subjects
$(y_{k_1},x_{k_1}), (y_{k_2},x_{k_2}), ..., (y_{k_s},x_{k_s})$ such
that $\sum_{i=1}^s \log p(y_{k_i}|x_{k_i}, f)$ has a unique
maximizer in $\mathcal {H}_0$.\\

Now we state our main theorem. \\

THEOREM 2.2. \emph{Under A.1, $\exists f_\lambda \in
\mathcal {H}_B$ such that
$
I^R_\lambda(f_\lambda) = \inf_{f\in \mathcal {H}_B} I^R_\lambda(f)
$.
}\\

Theorem 2.2 guarantees the existence of the RC-PLE estimate, which justifies the title of the paper.  In particular, if the null
space of the penalty functional $J(\cdot)$ contains only constants, then A.1 can be ignored.  In this case, the penalized likelihood estimate always exists.

Our proof of the theorem is based on lower-semicontinuity in the
weak topology.  We first recall some definitions.\\

DEFINITION 1.  A sequence $\{f_k\}_{k\in \mathbb{N}}$ in a Hilbert
space $\mathcal{H}$ is said to \textbf{converge weakly} to $f$ if $
\langle f_k, g \rangle \rightarrow \langle f,g \rangle $
for all $g \in \mathcal {H}$.  Here $\langle\cdot, \cdot\rangle$ denotes the inner product of $\mathcal {H}$.\\

DEFINITION 2.  Let $\mathcal{H}$ be a Hilbert space, a functional
$\gamma : \mathcal{H}\rightarrow \mathbb{R}$ is \textbf{(weakly)
sequentially lower semicontinuous} at $f\in \mathcal {H}$ if $
\gamma(f) \le \liminf \gamma(f_k) $ for any sequence $\{f_k\}_{k\in
\mathbb{N}}$ that (weakly) converges
to $f$.\\

DEFINITION 3.  Let $\mathcal{H}$ be a Hilbert space, a functional
$\gamma : \mathcal{H} \rightarrow \mathbb{R}$ is \textbf{positively coercive} if
$ ||f||_\mathcal{H}\rightarrow +\infty $ implies $\gamma(f) =
+\infty $. Here $||\cdot||_\mathcal {H}$ denotes the norm of $\mathcal {H}$.~\\

Theorem 2.2 can be shown by combining
Proposition 2.3 and Lemmas 2.4-2.6 below.  Note that Proposition 2.3 is
obtained from Theorem 7.3.7 in Kurdila and
Zabarankin (2005)\cite{Kurdila2005}, Page 217.  The proofs of lemmas are given in Appendix A. \\

PROPOSITION 2.3.  \emph{Let $\mathcal{H}$ be a Hilbert space.
Suppose that $\gamma :\mathcal{M}\subseteq \mathcal{H} \rightarrow
\mathbb{R}$ is positively coercive and weakly sequentially lower
semicontinuous over the closed and convex set $\mathcal{M}$, then
$\exists f_0 \in \mathcal{M}$ such that
$
\gamma(f_0) = \inf_{f \in \mathcal{M}} \gamma(f)
$.
}\\

LEMMA 2.4. \emph{Under A.1, the penalized
likelihood $I^R_\lambda(f)$ is positively coercive over $\mathcal {H}_B$. }\\

LEMMA 2.5.  \emph{The functional~~$\log \int_{\mathcal {X}_i} p(y_i|x_i, f) d
P_{i}{}:\mathcal {H}_B \rightarrow \mathbb{R}$~
is weakly sequentially continuous. } \\

LEMMA 2.6. \emph{The penalty functional ${J}(\cdot)$ is weakly sequentially lower semi-continuous.}\\

\textbf{Proof of Theorem 2.2.}  Consider the functional $I^R_\lambda :
\mathcal{H}_B \subseteq \mathcal {H}\rightarrow \mathbb{R}$.  Theorem 2.2 follows from Proposition 2.2, Lemma
2.4-2.6 and Proposition 2.3 ~~$\Box$.

\section{Randomized covariate penalized likelihood estimation (computation)}

In the preceding section, we theoretically extended penalized likelihood regression in RKHS to randomized covariate data, where $f$ was restricted on the Borel
measurable subspace $\mathcal {H}_B$.  In practical
applications, however, we often face the case that all functions in
the RKHS are Borel measurable.  In this case, we no longer need the
restriction mentioned in (\ref{HB}).  Thus, we would
like to proceed our discussion under the following condition\\

ASSUMPTION A.2.  Consider the Borel-$\sigma$ field of $\mathcal{H}$ (generated by the open sets).  Mapping:
\begin{eqnarray}
\mathscr{T} &\rightarrow & \mathcal {H} \nonumber \\
x  &\mapsto& K_x(\cdot) = K(\cdot, x) \nonumber
\end{eqnarray}
is Borel measurable for all $(\mathcal{X}_i, \mathcal {F}_i)$,
$i=1,...n$.  Here $K(\cdot, \cdot)$ denotes the reproducing kernel of $\mathcal{H}$. \\

Under A.2, by Theorem 90 of Berlinet and Thomas-Agnan
(2004)\cite{Berlinet2004}, Page 195, every function in $\mathcal
{H}$ is Borel measurable. It can be verified that
if the domain $\mathscr{T}\subseteq \mathbb{R}^d$ and every $\mathcal{F}_i$ is a
Borel $\sigma$-field, then A.2 is satisfied with
\begin{itemize}
\item Every
continuous kernel;
\item Kernels built from tensor sums or products of continuous kernels;
\item Any radial basis kernel $K(x,z) =
r(||x-z||_d)$ such that $r(\cdot)$ is continuous at 0.
Here $||\cdot||_d$ denotes the usual Euclidian norm.
\end{itemize}

\subsection{Quadrature penalized likelihood estimates}

As previously discussed, there is in general
 no finite dimensional subspace in which the RC-PLE estimate $f_\lambda$ is known {\it a priori} to lie, so direct computation is
 not attractive. In this case we shall find a finite dimensional
 approximating subspace and compute an estimator in this space.  We consider the following penalized likelihood:
\begin{equation}
\label{class}I_\lambda^{{Z}, \Pi}( f ) = -\frac{1}{n}
\sum_{i=1}^n \log \sum_{j=1}^{m_i} \pi_{ij} p(y_i|z_{ij}, f) +
\frac{\lambda}{2}{J}(f)
\end{equation}
where ${Z} = \{z_{11},...,z_{1m_1},z_{21},...,z_{nm_n}\} $ with $z_{ij}\in \mathscr{T}$ and $\Pi =
\{\pi_{11},...,\pi_{1m_1},\\ \pi_{21},...,\pi_{nm_n}\}$ with $\pi_{ij} >0$.  In words, when we evaluate the integrals on the right hand side of (\ref{plkr}), each $(\mathcal {X}_i,
\mathcal{F}_i, P_{i})$ is replaced by a discrete
probability distribution defined over
$\{z_{i1},z_{i2},...,z_{im_i}\}$ with probability mass function
$ P(x_i = z_{ij}) = \pi_{ij},~ j=1,...,m_i$.
Thus $z_{ij}, 1\le j\le m_i$ and $\pi_{ij}, 1\le j\le m_i$ are referred to as \textbf{nodes} and \textbf{weights} of a \textbf{quadrature rule} for probability measure $P_i$.

In (\ref{class}), $f$ is only evaluated on a finite number of quadrature nodes.  Under A.1, it can be
seen from Theorem 2.2 and the arguments in Kimeldorf and Wahba
(1971)\cite{Kimeldorf1971} that the minimizer of
$I_\lambda^{{Z}, \Pi}( f )$ in $\mathcal {H}$ is in a finite
dimensional subspace $\mathcal{H}_{Z}$ spanned by $\mathcal {H}_0 $ and $\{K(\cdot,
z_{ij}):z_{ij}\in {Z}\}$.  Thus, $I_\lambda^{{Z}, \Pi}( f )$ can be formulated as a parametric penalized likelihood.  Green (1990)\cite{Green1990} gave a general discussion on the use of the EM algorithm for parametric penalized likelihood estimation with incomplete data.  His method can be extended to minimize $I_\lambda^{{Z}, \Pi}( f )$.  It
can be shown that the E-step at iteration $t+1$ has the form of
\begin{equation}
\label{Q}Q(f|f^{(t)}) = \frac{1}{n}\sum_{i=1}^n \sum_{j=1}^{m_i}
w_{ij}^{(t)}\cdot \log p(y_i|z_{ij}, f) -\frac{\lambda}{2}{J}(f)
\end{equation}
where $f^{(t)}$ is estimated at iteration $t$ and the weight
\begin{equation}\label{weight1}
w_{ij}^{(t)} = \frac{\pi_{ij}p(y_i|z_{ij}, f^{(t)})}{ \sum_{k}\pi_{ik}p(y_i|z_{ik}, f^{(t)})}
\end{equation}
indicates the conditional probability of $[z_{ij}|y_i,f^{(t)}]$.
The M-step updates $f$ by maximizing $Q(f|f^{(t)})$ in $\mathcal{H}$.  This is straightforward because $-Q(f|f^{(t)})$ is seen to be a weighted complete data
penalized likelihood.

When the EM algorithm converges, we will obtain an estimator $\hat{f}_\lambda$ which approximates the RC-PLE estimate $f_\lambda$. Note that $\hat{f}_\lambda$ can be interpreted as a minimizer of $I^R_\lambda(f)$ when the integrals are approximated by quadrature rules.  Hence, this computational technique is referred to as \textbf{quadrature penalized likelihood estimation} or \textbf{QPLE}.  The motivation behind this approach is that an efficient quadrature rule often requires only a few nodes for a good approximation to the integral.  This convenient property eases the computation burden at each M-step.

\subsection{Construction of quadrature rules}

Construction of quadrature rules is a practical issue.  In order to derive more applicable results, we further assume that each $x_i = (x_{i1},...,x_{id})^T$ is a random vector, i.e., $\mathscr{T} \subset \mathbb{R}^d$.
\subsubsection{Univariate quadrature rules}
Suppose that $x_i$ is univariate (i.e., $d=1$).  In this case, if $x_i$ is a categorical random variable or exactly observed, then $(\mathcal {X}_i, P_{i})$ itself can be used as a quadrature rule.  Otherwise, if $x_i$ is a continuous random variable, we will construct a \textbf{Gaussian quadrature rule}.  Development of computational methods and routines of Gaussian quadrature integration formulae for probability measures
is a mathematical research topic.  We will not survey the general literature here, other than to say that the methods considered in this paper can be obtained from, Golub and Welsch (1969)\cite{Golub1969}, Fernandes and Atchley (2006)\cite{Fernandes2006}, Bosserhoff (2008)\cite{Bosserhoff2008} and Rahman (2009)\cite{Rahman2009}.  Though a $k$-node Gaussian quadrature rule typically requires the first $2k$ moments of the measure $P_{i}$ to be finite, this convention can be satisfied by most popular probability distributions including normal, uniform,  exponential,  gamma, beta and others.  Besides Gaussian quadrature rules, if $x_i$ has a density with respect to the Lebesgue measure,  we also consider a quadrature rule with equally-spaced points.  More specifically, suppose that $x_i$ ranges over $[a, b]$, then we take equally-spaced points in $[a,b]$ as quadrature nodes while  the quadrature weights are proportional to the density evaluated at the chosen nodes.  Note that if $a=-\infty$ (or $b=+\infty$), we set $a = \mu_i-3\sigma_i$ (or $b = \mu_i+3\sigma_i$)  where $\mu_i$ and $\sigma_i$ denote the first and second moments of $P_{i}$.  We refer to this simple quadrature rule as the \textbf{grid quadrature rule}.

\subsubsection{Multivariate quadrature rules}
Suppose that $x_i= (x_{i1},...,x_{id})^T$ is a multivariate random vector (i.e., $d>1$).  In this case, a quadrature rule can be generated recursively with one-dimensional conditional quadrature rules.  The algorithm is summarized as follows:
    \begin{itemize}
    \item[1.] Set $s=1$.  Compute the marginal distribution of $x_{i1}$ and generate a quadrature rule for $x_{i1}$ by using the method for univariate random variables.
    \item[2.] Let $\{z^{(s)}_{1},...,z^{(s)}_{m_s}\}$ and $\{\pi^{(s)}_{1},...,\pi^{(s)}_{m_s}\}$ be the quadrature rule generated for the marginal distribution of $(x_{i1},...,x_{is})^T$.  For each $z^{(s)}_{j}, 1\le j \le m_s$, compute the one-dimensional conditional distribution of
        \begin{equation}\nonumber
        [x_{i(s+1)} |(x_{i1},...,x_{is})^T = z^{(s)}_{j}]
        \end{equation}
        Then generate a quadrature rule for this distribution, denoted by $\{z^{*}_{j1},...,z^{*}_{j n_j}\}$ and $\{\pi^{*}_{j1},...,\pi^{*}_{j n_j}\}$.  Then $\{ ((z^{(s)}_{j})^T, z^{*}_{jr})^T, 1\le r \le n_j, 1\le j \le m_s\}$ and $\{ \pi ^{(s)}_{j}\cdot \pi^{*}_{jr} , 1\le r \le n_j, 1\le j \le m_s\}$ compose a quadrature rule for the marginal distribution of $(x_{i1},...,x_{is}, x_{i(s+1)})^T$.
    \item[3.] Set $s=s+1$.  Repeat step 2 until $s=d$.
    \end{itemize}
    The order that $x_{ij}$'s jump into the algorithm is not important.  One may rearrange the order to simplify the computation of the quadrature rules.
From our experience, a quadrature rule with 7 to 12 nodes for each component of $x_i$ usually yields a very good approximation.  In this case, the above EM algorithm usually converges very rapidly.

\subsection{Choice of the smoothing parameter}
\subsubsection{The comparative KL distance and leaving-out-one-subject CV}
So far the smoothing parameter $\lambda$, is assumed to be fixed.  Choice of $\lambda$ is a key problem in the penalized likelihood regression.  For non-Gaussian data, Kullback-Leibler (KL) distance is commonly used as the risk function for the estimator ${}{f}_\lambda$
\begin{equation}\label{KL}
\text{KL}(f^*, {}{f}_\lambda) = \frac{1}{n}\sum_{i=1}^{n}E_{y_i^0|f^*} \left\{
\log\frac{p(y_i^0|f^* )}{p(y_i^0| {}{f}_\lambda)}\right\}
\end{equation}
where $f^*$ denotes the true regression function and the expectation is taken over $y_i^0 \sim
p(y|f^*)$ independent of $y_i$.  In order to estimate $\text{KL}(f^*, {}{f}_\lambda)$,  Xiang and Wahba (1996)\cite{Xiang1996} proposed \textbf{generalized approximate cross validation} (GACV) beginning with a leaving-out-one argument to choose the smoothing parameter, which works well for Bernoulli data. Lin, Wahba, Xiang, Gao, Klein and Klein (2000)\cite{Lin2000} derived a randomized version of GACV (ranGACV) which is more computationally friendly for large data sets.  In this section we obtain a convenient form of leaving-out-one-subject CV for randomized covariate data and extend GACV and randomized GACV to randomized covariate data in subsequent sections.

In the situation when each observed covariate is actually a probability space $(\mathcal {X}_i, \mathcal{F}_i, P_{i})$, $[y_i^0|f]$ has a density of
\begin{equation}\label{denRCD}
p(y_i^0|f) = \int_{\mathcal {X}_i} p(y_i^0|x_i, f) d P_{i}{}.
\end{equation}
Following (\ref{KL}) and leaving out the quantities which do not depend on $\lambda$, the comparative KL (CKL) distance can be written as
\begin{equation}\label{CKL}
\text{CKL}(\lambda) = -\frac{1}{n}\sum_{i=1}^{n}E_{y_i^0|f^*} \left\{
\log  \int_{\mathcal {X}_i} \exp\left\{ y_i^0 {}{f}_\lambda({x_i}) - b({}{f}_\lambda({x_i})) \right\} d P_{i}{} \right\}.
\end{equation}
To simplify the notation, let's denote
\begin{equation}\label{L}
L(y,f, P_{i}) = \log  \int_{\mathcal {X}_i} \exp\left\{ y {f}({x_i}) - b({f}({x_i})) \right\} d P_{i}{}
\end{equation}
the log-likelihood function for randomized covariate data.  Using first order Taylor expansion to expand ${L}$ at the point $y_i$, we have that
\begin{eqnarray}\label{TaylorL}
\label{L}
{L}(y_i^0, {}{f}_\lambda, P_{i}) \approx \label{L}
{L}(y_i, {}{f}_\lambda, P_{i}) + (y_i^0 - y_i) \frac{\partial {L} }{\partial{y}}(y_i, {}{f}_\lambda, P_{i}).
\end{eqnarray}
Direct calculation yields
\begin{eqnarray}
 \frac{\partial {L} }{\partial{y}}(y_i, {}{f}_\lambda, P_{i}) &=& \frac{ \int_{\mathcal {X}_i} {}{f}_\lambda({x_i})  \exp\left\{ y_i {}{f}_\lambda({x_i}) - b({}{f}_\lambda({x_i})) \right\}  d P_{i}{}}{\int_{\mathcal {X}_i} \exp\left\{ y_i {}{f}_\lambda({x_i}) - b({}{f}_\lambda({x_i})) \right\} d P_{i}{}} \label{partialL1} \nonumber \\
 &=& E_{x_i|y_i, {}{f}_\lambda} {}{f}_\lambda(x_i)\label{partialL2}.
\end{eqnarray}
Plugging (\ref{TaylorL}) and (\ref{partialL2}) into (\ref{CKL}), we have that
\begin{eqnarray}
\text{CKL}(\lambda) &\approx& \text{OBS}(\lambda) + \frac{1}{n}\sum_{i=1}^{n}E_{y_i^0|f^*}(y_i - y_i^0) E_{x_i|y_i, {}{f}_\lambda} {}{f}_\lambda(x_i) \nonumber \\
& = & \text{OBS}(\lambda) + \frac{1}{n}\sum_{i=1}^{n}(y_i - \mu_i^*) E_{x_i|y_i, {}{f}_\lambda} {}{f}_\lambda(x_i) \label{APPCKL}
\end{eqnarray}
where $\mu_i^* = E_{y_i^0|f^*}y_i^0$ is the true mean response and
\begin{eqnarray} \label{OBS}
\text{OBS}(\lambda) = -\frac{1}{n} \sum_{i=1}^{n} \log  \int_{\mathcal {X}_i} \exp\left\{ y_i {}{f}_\lambda({x_i}) - b({}{f}_\lambda({x_i})) \right\} d P_{i}{}
\end{eqnarray}
is the observed log-likelihood.  Denote ${}{f}^{[-i]}_\lambda$ the leaving-out-one estimator, i.e., the minimizer of $I^R_\lambda(f)$ with the $i$th subject omitted.  Since $E_{x_i|y_i, {}{f}_\lambda} {}{f}_\lambda(x_i)$ is the posterior mean estimate of $f^*(x_i)$, following Xiang and Wahba (1996)\cite{Xiang1996}, we may replace
$\mu_i^* E_{x_i|y_i, {}{f}_\lambda} {}{f}_\lambda(x_i)$ by  $y_i E_{x_i|y_i, {}{f}^{[-i]}_\lambda} {}{f}^{[-i]}_\lambda(x_i)$ and define the leaving-out-one-subject cross validation (CV) by
\begin{equation}\label{CV}
\text{CV}(\lambda) = \text{OBS}(\lambda) + \frac{1}{n}\sum_{i=1}^{n} y_i ( E_{x_i|y_i, {}{f}_\lambda} {}{f}_\lambda(x_i) - E_{x_i|y_i, {}{f}^{[-i]}_\lambda} {}{f}^{[-i]}_\lambda(x_i)).
\end{equation}
It can be seen that (\ref{APPCKL}) and (\ref{CV}) generalize the complete data CKL and CV formulas proposed in Xiang and Wahba (1996)\cite{Xiang1996}. If $\hat{f}_\lambda$ denotes the QPLE estimate, then we may further approximate (\ref{CV}) by quadrature rules.  More specifically, $\text{OBS}(\lambda)$ can be evaluated by
\begin{equation} \label{OBS2}
\widehat{\text{OBS}}(\lambda) = -\frac{1}{n} \sum_{i=1}^{n} \log \sum_{j=1}^{m_i} \pi_{ij}\exp\left\{ y_i \hat{f}_\lambda(z_{ij}) - b(\hat{f}_\lambda(z_{ij}))\right\}
\end{equation}
where $z_{ij}$'s and $\pi_{ij}$'s represent nodes and weights of the quadrature rules given in the preceding section. Define the weight functions
\begin{equation}\label{Wfun}
w_{ij}(\tau) = \frac{\pi_{ij}\exp\left\{ y_i \tau_j - b(\tau_j)\right\} }{\sum_{k} \pi_{ik}\exp\left\{ y_i \tau_k - b(\tau_k)\right\}}, ~j=1,...,m_i
\end{equation}
where $\tau=(\tau_1,...,\tau_{m_i})^T$ is an arbitrary vector of length $m_i$. Let us use the notations
\begin{eqnarray}
&&\vec{f}_{\lambda i} = (\hat{f}_\lambda(z_{i1}), ..., \hat{f}_\lambda(z_{im_i}))^T \label{vf1}\\
&&\vec{f}^{~[-i]}_{\lambda i} = (\hat{f}^{[-i]}_\lambda(z_{i1}), ..., \hat{f}^{[-i]}_\lambda(z_{im_i}))^T \label{vf2}.
\end{eqnarray}
Then (\ref{partialL1}) yields
\begin{eqnarray}
&&E_{x_i|y_i, \hat{f}_\lambda} \hat{f}_\lambda(x_i) \approx  \sum_{j=1}^{m_i}w_{ij}(\vec{f}_{\lambda i})\hat{f}_\lambda(z_{ij}) =  \sum_{j=1}^{m_i}w_{\lambda, ij}\hat{f}_\lambda(z_{ij}) \label{EW1} \\
&&E_{x_i|y_i, \hat{f}^{[-i]}_\lambda} \hat{f}^{[-i]}_\lambda(x_i)  \approx  \sum_{j=1}^{m_i} w_{ij}(\vec{f}^{~[-i]}_{\lambda i})\hat{f}^{[-i]}_\lambda(z_{ij}) = \sum_{j=1}^{m_i} w^{[-i]}_{\lambda,ij}\hat{f}^{[-i]}_\lambda(z_{ij})  \label{EW2}
\end{eqnarray}
where $w_{\lambda, ij} =  w_{ij}(\vec{f}_{\lambda i})$ and $w^{[-i]}_{\lambda,ij} = w_{ij}(\vec{f}^{[-i]}_{\lambda i}) $
equal the weights at the final iteration of the EM algorithm, respectively, when $\hat{f}_\lambda$ and $\hat{f}^{[-i]}_\lambda$ were computed.  Therefore a more convenient version of CV can be obtained as
\begin{equation}\label{CV2}
\text{CV}(\lambda) \approx \widehat{\text{OBS}}(\lambda) + \frac{1}{n}\sum_{i=1}^{n} y_i \sum_{j=1}^{m_i}(w_{\lambda, ij}\hat{f}_\lambda(z_{ij}) - w^{[-i]}_{\lambda,ij}\hat{f}^{[-i]}_\lambda(z_{ij})).
\end{equation}

\subsubsection{Parametric formulation of $I_\lambda^{{Z}, \Pi}$}
Based on (\ref{CV2}) and by using several first order Taylor expansions, a generalized approximate cross validation (GACV) can be derived for randomized covariate data.  Before we proceed, we would like to establish some notations.

As we previously discussed, $I_\lambda^{{Z}, \Pi}( f )$ can be formulate parametrically as
\begin{equation}\label{class2}
I_\lambda^{{Z}, \Pi}(\vec{y}, \vec{f}~) =  -\frac{1}{n}
\sum_{i=1}^n \log \sum_{j=1}^{m_i} \pi_{ij} p(y_i|f_{ij}) + \frac{\lambda}{2} \vec{f}^{~T} \Sigma_\lambda \vec{f}
\end{equation}
where $\vec{f} = (f_{11},...,f_{1m_1}, f_{21},...,f_{nm_n})^T$ denotes the vector of $f$ evaluated at $\{z_{ij}, 1\le i\le n, 1\le j \le m_i\}$, $\vec{y} = (\vec{y}^{~T}_1...,\vec{y}^{~T}_n)^T$  with $\vec{y}_i = (y_i,...,y_i)^T$ being $m_i$ replicates of $y_i$ and $\Sigma_\lambda$ is the positive semi-definite matrix satisfying $\lambda{J}(f) = \vec{f}^{~T} \Sigma_\lambda \vec{f}$.  Note that minimizing $I_\lambda^{{Z}, \Pi}( f )$ in $\mathcal{H}$ is equivalent to minimizing $I_\lambda^{{Z}, \Pi}(\vec{y}, \vec{f}~)$ in $\mathbb{R}^{m_1+\cdots+m_n}$.  Hence $\vec{f}_\lambda = (\hat{f}_\lambda(z_{11}),...,\hat{f}_\lambda(z_{1m1}),\hat{f}_\lambda(z_{21}),...,\\ \hat{f}_\lambda(z_{nm_n}))^T$ minimizes (\ref{class2}).  Similarly, we can denote $\vec{f}_\lambda^{~[-i]} = (\hat{f}^{[-i]}_\lambda(z_{11}),...,\\ \hat{f}^{[-i]}_\lambda(z_{1m_1}),\hat{f}^{[-i]}_\lambda(z_{21}),...,\hat{f}^{[-i]}_\lambda(z_{nm_n}))^T$ the minimizer of (\ref{class2}) with $i$th subject omitted.

\subsubsection{Generalized average of submatrices, randomized estimator}
To define the GACV and randomized GACV we use the concept of {\bf generalized average} of submatrices and its {\bf randomized estimator} introduced in Gao, Wahba, Klein and Klein (2001)\cite{Gao2001} for the multivariate outcomes case.  Let $A$ be a square matrix with submatrices $A_{ii}, 1\le i\le n$ on the diagonal. Denote $A_{ii}  =  (a^i_{st})_{m_i\times m_i}, 1\le s, t\le m_i$.  Because $A_{ii}$'s may have different dimensions, we calculate for each $A_{ii}$
\begin{equation}\label{delta}
\delta_i   =  \frac{1}{nm_i} \sum_{k=1}^{n}\sum_{j=1}^{m_k} a^k_{jj} = \frac{1}{nm_i} tr(A)
\end{equation}
and
\begin{equation}\label{gamma}
\gamma_i = \left\{
             \begin{array}{ll}
               0, & \text{ if } m_i=1  \\
               1/(nm_i(m_i-1))\sum_{k=1}^{n} \sum_{s\neq t} a^k_{st}, & \text{ if } m_i>1.
             \end{array}
           \right.
\end{equation}
Then the generalized average of $A_{ii}$ is defined by
\begin{equation} \label{AMatrix}
\bar{A}_{ii} = (\delta_i - \gamma_i) I_{m_i\times m_i} + \gamma_i\cdot e_{i}e_{i}^T = \left(
                                                                              \begin{array}{cccc}
                                                                                \delta_i & \gamma_i & \cdots & \gamma_i \\
                                                                                \gamma_i & \delta_i & \cdots & \gamma_i \\
                                                                                \vdots & \vdots & \ddots & \vdots \\
                                                                                \gamma_i & \gamma_i & \cdots & \delta_i \\
                                                                              \end{array}
                                                                            \right)
\end{equation}
where $e_{i} = (1,1...,1)^T$ is the unit vector of length $m_i$.  In this case, the inverse of $\bar{A}_{ii}$ can be easily obtained by
\begin{equation}\label{IAmatrix}
\bar{A}_{ii}^{-1} = \frac{1}{\delta_i -\gamma_i}I_{m_i\times m_i} - \frac{\gamma_i}{(\delta_i - \gamma_i)(\delta_i+(m_i-1)\gamma_i) }~ e_ie_i^T.
\end{equation}

Now we discuss how to obtain a randomized estimator of $\bar{A}_{ii}$.  Let $\epsilon = (\epsilon_{1}^T,...,\epsilon_n^T)^T$, where $\epsilon_i = (\epsilon_{i1},...,\epsilon_{im_i})^T$ with each $\epsilon_{ij}$ generated independently from $N(0,\sigma^2)$.  Denote $\bar{\epsilon} = (\bar{\epsilon}_1,...,\bar{\epsilon}_1,\bar{\epsilon}_2,...,\bar{\epsilon}_n)^T$ the corresponding mean vector with $m_i$ replicates of $\bar{\epsilon}_i$ for each $1\le i \le n$, where $\bar{\epsilon}_i = 1/\sqrt{m_i}\sum_{j=1}^{m_i}\epsilon_{ij}$.  Then we observe the following facts
\begin{eqnarray}
&& E\epsilon^T A \epsilon  =  \sigma\cdot tr(A)\\
&& E\left\{\bar{\epsilon}^T A \bar{\epsilon} - \epsilon^T A \epsilon \right\}= \sigma \cdot \sum_{k=1}^{n} \sum_{s\neq t} a^k_{st}.
\end{eqnarray}
Thus, a randomized estimate of $\bar{A}_{ii}$ can be obtained by replacing $\delta_i$ and $\gamma_i$ with their unbiased estimates $\frac{1}{nm_i\sigma}\epsilon^T A \epsilon$ and $\frac{1}{nm_i(m_i-1)\sigma}(\bar{\epsilon}^T A \bar{\epsilon} - \epsilon^T A \epsilon)$.

\subsubsection{The GACV and randomized GACV}

We now present the result of GACV as follow. Details of the derivation can be found in Appendix B.   Denote $H$ the influence matrix of (\ref{class2}) with respect to $\vec{f}$ evaluated at $\vec{f}_\lambda$.  Write
\begin{equation}\label{H}
H = \left(
      \begin{array}{cccc}
          H_{11} & * & * & * \\
          * & H_{22} & \cdots & * \\
          \vdots & \vdots & \ddots & \vdots \\
          * & * & \cdots & H_{nn} \\
      \end{array}
    \right)_{\sum m_i\times \sum m_i}
\end{equation}
where each $H_{ii}$ is a $m_i\times m_i$ submatrix matrix on the diagonal with respect to to $(f_{i1},...,f_{im_i})^T$.  Define ${W}_i = \text{diag}(b''(\hat{f}_\lambda(z_{i1})),...,b''(\hat{f}_\lambda(z_{im_i})))$ the diagonal matrix of estimated variances. Let ${W} =\text{diag}({W}_1,...,{W}_n)$ be the ``big'' variance matrix for all the observations.  Denote $G = I - H{W}$ with submatrices $G_{ii} = I_{m_i\times m_i} - H_{ii}{W}_i, 1\le i\le n$ on the diagonal.    Now let $\bar{H}_{ii}$ and $\bar{G}_{ii}$ denote the generalized average of submatrices $H_{ii}$ and $G_{ii}$.  Then the generalized approximate cross validation (GACV) can be written as
\begin{equation}
\text{GACV}(\lambda) = \widehat{\text{OBS}}(\lambda) + \frac{1}{n}\sum_{i=1}^{n} y_i (d_{i1},...,d_{im_i}) \bar{G}_{ii}^{-1} \bar{H}_{ii} \left(
                                                                                                                                     \begin{array}{c}
                                                                                                                                      y_i-\hat{\mu}_{\lambda} (z_{i1}) \\
                                                                                                                                      \vdots \\
                                                                                                                                       y_i- \hat{\mu}_{\lambda} (z_{im_i})  \\
                                                                                                                                     \end{array}
                                                                                                                                   \right)
\end{equation}
where $\hat{\mu}_{\lambda} (z_{ij}) = b'(\hat{f}_\lambda(z_{ij}))$ denote the estimated mean response and
\begin{equation}\label{WFDV}
d_{ij} = w_{\lambda, ij} \left[ (y_i - \hat{\mu}_{\lambda}(z_{ij}))(\hat{f}_\lambda(z_{ij}) - \sum_{k=1}^{m_i}w_{\lambda, ik}\hat{f}_\lambda(z_{ik}))+1\right].
\end{equation}

In practice, however, computation of the influence matrix $H$ for large data sets is expensive and may be unstable.  Note that, in order to compute $\bar{H}_{ii}$ and $\bar{G}_{ii}$, we only need the sum of traces and the sum of off-diagonal entries of $H_{ii}$'s and $G_{ii}$'s.  Therefore, the exact computation of $H$ and $G$ can be avoided using randomized estimates of $\bar{H}_{ii}$ and $\bar{G}_{ii}$.  To do this, we first generate a random perturbation vector $\epsilon = (\epsilon_{1}^T,...,\epsilon_{n}^T)^T$, where $\epsilon_i = (\epsilon_{i1},...,\epsilon_{im_i})^T$ and $\epsilon_{ij}$'s are iid from $N(0,\sigma^2)$.  Then compute the mean vector $\bar{\epsilon}= (\bar{\epsilon}_1,...,\bar{\epsilon}_1,\bar{\epsilon}_2,...,\bar{\epsilon}_n)^T$ where $\bar{\epsilon}_i = 1/\sqrt{m_i}\sum_{j=1}^{m_i}\epsilon_{ij}$.  Denote $\vec{f}^{~\vec{y}+\epsilon}_\lambda$ and $\vec{f}^{~\vec{y}+\bar{\epsilon}}_\lambda$ the minimizers of (\ref{class2}) with the perturbed data $\vec{y}+\epsilon$ and $\vec{y}+\bar{\epsilon}$. Similarly, denote $\vec{f}^{~\vec{y}}_\lambda(=\vec{f}_\lambda)$ the minimizer with the original data. To ease the computational burden, we can set $\vec{f}^{~\vec{y}}_\lambda$ as the initial value for the EM algorithm of $\vec{f}^{~\vec{y}+\epsilon}_\lambda$ and $\vec{f}^{~\vec{y}+\bar{\epsilon}}_\lambda$.  Because $H$ is the influence matrix, we have that
\begin{equation}
\vec{f}^{~\vec{y}+\epsilon}_\lambda \approx  \vec{f}^{~\vec{y}}_\lambda + H\epsilon,~~
\vec{f}^{~\vec{y}+\bar{\epsilon}}_\lambda \approx  \vec{f}^{~\vec{y}}_\lambda + H\bar{\epsilon}.
\end{equation}
This yields
\begin{equation}
  \epsilon^T H\epsilon \approx \epsilon^T(\vec{f}^{~\vec{y}+\epsilon}_\lambda - \vec{f}^{~\vec{y}}_\lambda),~~
\bar{\epsilon}^T H\bar{\epsilon} \approx  \bar{\epsilon}^T(\vec{f}^{~\vec{y}+\bar{\epsilon}}_\lambda -\vec{f}^{~\vec{y}}_\lambda).
\end{equation}
Thus, a randomized estimate of $\bar{H}_{ii}$ can be obtained as we previously described.  Also it is straightforward to show that
\begin{equation}
  \epsilon^T G \epsilon \approx \epsilon^T\epsilon - \epsilon^T {W} (\vec{f}^{~\vec{y}+\epsilon}_\lambda - \vec{f}^{~\vec{y}}_\lambda),~~
\bar{\epsilon}^T G \bar{\epsilon} \approx  \bar{\epsilon}^T\bar{\epsilon} - \bar{\epsilon}^T {W} (\vec{f}^{~\vec{y}+\bar{\epsilon}}_\lambda - \vec{f}^{~\vec{y}}_\lambda)
\end{equation}
which implies a randomized estimate of $\bar{G}_{ii}$.  In order to reduce the variance of randomized trace estimates, one may draw $R$ independent perturbation vectors $\epsilon^{1},...,\epsilon^{R}$ and compute for each $\epsilon^{r}$ the randomized estimates $\hat{\bar{H}}_{ii}^{r}, 1\le i\le n$ and $\hat{\bar{G}}_{ii}^{r}, 1\le i\le n$.  Then the ($R$-replicated) ranGACV function is\\

\begin{equation}\label{ranGACV}
\text{ranGACV}(\lambda) =  \widehat{\text{OBS}}(\lambda) + \frac{1}{nR}\sum_{r=1}^{R}\sum_{i=1}^{n} y_i (d_{i1},...,d_{im_i}) (\hat{\bar{G}}_{ii}^{r})^{-1}  \hat{\bar{H}}_{ii}^{r} \left(
                                                                                                                                     \begin{array}{c}
                                                                                                                                      y_i-\hat{\mu}_{\lambda} (z_{i1}) \\
                                                                                                                                      \vdots \\
                                                                                                                                       y_i- \hat{\mu}_{\lambda} (z_{im_i})  \\
                                                                                                                                     \end{array}
                                                                                                                                   \right).
\end{equation}

\section{Covariate measurement error (model)}
Covariate measurement error is a common occurrence in many experimental settings including surveys, clinical trials and medical studies.  Suppose that $x_i=(x_{i1},...,x_{id})^T$ takes values in the real space $\mathbb{R}^d$.  In the presence of measurement error, $x_i$ is not directly observed but instead
$
x^{err}_i = x_i + {u}_i
$
is observed, where ${u}_i,1\le i\le n$ are iid random errors, independent of $(y_i, x_i)$.  To estimate the regression function, our idea is to treat measurement error as a special case of randomized covariates.  More specifically, each $x_i$ is considered as a random vector distributed as $x^{err}_i-{u}_i$.  When the error distribution is known, the distribution for $x_i$ can be obtained immediately, and therefore RC-PLE can be directly employed without any extra effort.

However, in practical applications, we often face the case that the error distribution is unknown. One common approach in the measurement error literature is to assume a parametric model for the error density and to estimate the unknown parameters from the data.  Let $p({u}_i|\theta)$ denote the specified error density indexed by a real vector $\theta$ ranging over $\Theta \subseteq \mathbb{R}^q$ and let $F({u}_i|\theta)$ denote the corresponding c.d.f. function.  Since our goal is to estimate the regression function, $\theta$ is treated as a nuisance parameter.  Given $(f, \theta)$,  $y_i$ has a marginal density of
\begin{equation}\label{dME}
p(y_i|f, \theta) = \int_{\mathbb {R}^d} p(y_i|x^{err}_i-{u}_i, f)p({u}_i|\theta)d{u}_i.
\end{equation}
Thus RC-PLE can be extended by
\begin{equation}\label{plke}
I_{\lambda}^{E} (f, \theta)=  -\frac{1}{n}\sum_{i=1}^n \log
\int_{\mathbb {R}^d} p(y_i|x^{err}_i-{u}_i, f)p({u}_i|\theta)d{u}_i+
\frac{\lambda}{2}{J}(f).
\end{equation}
In this case, we still need Assumption A.1 to obtain the existence of the penalized likelihood estimate.  In addition we state the following extra assumption which can be satisfied with most parametric models for the error distribution. \\

ASSUMPTION B.1. The c.d.f. function $F({u}|\theta)$ is continuous in $\theta$ for any ${u} \in \mathbb{R}^d$ and the parameter space $\Theta$ is compact.\\


Now we can show the existence of penalized likelihood estimate by the following Theorem which is actually a corollary to Theorem 2.2.\\

THEOREM 4.1.  \emph{Under A.1, A.2 and B.1, there exist
$f_\lambda \in \mathcal{H}$ and $ \theta_\lambda \in \Theta$ such
that
$
I^E_\lambda(f_\lambda, \theta_\lambda) = \inf_{f\in \mathcal {H},
\theta \in \Theta} I^E_\lambda(f,\theta)
$.
}\\

\textbf{Proof} See Appendix A. $~\Box$

\section{Covariate measurement error (computation)}
In order to compute an estimator, we extend QPLE described in Section 3.1 as follows. Denote $(f^{(t)}, \theta^{(t)})$ the parameters estimated at iteration $t$.  Let $z_{j}^{(t)},  1\le j\le m$ and $\pi_{j}^{(t)}, 1\le j\le m$
denote the quadrature rule based on the density function $p({u}|\theta^{(t)})$.  Note that the quadrature rules can be generated using the method introduced in Section 3.1.  It is not hard to see that the E-step at iteration $t+1$ is to compute the expectation of the penalized likelihood $-\frac{1}{n}\sum_{i=1}^n \log p(y_i|x^{err}_i-{u}_i, f)p({u}_i|\theta)+
\frac{\lambda}{2}{J}(f) $ with respect to the conditional distributions $[{u}_i | y_i, x^{err}_i, f^{(t)}, \theta^{(t)}], 1\le i \le n$.  Using the quadrature rule, each $[{u}_i | y_i, x^{err}_i, f^{(t)}, \theta^{(t)}]$ can be approximated by a discrete distribution with support $\{z_{j}^{(t)},  1\le j\le m\}$ and mass function $P({u}_i = z_{j}^{(t)}) = w^{(t)}_{ij}$, where
\begin{equation}\label{WeightE}
w^{(t)}_{ij} = \frac{\pi_{j}^{(t)} p(y_i|x^{err}_i-z^{(t)}_{j}, f^{(t)})}{\sum_{k}\pi_{k}^{(t)} p(y_i|x^{err}_i-z^{(t)}_{k}, f^{(t)})}.
\end{equation}
Thus the E-step can be written as
\begin{eqnarray}
Q(f,\theta|f^{(t)}, \theta^{(t)}) = \frac{1}{n}\sum_{i=1}^n \sum_{j=1}^{m}
w_{ij}^{(t)}\cdot \log p(y_i|x^{err}_i - z^{(t)}_{j}, f) -\frac{\lambda}{2}{J}(f) &&\nonumber \\
+ \frac{1}{n}\sum_{i=1}^n \sum_{j=1}^{m}
w_{ij}^{(t)}\cdot \log p(z^{(t)}_{j}|\theta). && \label{QE}
\end{eqnarray}
Then the M-step maximizes $Q(f,\theta|f^{(t)}, \theta^{(t)})$, which can be done by separately maximizing a complete data penalized likelihood of $f$
\begin{equation}
\frac{1}{n}\sum_{i=1}^n \sum_{j=1}^{m}
w_{ij}^{(t)}\cdot \log p(y_i|x^{err}_i - z^{(t)}_{j}, f) -\frac{\lambda}{2}{J}(f)
\end{equation}
and a complete data log likelihood of $\theta$
\begin{equation}
\frac{1}{n}\sum_{i=1}^n \sum_{j=1}^{m}
w_{ij}^{(t)}\cdot \log p(z^{(t)}_{j}|\theta).
\end{equation}
Therefore the M-step becomes a standard problem which can be solved by much existing software.  When the EM algorithm converges, we will obtain the QPLE estimate $(\hat{f}_\lambda, \hat{\theta}_\lambda)$.

Finally, we show how to select the smoothing parameter $\lambda$ in the case of covariate measurement error.  Note that our goal is to construct a good estimator of $f$, and $\theta$ is treated as a nuisance parameter.  In other words, we only care about the goodness of fit of the $\hat{f}_\lambda$.  Therefore $\lambda$ can be selected in the same way as randomized covariate data.  To do this, we first estimate the error distribution by $p({u}_i|\hat{\theta}_\lambda)$ and then determine each covariate distribution $P_{i}$ according to the relation $x_i = x^{err}_i - {u}_i$.  After that, the method introduced in Section 3.3 can employed directly for the choice of $\lambda$.

Correcting for measurement error is a broad statistical research topic.  In the interest of space, we only discuss the situation when we have a parametric model for the error distribution.  It would be possible to extend our method to other situations of measurement error.  For example, when additional data is available, such as a sample from the error distribution or repeated observations for some $x_i$, we may estimate the error distribution more accurately by using other approaches.  Also, sometimes, the parametric model $p({u}_i|\theta)$ may not be available and in this case, we may want to estimate the error distribution nonparametrically.  These are interesting topics for future research.

\section{Missing covariate data (model)}

Now we describe penalized likelihood regression with missing covariate data.  We assume the missing mechanism to be missing at
random.

\subsection{Notations and model}
Let $x_i=(x_{i1},...,x_{id})$ denote the vector of covariates ranging over a subspace of $\mathbb{R}^d$.
By the idea of Ibrahim's method of weights (Ibrahim, 1990\cite{Ibrahim1990} and Ibrahim, Lipsitz and Chen, 1999\cite{Ibrahim1999}), we first assume a parametric
model for the marginal density of $x_i$, denoted as $p(x_i|\theta)>0$, where
$\theta\in \Theta \subseteq \mathbb{R}^q$ is a real vector of
indexing parameters.  Here $\theta$ is treated as a nuisance parameter.

Write $x_i =( x_i^{obs}, x_i^{mis})$ where $x_i^{obs}$ is a vector of observed
components and $x_i^{mis}$ is a $d_i\times 1$ vector of
missing components.  Following Little and Rubin, (2002)\cite{Little2002}, the likelihood of $(f, \theta)$ can be obtained by integrating or summing out the missing components in the joint
density for $(y_i, x_i)$
\begin{eqnarray}
\label{lkm}L(f, \theta) = \sum_{i=1}^n \log \int_{\mathbb{R}^{d_i}}
p(y_i|x_i, f)p(x_i|\theta) d x_i^{mis}
\end{eqnarray}
where $\int_{\mathbb{R}^{d_i}}
p(y_i|x_i, f)p(x_i|\theta) d x_i^{mis}\equiv p(y_i|x_i, f)p(x_i|\theta)$ if $x_i$ is completely observed.
Then $(f, \theta)$ can be estimated by minimizing the following missing data penalized likelihood:
\begin{eqnarray}
\label{plkm}I_\lambda^M(f,\theta) = -\frac{1}{n} \sum_{i=1}^n \log \int_{\mathbb{R}^{d_i}}
p(y_i|x_i, f)p(x_i|\theta) d x_i^{mis} +
\frac{\lambda}{2}{J}(f).
\end{eqnarray}

We note that this method can be viewed as an extension of RC-PLE.
%
Define $P_{i,mis}^{\theta}$ the probability measure over $\mathbb{R}^{d_i}$, with respect to the conditional density of $[x_i^{mis}|x_i^{obs}]$
\begin{equation}\label{measure}
p(x_i^{mis}|\theta, x_i^{obs}) = \frac{p( x_i |\theta)}{\int_{\mathbb{R}^{d_i}}p(x_i|\theta)dx_i^{mis}},~~x_i^{mis} \in \mathbb{R}^{d_i}.
\end{equation}
Note that
(\ref{measure}) is well-defined since
$\int_{\mathbb{R}^{d_i}}p(x_i|\theta)dx_i^{mis}<\infty$ from the Fubini's
Theorem.  Let
\begin{equation}\label{dirac}
\delta_{x_i^{obs}}(A) = \left\{
                       \begin{array}{ll}
                         1 & \text{if } x_i^{obs} \in A \\
                         0 & \text{if } x_i^{obs} \notin A
                       \end{array}
                     \right.
\end{equation}
denote the dirac measure defined for $x_i^{obs}$.  Consider the product measure $P_{i}^{\theta} = \delta_{x_i^{obs}} \times P_{i,mis}^{\theta}$ which satisfies that for any Borel sets $A_1\subset \mathbb{R}^{d-d_i}$, $A_2\subset \mathbb{R}^{d_i}$ and their Cartesian product $A_1\times A_2$, we have
\begin{equation}\label{measureG}
P_{i}^{\theta} (A_1\times A_2) = \delta_{x_i^{obs}}(A_1) \cdot P_{i,mis}^{\theta}(A_2).
\end{equation}
Then it is not hard to see that
\begin{equation}
\label{plkm2}I_\lambda^M(f,\theta) = -\frac{1}{n}\sum_{i=1}^n\log
\int_{\mathbb{R}^d} p(y_i|x_i, f) dP_{i}^\theta  +
\frac{\lambda}{2}{J}(f) - \frac{1}{n} \sum_{i=1}^n \log
\int_{\mathbb{R}^{d_i}}p(x_i|\theta)dx_i^{mis}
\end{equation}
is composed of a randomized covariate penalized likelihood of $f$ and a log-likelihood of $\theta$.  Hence missing covariate data can be treated as a special case of
randomized covariate data, allowing covariate distributions to be flexible.

\subsection{Existence of the estimator}

The following assumptions can
be easily satisfied in the most experimental settings. \\

ASSUMPTION M.1. $\mathcal{D}^\theta_i = \{x_i^{mis} \in \mathbb{R}^{d_i}~:~p(x_i|\theta) >0 \}$ is compact for all $1\le i\le n$ and $\theta \in \Theta$.\\

ASSUMPTION M.2. The density function $p(x|\theta)$ is continuous in $\theta$ for any $x\in \mathbb{R}^d$ and the
parameter space $\Theta$ is compact.\\

The existence of the penalized likelihood estimate can be guaranteed by the following Theorem which is actually a corollary to Theorem 2.2. \\

THEOREM 6.1.  \emph{Under A.1, A.2, M.1 and M.2, there exist
$f_\lambda \in \mathcal{H}$ and $ \theta_\lambda \in \Theta$ such
that
$
I^M_\lambda(f_\lambda, \theta_\lambda) = \inf_{f\in \mathcal {H},
\theta \in \Theta} I^M_\lambda(f,\theta)
$.
}\\

\textbf{Proof} See Appendix A. $~\Box$

\section{Missing covariate data (computation)}

In order to compute an estimator, we can extend QPLE in the same way as covariate measurement error.   Denote $(f^{(t)}, \theta^{(t)})$ the parameters estimated at iteration $t$.  Let $z_{ij}^{(t)},  1\le j\le m_i$ and $\pi_{ij}^{(t)}, 1\le j\le m_i$ denote the quadrature rule based on the probability measure $P_{i}^{\theta^{(t)}}$ defined in (\ref{measureG}).  Then the E-step at iteration $t+1$ can be written as
\begin{eqnarray}
Q(f,\theta|f^{(t)}, \theta^{(t)}) = \frac{1}{n}\sum_{i=1}^n
\sum_{j=1}^{m_i} w_{ij}^{(t)}\cdot \log p(y_i|z_{ij}^{(t)}, f)
-\frac{\lambda}{2}{J}(f) \nonumber && \\ + \frac{1}{n} \sum_{i=1}^n \sum_{j=1}^{m_i}
w_{ij}^{(t)}\cdot \log p(z_{ij}^{(t)}|\theta) && \label{Qfun2}
\end{eqnarray}
where
\begin{equation}
w_{ij}^{(t)} =\frac{\pi_{ij}^{(t)}
p(y_i|z_{ij}^{(t)}, f^{(t)})}{\sum_{k}\pi_{ik}^{(t)}
p(y_i|z_{ik}^{(t)}, f^{(t)})}.
\end{equation}
Then the M-step can be done by separately maximizing
\begin{equation}
\label{M2}\frac{1}{n}\sum_{i=1}^n \sum_{j=1}^{m_i} w_{ij}^{(t)}\cdot
\log p(y_i|z_{ij}^{(t)}, f) -\frac{\lambda}{2}{J}(f)
\end{equation}
and
\begin{equation}
\frac{1}{n} \sum_{i=1}^n\sum_{j=1}^{m_i} w_{ij}^{(t)}\cdot \log
p(z_{ij}^{(t)}|\theta)
\end{equation}
which is computationally straightforward assuming the log-concavity
of $p(x|\theta)$ as a function of $\theta$.  Again, when the EM algorithm converges, the QPLE estimate $(\hat{f}_\lambda, \hat{\theta}_\lambda)$ can be obtained.

In order to select the smoothing parameter, we note that $\theta$ is a nuisance parameter and the choice of $\lambda$
only depends on the goodness of fit of $\hat{f}_\lambda$. Therefore,
we may select $\lambda$ in the same way as randomized covariate data.  This is straightforward, since we can take $P_{i}^{\hat{\theta}_\lambda}$ defined in (\ref{measureG}) as the covariate distribution.  After that the method in Section 3.3 can employed directly.

Following Ibrahim, Lipsitz and Chen (1999)\cite{Ibrahim1999}, our method can also be extended to the non-ignorable missing data mechanism.  In this case, we may specify a parametric model for the missing data mechanism and incorporate it into the penalized likelihood.  The extension is similar but more complicated.  Thus this is another topic for future research.

\section{Numerical Studies}\label{numer}

In this section, we illustrate our method by several simulated examples with covariate measurement error and missing covariates.  For each simulated data set, we will compare: (a) QPLE; (b) full data analysis before measurement error or missing covariates;  and (c) naive estimator that ignores measurement error or leaves out the observations with missing covariates.  Note that the choice of the smoothing parameter has strong effect on the penalized likelihood estimator.  Hence in order to show the potential gain of our method, for each data set, $\lambda$ is selected by both ranGACV and the optimal value that minimizes the Theoretical Kullback-Leibler distance (TKL), which does not depend on the nuisance parameter $\theta$.
\begin{equation}\label{TKL}
\text{TKL}  = \frac{1}{n}\sum_{i=1}^{n}E_{y_i^0|x_i, f^*} \left\{
\log\frac{p(y_i^0|x_i, f^* )}{p(y_i^0|x_i, \hat{f})}\right\}
\end{equation}
where $f^*$ is the true regression function, $\hat{f}$ denotes its estimator and $x_i$ denotes the true covariate vector before measurement error or 'missing'.  Note that tuning by minimizing TKL is only available in a simulation study when the "truth" is known.

Our numerical studies focus on Poisson distribution and Bernoulli distribution which are also the cases in our real data set.
The goal is to illustrate:
\begin{itemize}
\item the gain of QPLE;
\item the performance of ranGACV;
\item the robustness of QPLE to the choice of quadrature rules.
\end{itemize}
All the simulations are conducted using R-2.9.1 installed in Red Hat Enterprise Linux 5.

\subsection{Examples of measurement error}

Cubic spline regression is perhaps the most popular case of penalized likelihood regression.  We consider the following examples from Binomial and Poisson distributions:

\begin{itemize}
\item [(i)] $p(y|x) = \left(
                           \begin{array}{c}
                             2 \\
                             y
                           \end{array}
                         \right)
p(x)^y(1-p(x))^{2-y},  ~y=0,1,2$, where
\begin{equation}
p(x) = 0.63 x \cos(2\pi x) + 0.36\nonumber;
\end{equation}
\item [(ii)] $p(y|x) = \Lambda(x)^y e^{-\Lambda(x)}/y!,~y =  0,1,2...$, where
\begin{equation}
\Lambda(x) =  16e^{-18(x-0.4)^2} - 5e^{-7(x-0.5)^2} + 5\nonumber;
\end{equation}
\item [(iii)] Same distribution as (ii) except
\begin{equation}
\Lambda(x) = 10^6(x^{11}(1-x)^6)+10^4(x^3(1-x)^{10}) + 2\nonumber
\end{equation}
which is a modification of Example 5.5 of Gu (2002)\cite{Gu2002}.
\end{itemize}
In each case, we take $X\sim U[0,1]$ and generate a sample of $n=101$ $(x,y)$ pairs.
For each sample generated, measurement errors are created with the following scheme. We first randomly select five $(x,y)$ pairs as complete observations and then in the rest of the 96 pairs, random errors are generated by $x_i+{u}_i$, where ${u}_i$'s are iid either $\text{N}(0,\sigma^2)$ or $\text{U}[-\delta,\delta]$ for various values of the noise-to-signal ratio $\text{var}({u})/\text{var}(X)$.
For each generated data set, QPLE is conducted using either the Gaussian quadrature rule or the grid quadrature rule, where the Gaussian quadrature rule is computed by the \emph{statmod} package in R-2.9.1.  Note that we generate the same number of nodes for each noisy $x_i$.  Simulation results are summarized by the following figures.

Figure \ref{fig1} shows the estimated curves from one simulated data set of case (i) with normal error and $\text{var}({u})/\text{var}(X) = 0.25$.
QPLE is computed via Gaussian quadrature where 11 nodes are created for each noisy $x_i$.  Panel (c) plots for each regression method the box plot of theoretical Kullback-Leibler distances (\ref{TKL}) calculated from 100 repeated simulations. We also report in (d) the TKL distances calculated in the same simulation setting except that ${u}$ is uniform (with the same noise-to-signal ratio). \\

\begin{figure}[htbp]
\centerline{
\includegraphics[width=2.8in,height=1.8in]{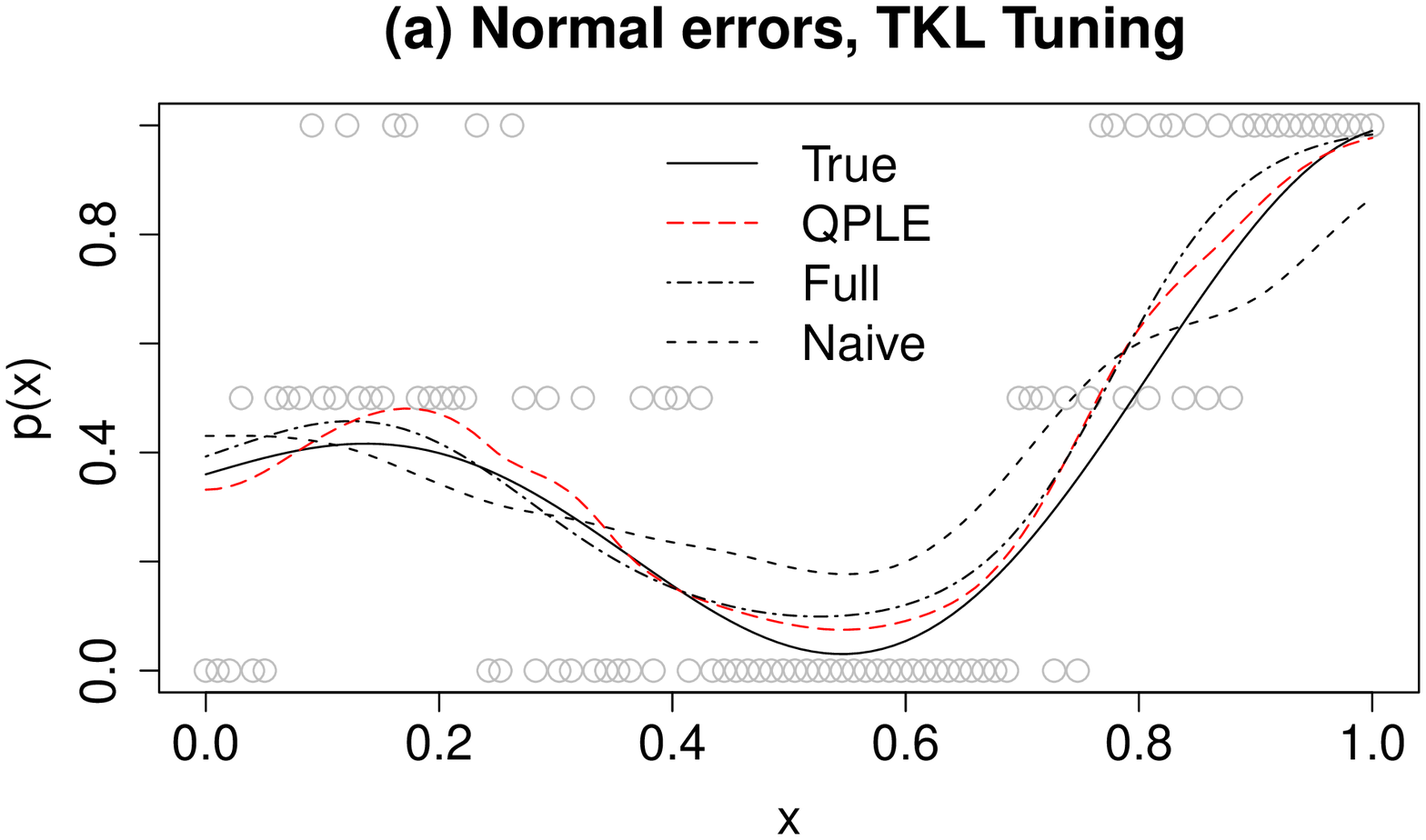}\includegraphics[width=2.8in,height=1.8in]{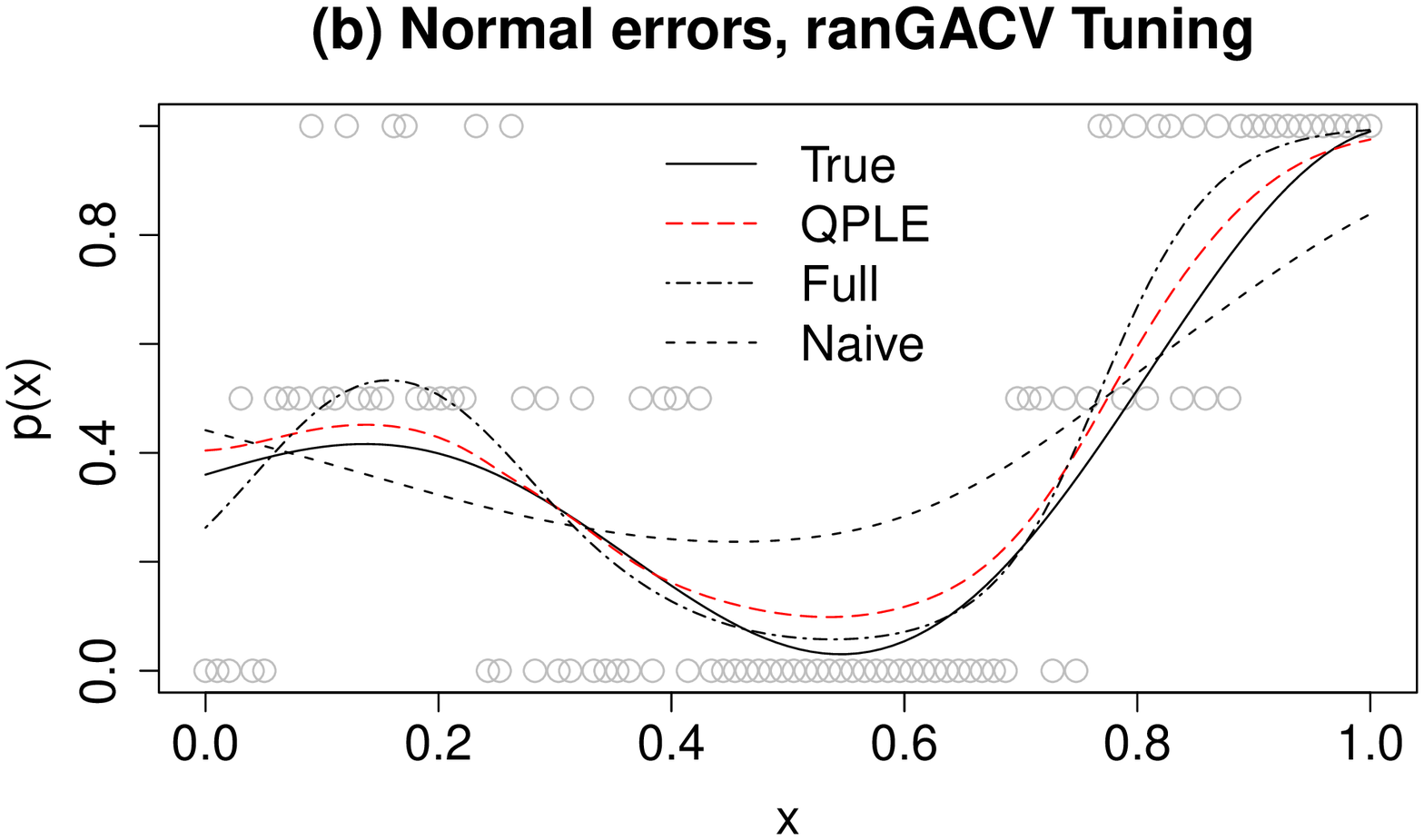}
}
\centerline{
\includegraphics[width=2.8in,height=1.8in]{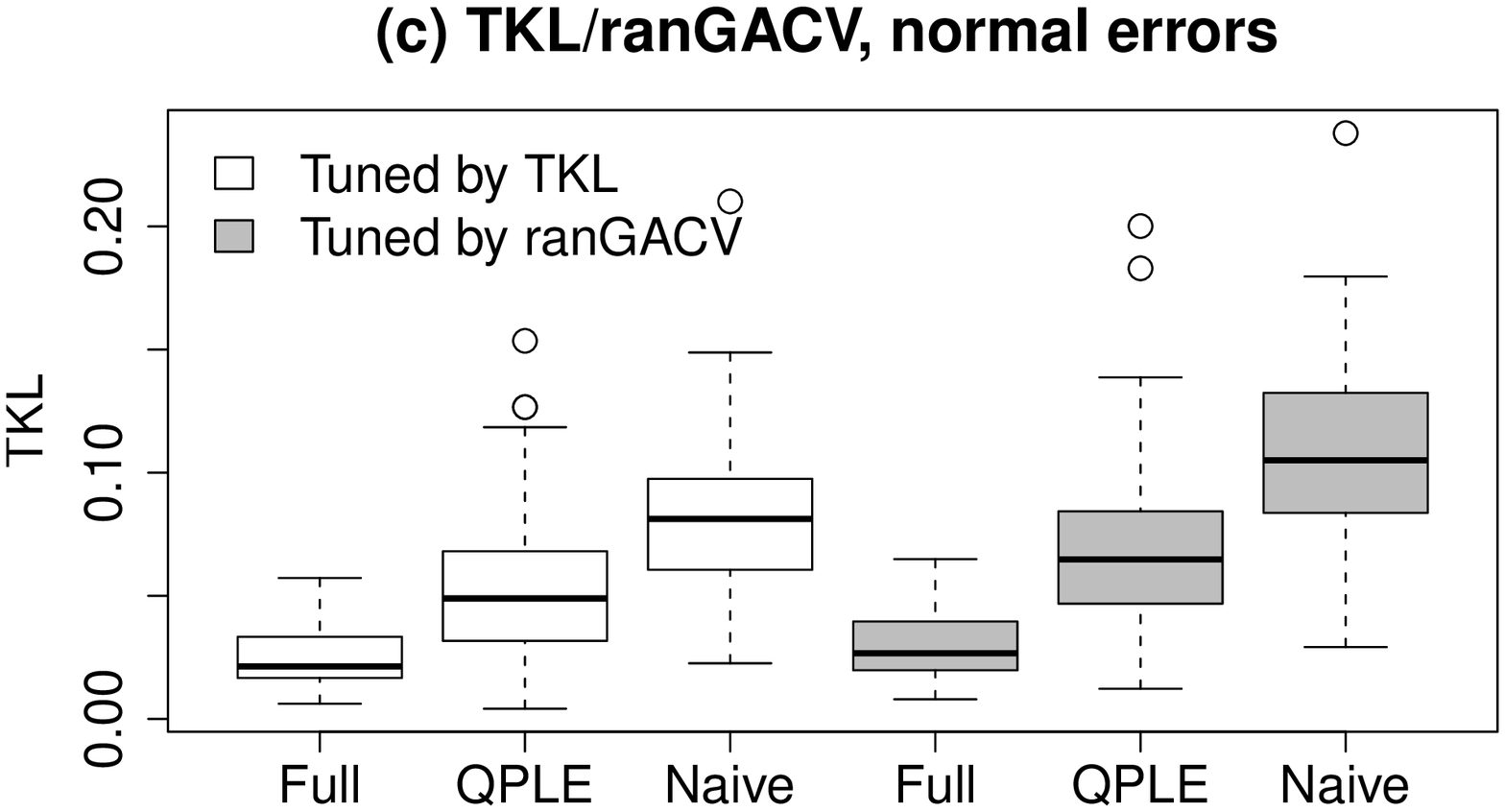}\includegraphics[width=2.8in,height=1.8in]{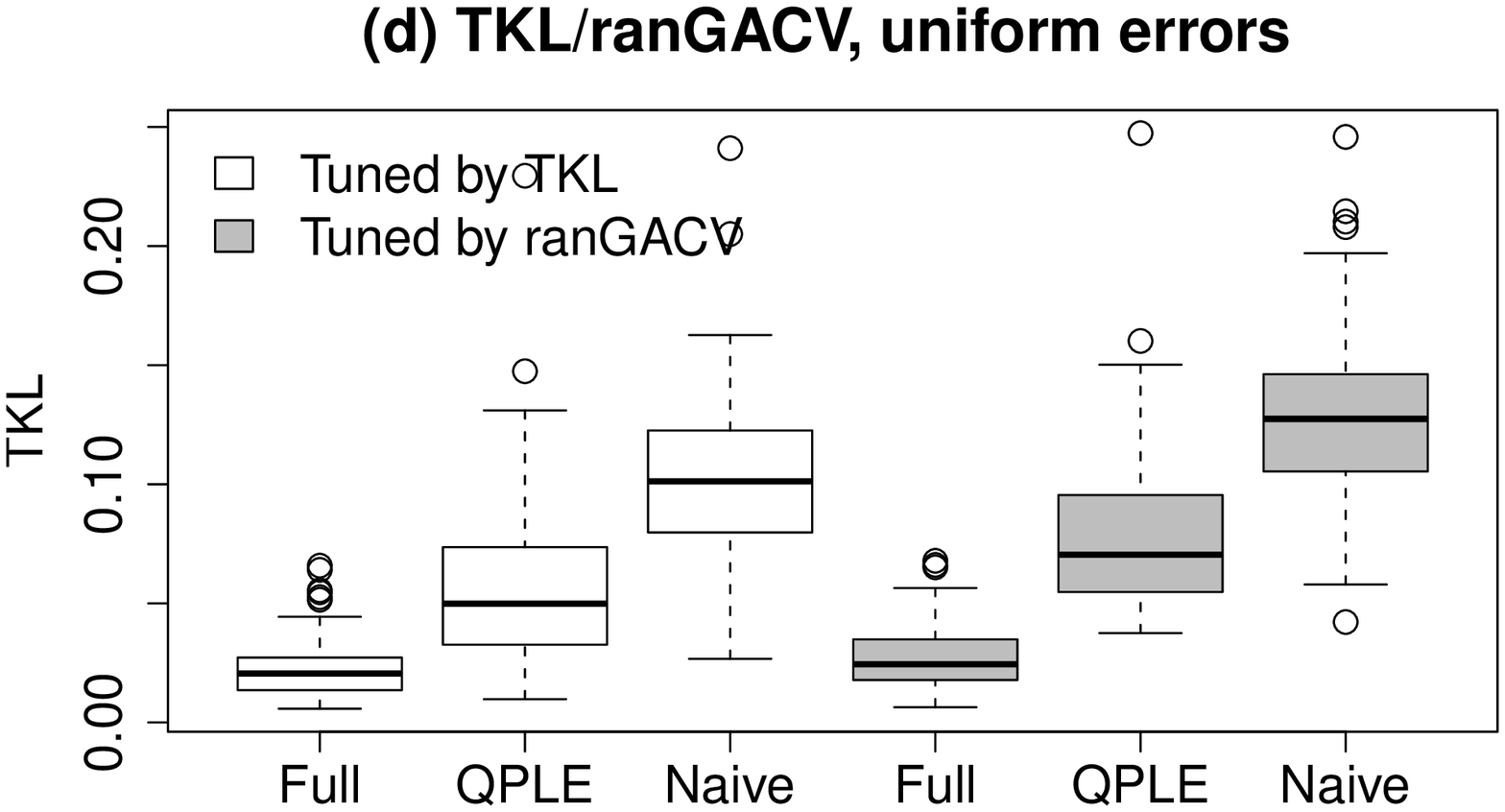}
}
\caption{Estimated curves and TKL distances for case (i). Panels (a) and (b) compare the target (True) curve, and three estimated curves obtained from the full data analysis (Full), the QPLE estimate, and the Naive estimate. (a) Tuning: TKL, (b) Tuning: ranGACV. In (a) and (b) ${u}\sim N(0, 0.145^2)$, assumed known. Panels (c) and (d) provide  plots of TKL distances. (c) ${u}\sim N(0,0.145^2)$, assumed known. (d) ${u}\sim U[-0.25,0.25]$, assumed known.}\label{fig1}
\end{figure}

\textbf{Remark 1} \emph{Throughout Section 8,
the choice of the curves to display from the
various 100 simulations is primarily subjective but deemed to be typical of
the bulk of the visual images of the comparisons between the estimates.
An idea of the scatter in the TKL distances
over the 100 simulations may be seen in the box plots.}\\

\begin{figure}[htbp]
\centerline{
\includegraphics[width=2.8in,height=1.8in]{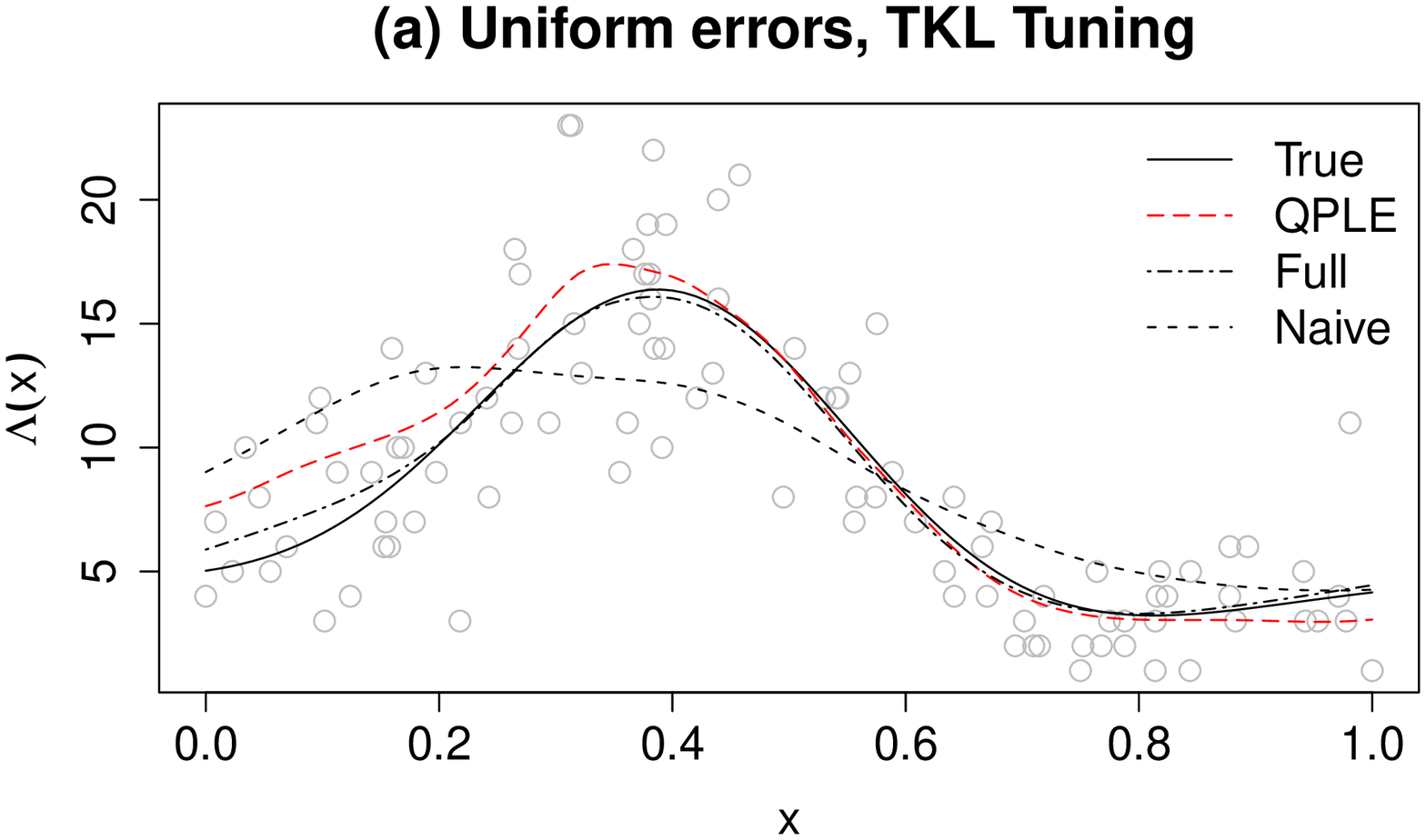}\includegraphics[width=2.8in,height=1.8in]{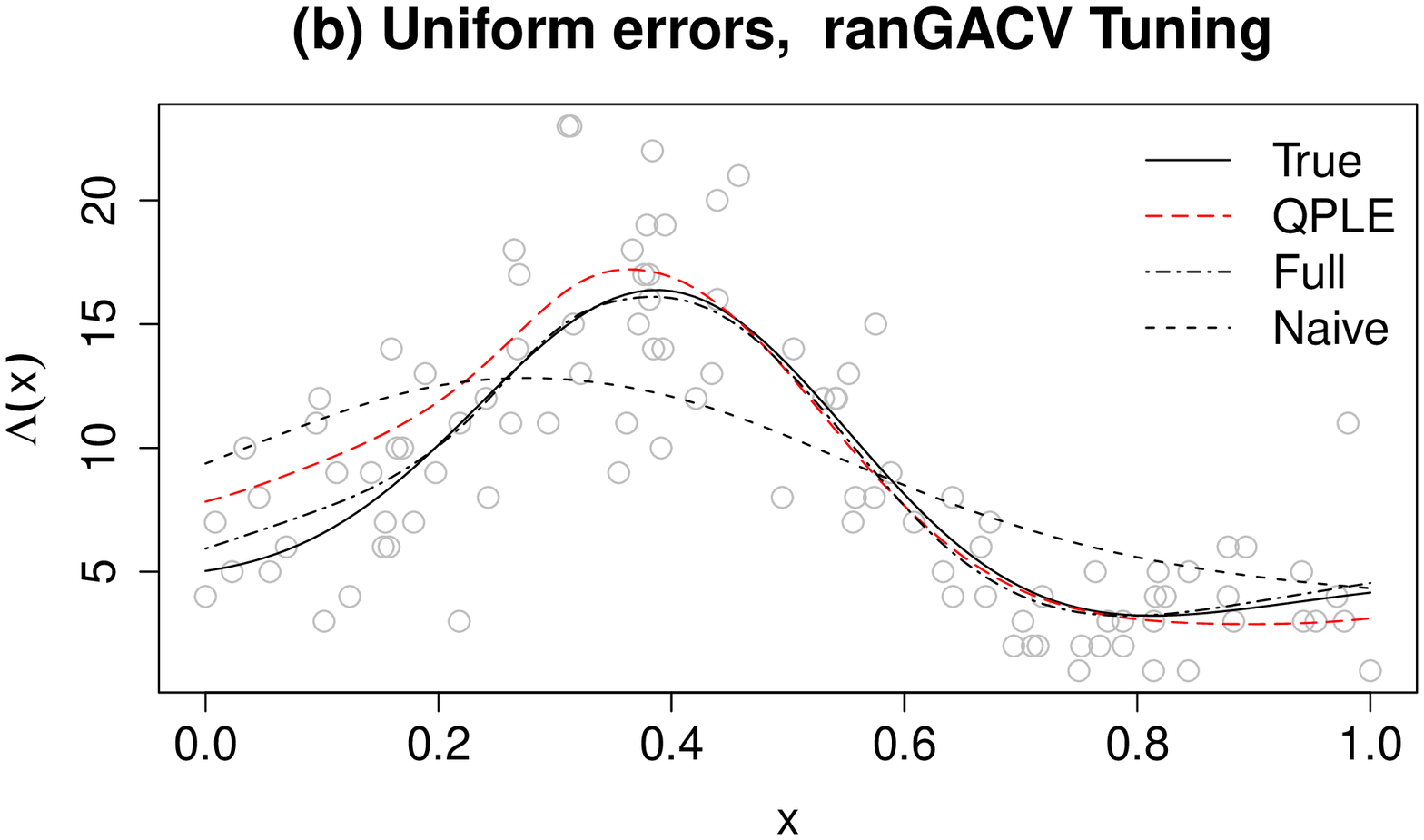}
}
\centerline{
\includegraphics[width=2.8in,height=1.8in]{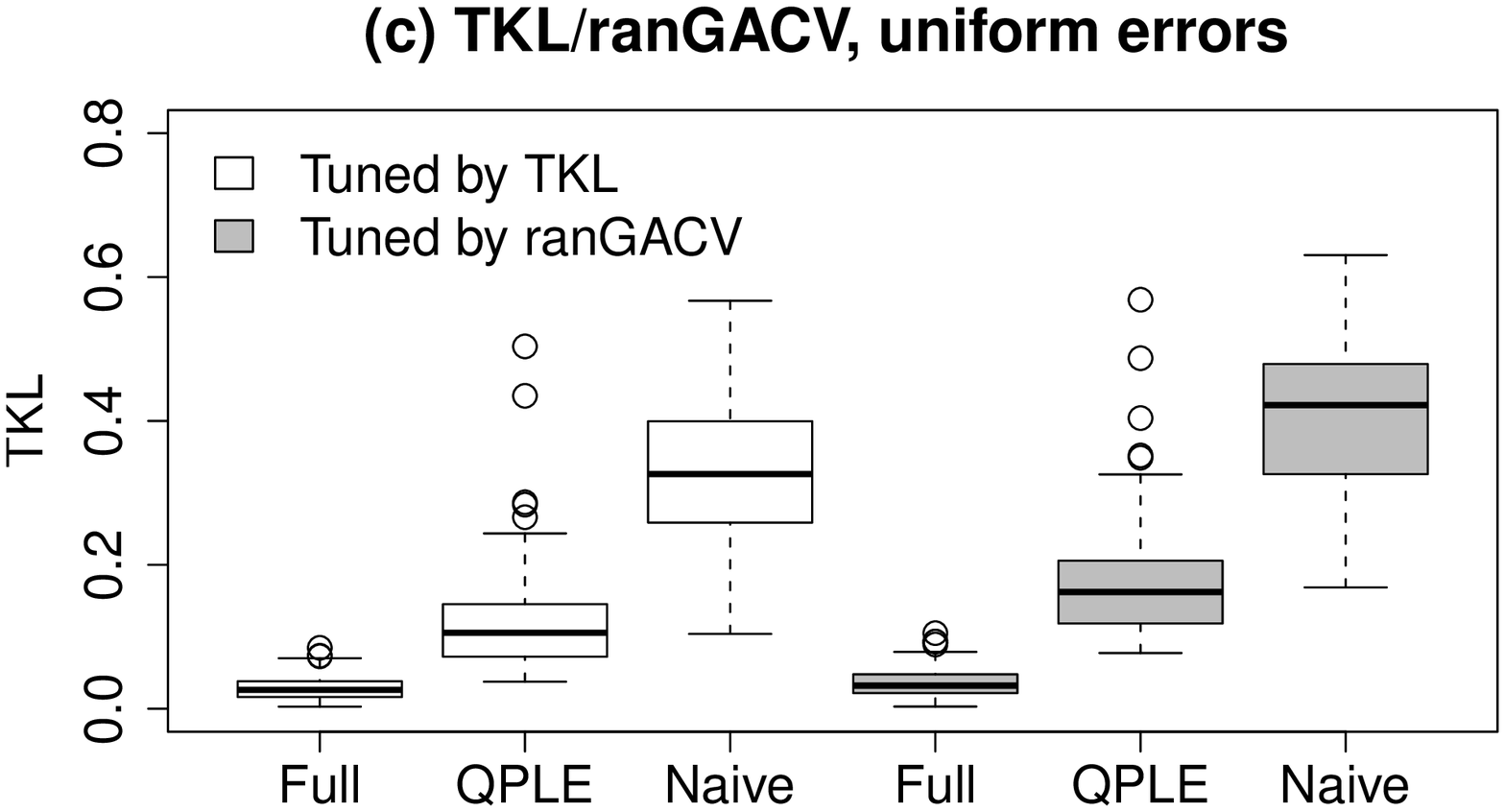}\includegraphics[width=2.8in,height=1.8in]{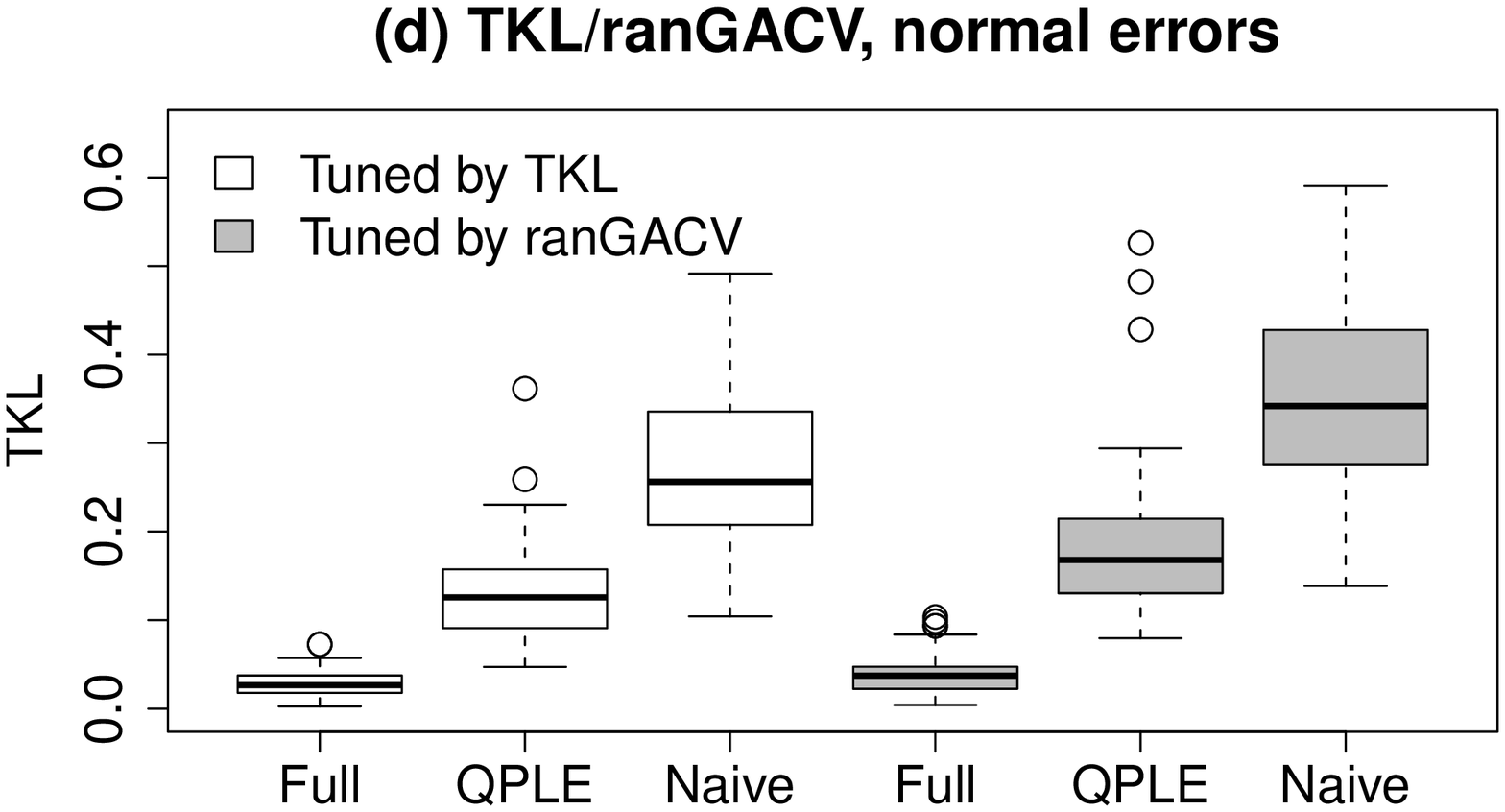}
}
 \caption{Estimated curves and TKL distances for case (ii).
Panels (a) and (b) compare the target (True) curve,
and three estimated curves obtained from the full data
analysis (Full), the QPLE estimate, and the Naive estimate.
(a) Tuning: TKL, (b) Tuning: ranGACV. In (a) and (b)
${u}\sim U[-0.273, 0.273]$, $\delta = 0.273$ assumed unknown.
Panels (c) and (d) provide  plots of TKL distances.
(c) ${u}\sim U[-0.273,0.273]$, $\delta = 0.273$ assumed unknown.
(d) ${u}\sim N(0,0.158^2)$, $\sigma=0.158$ assumed unknown.}\label{fig2}
\end{figure}

Figure \ref{fig2} shows the estimated curves from one simulation for case (ii) with uniform error and $\text{var}({u})/\text{var}(X) = 0.3$.  We assume that $\delta$ is unknown when QPLE is conducted.  At each EM iteration, we use Gaussian quadrature and create 9 nodes for each noisy $x_i$.  Panel (c) shows the TKL distances from 100 simulations. Panel (d) is obtained in the same simulation setting except that ${u}$ is normal (with the same noise-to-signal ratio), $\sigma$ is unknown.

Our results indicate the significant gain of QPLE, when the smoothing parameter is selected by either TKL or ranGACV. As we previously discussed, QPLE incorporates the information about the error distribution and hence is more informative.  Generally speaking, when measurement errors are ignored, the estimated curve of naive method tends to be oversmoothed and more biased near the modes and boundaries.  Similar phenomenon has been noted for other nonparametric regression methods, for example, Local polynomial estimate, as in Delaigle, Fan and Carroll (2009)\cite{Delaigle2009}.  For the choice of smoothing parameter, the proposed ranGACV inherits the property of traditional ranGACV. As simulations suggest, it is capable of picking $\lambda$ close to its optimal value even when $\theta$ is estimated.

\begin{figure}[htbp]
\centerline{
\includegraphics[width=2.8in,height=1.8in]{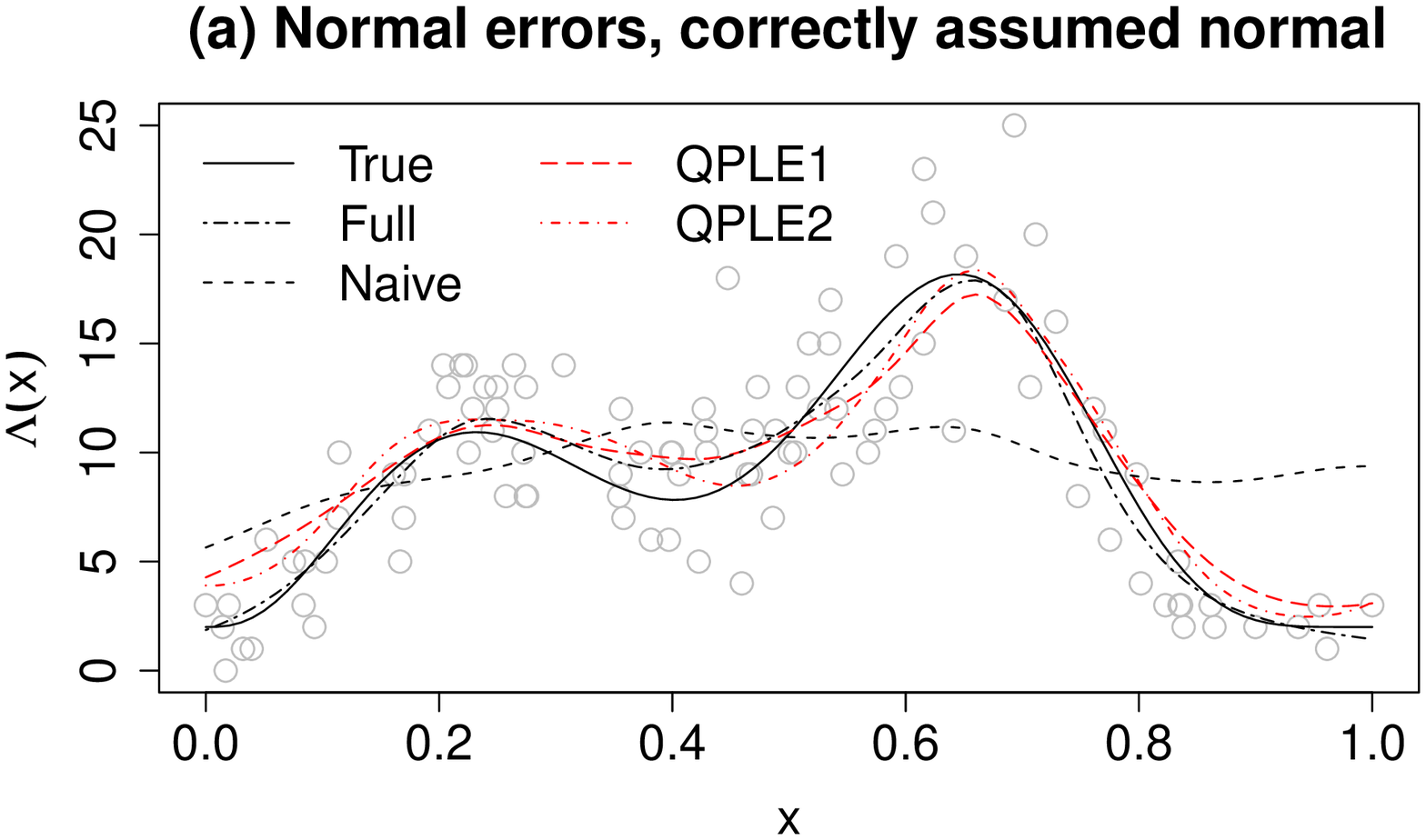}\includegraphics[width=2.8in,height=1.8in]{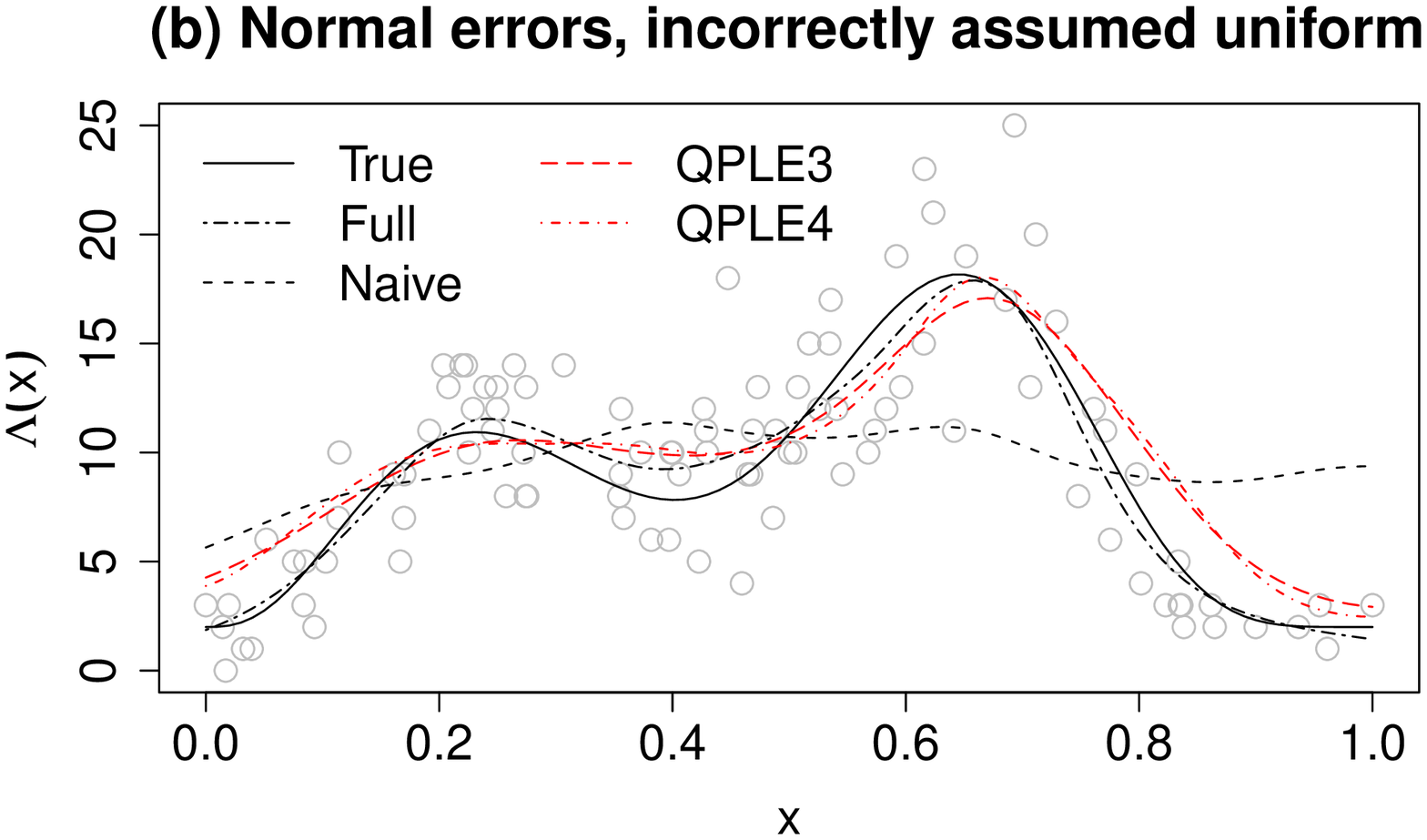}
}
\centerline{
\includegraphics[width=2.8in,height=1.8in]{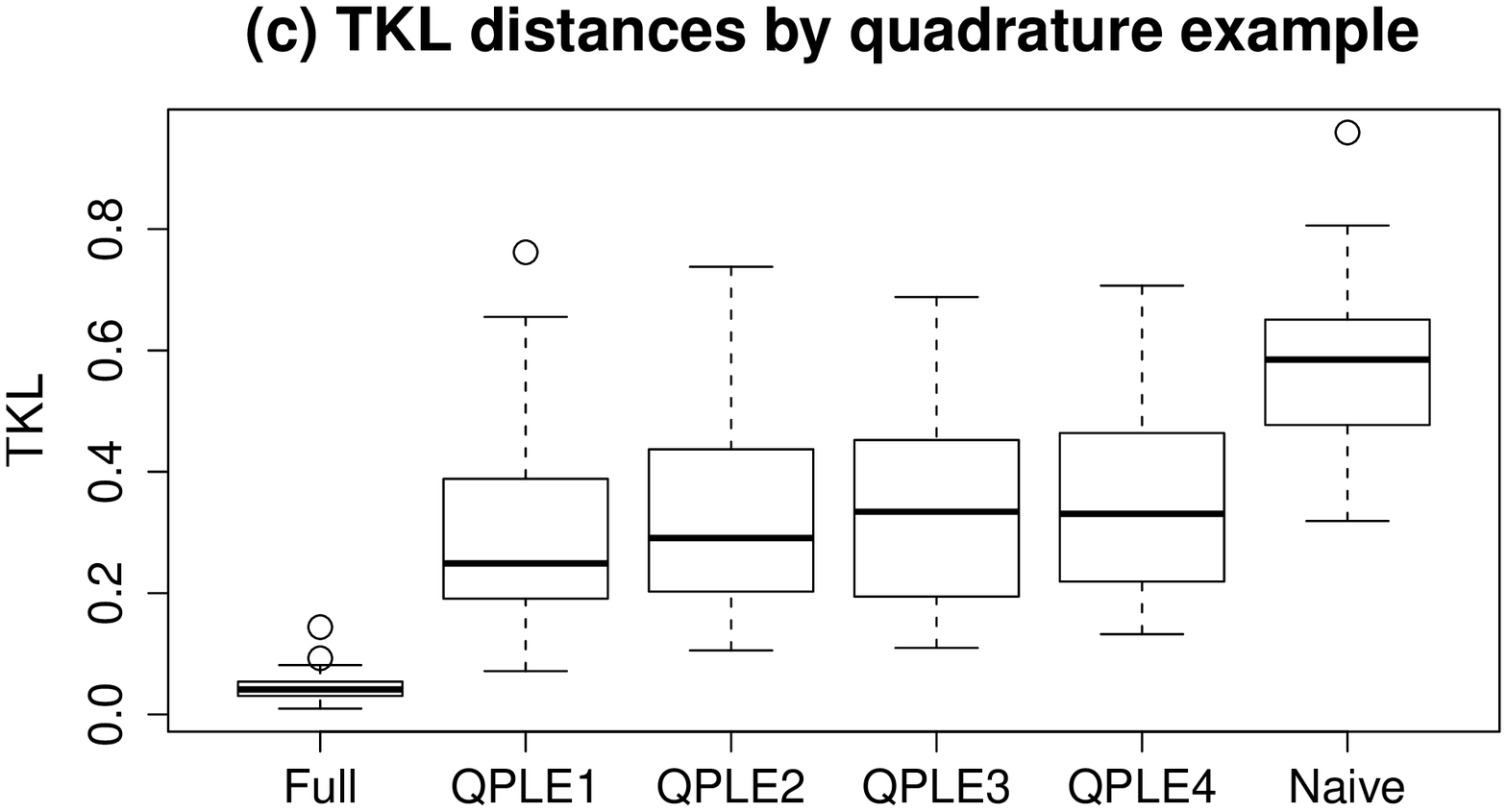}\includegraphics[width=2.8in,height=1.8in]{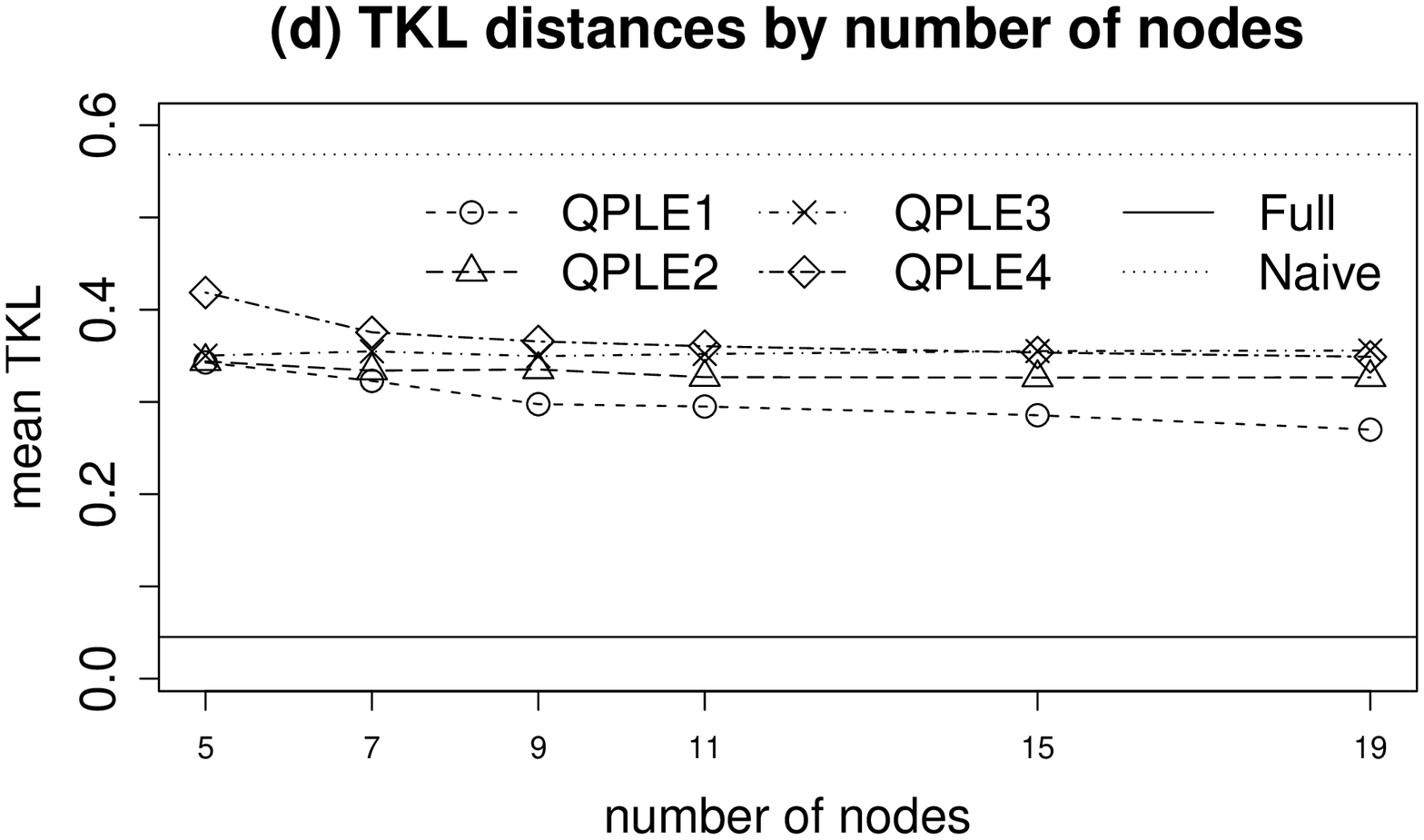}
}
\caption{Estimated curves and TKL distances for case (iii).
$u~\sim N(0, 0145^2)$, assumed unknown. Tuning: TKL.
Panels (a) and (b) give the target curve, and estimated curves from Full and Naive estimate.
Panel (a) compares the Gaussian quadrature (QPLE1) and
the grid quadrature (QPLE2) when the errors are correctly assumed
to be zero-mean normal (with unknown variance), and panel (b) compares the
Gaussian quadrature (QPLE3) and the grid quadrature (QPLE4)
when the errors are incorrectly assumed to be uniform
(with unknown range); (a) and (b) use 11 nodes.
Panel (c) plots TKL distances, using 11 nodes.
Panel (d) plots mean TKL versus number of nodes.
The dotted upper and solid lower lines represent the mean TKL
for the naive method and the full data analysis.}\label{fig3}.
\end{figure}

We summarize the influence of quadrature rules on QPLE at Figure \ref{fig3},  using case (iii) with normal error and $\text{var}({u})/\text{var}(X)= 0.25$.  In the computation, $\text{var}({u})$ is assumed to be unknown and $\lambda$ is selected by TKL.  We consider four QPLE estimators (QPLE1, QPLE2, QPLE3 and QPLE4) computed via, respectively, Gaussian quadrature, grid quadrature, Gaussian quadrature when ${u}$ is wrongly assumed to be uniform and grid quadrature when ${u}$ is wrongly assumed to be uniform.  We first compare these quadrature rules by setting the number of nodes (for each noisy $x_i$) to be 11.  The top two panels show the estimated curves from one simulation and panel (c) reports the TKL distances calculated from 100 simulations.  Then we study the influence of the number of the nodes.  On panel (d), we plot for each quadrature the mean TKL distance (based on 100 simulations) versus the number of nodes.  From the simulation results, we observed no significant difference between Gaussian quadrature and grid quadrature, though, as we expected, Gaussian quadrature is more efficient. Surprisingly, even with a wrong error distribution prespecified, the potential gain of QPLE is still significant.  Hence we may say that QPLE is robust to the choice of the quadrature.  We also note that QPLE does not require a large number of quadrature nodes to compute a good estimator.  There is not much gain to create more nodes if we already have enough.  Hence, in our numerical experiments, we generally compute 7-12 nodes for each noisy or missing component in the covariates.

\subsection{Examples of missing covariate data}

\begin{figure}[htbp]
\centerline{
\includegraphics[width=4.2in,height=2.7in]{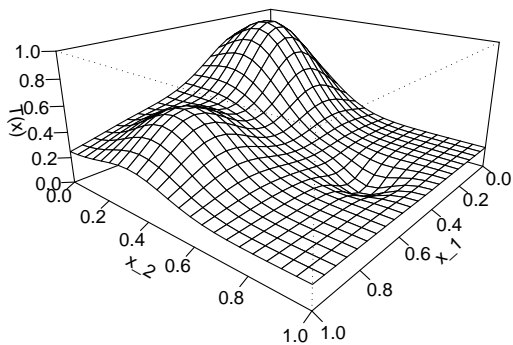}
}
 \caption{Franke's principal test function}\label{figFranke}
\end{figure}

In this section, we consider Franke's ``principal test function"
\begin{eqnarray}\label{Franke}
{T}(x) = \frac{3}{4} e^{-((9x_1-2)^2 + (9x_2-2)^2)/4} +
                      \frac{3}{4} e^{-((9x_1+1)^2/49 + (9x_2+1)^2/10)} && \nonumber \\
+\frac{1}{2}e^{-((9x_1-7)^2 + (9x_2-3)^2)/4} -
                                  \frac{1}{5} e^{-((9x_1-4)^2 + (9x_2-7)^2)}&&
\end{eqnarray}
which was used as a test function of smoothing splines in Wahba (1983)\cite{Wahba1983}.  $T(x)$ is shown in Figure \ref{figFranke}.
Consider the following examples
\begin{itemize}
\item[(i)] Binomial distribution: $p(y|x) =
\Bigl(
  \begin{array}{c}
    5 \\
    y \\
  \end{array}
\Bigr)
p(x)^y (1-p(x))^{5-y}
$,
where
\begin{equation}
p(x) =  \frac{1}{1.24}({T}(x) + 0.198);
\end{equation}
\item[(ii)] Poisson distribution: $p(y|x) = \Lambda(x)^y e^{-\Lambda(x)}/y!$, where
\begin{equation}
\Lambda(x) = 15{T}(x)+ 3.
\end{equation}
\end{itemize}
In each case, we take $X = (X_1, X_2)\sim U[0,1]\times[0,1]$ and generate a sample of $n=300$ observations from the distribution of $(Y, X)$.  Afterwards, a missing data is created in a way that if $y>3$ in case (i) or $y>10$ in case (ii), we randomly take one of the following actions with equal probability: (1) delete $x_1$ only; (2) delete $x_2$ only and (3) delete both $x_1$ and $x_2$.  On average, we create 47 incomplete observations (out of 300) in case (i) and 61 incomplete observations in case (ii).
\begin{figure}[htbp]
\centerline{
\includegraphics[width=2.8in,height=1.8in]{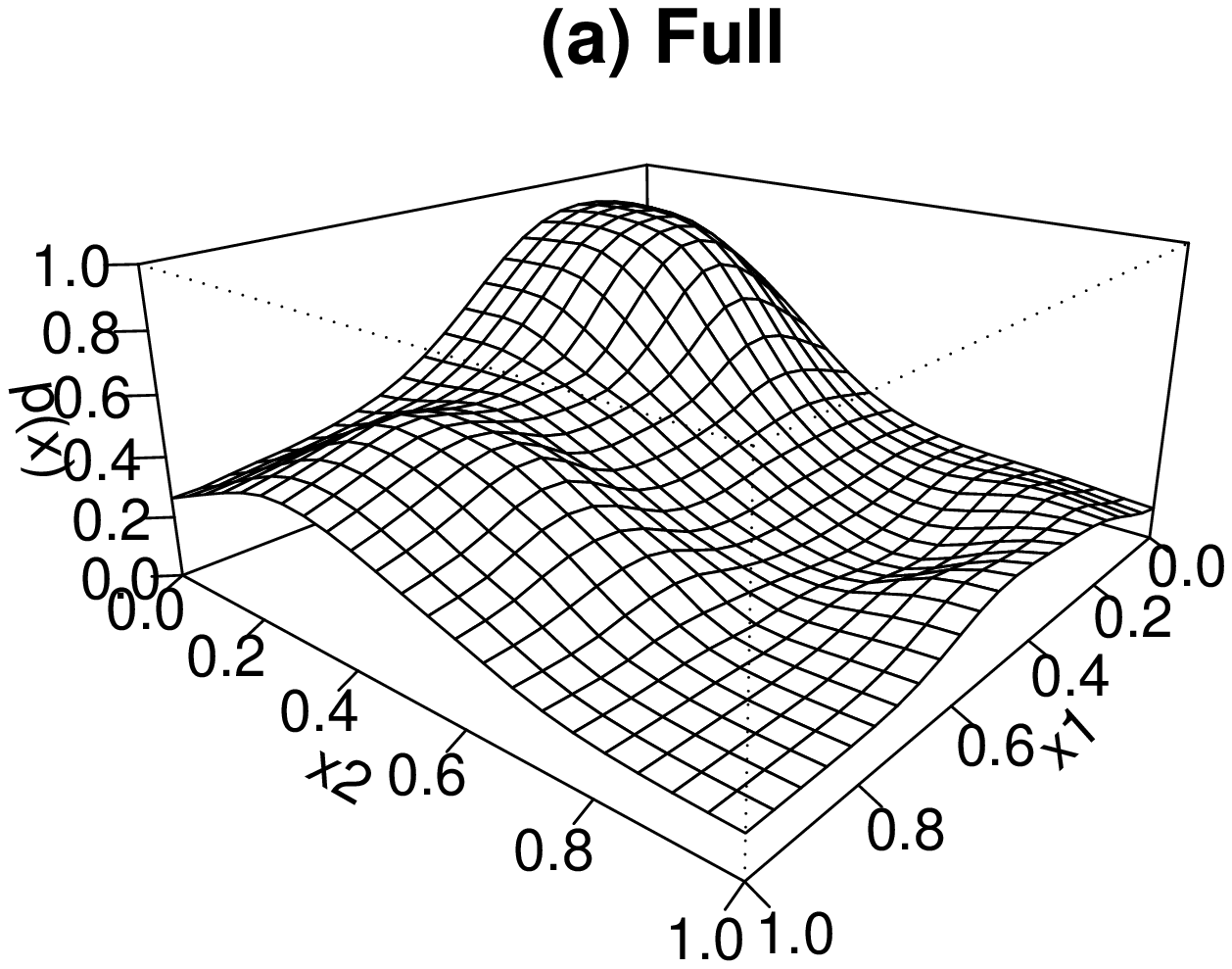}\includegraphics[width=2.8in,height=1.8in]{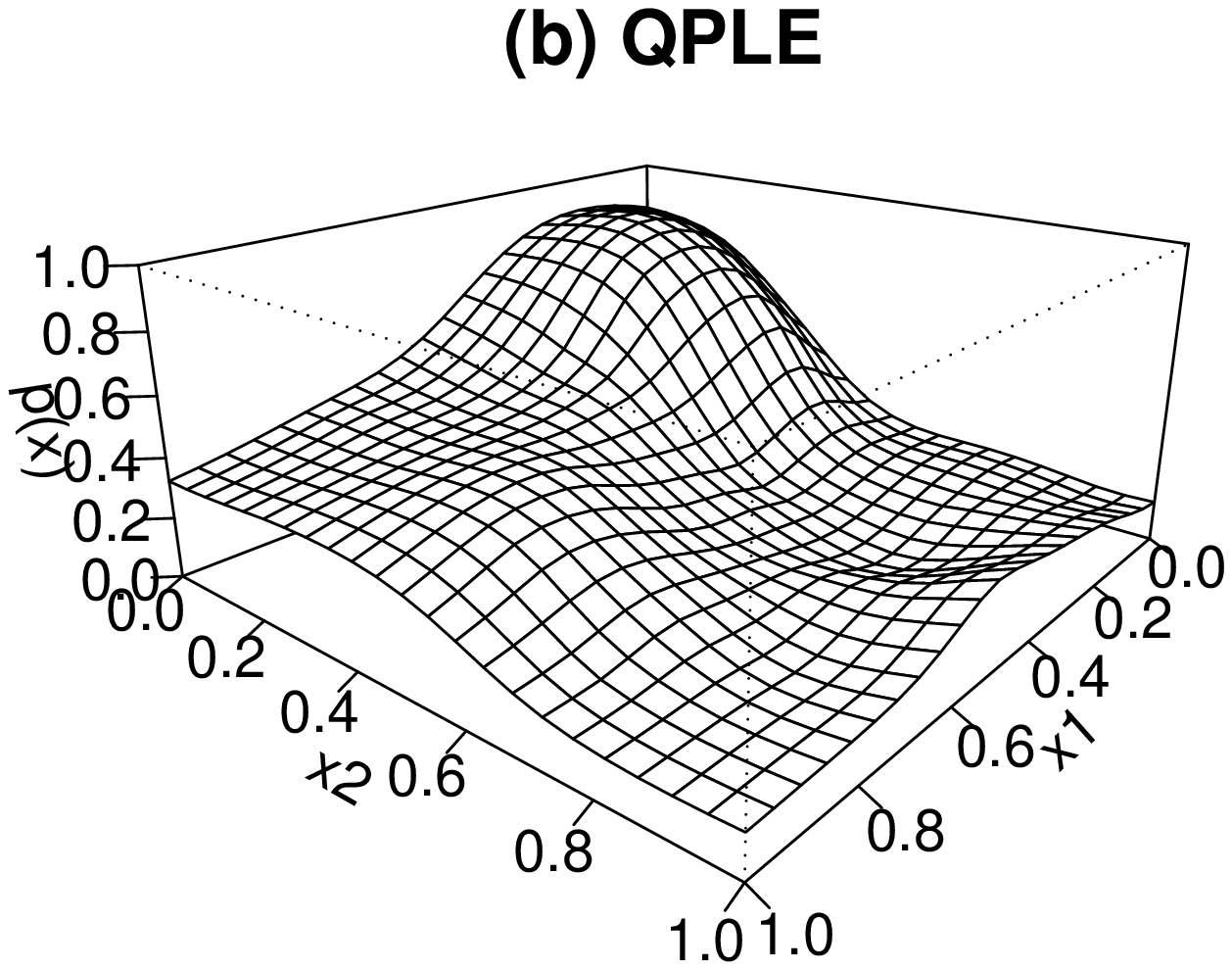}
}
\centerline{
\includegraphics[width=2.8in,height=1.8in]{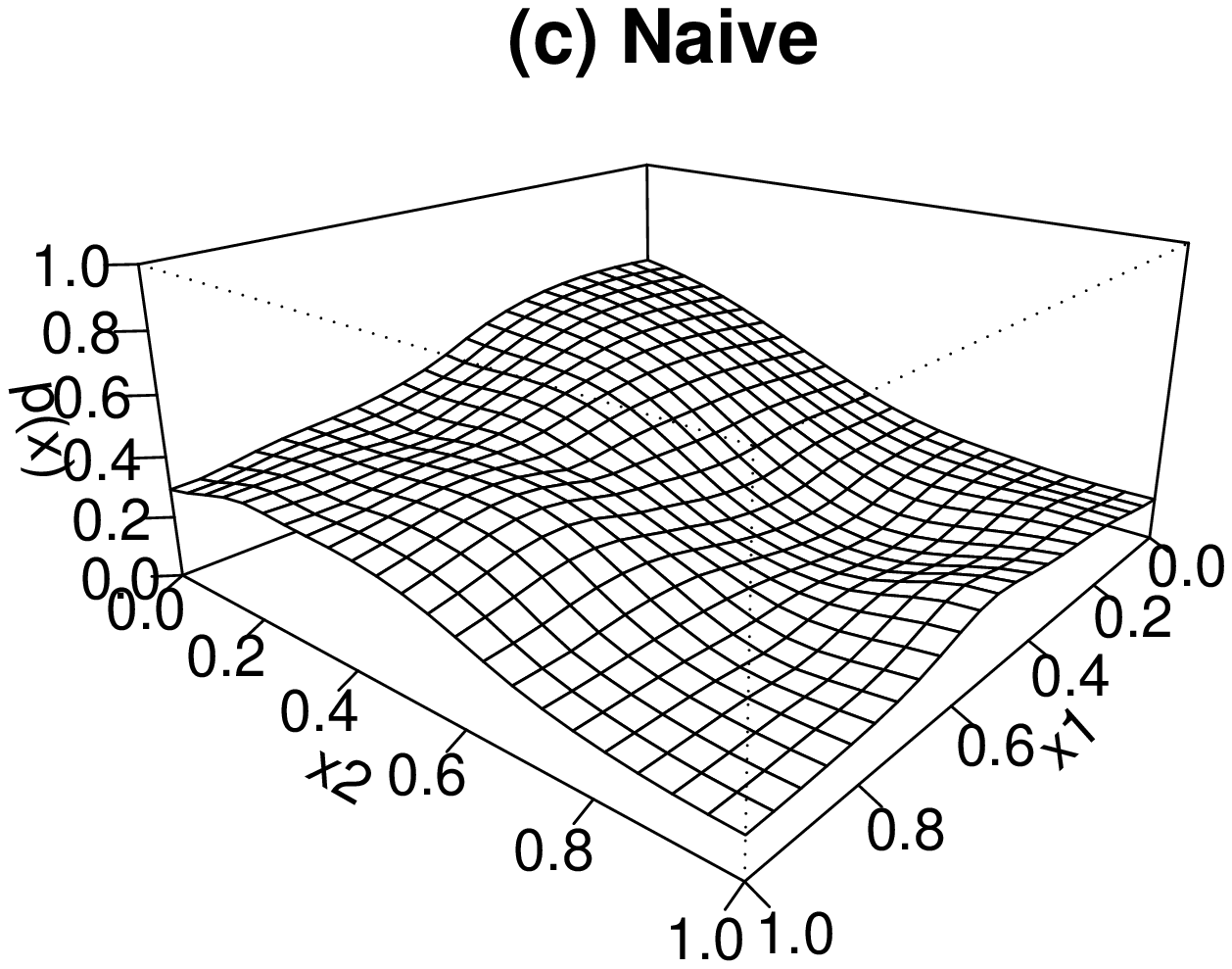}\includegraphics[width=2.8in,height=1.8in]{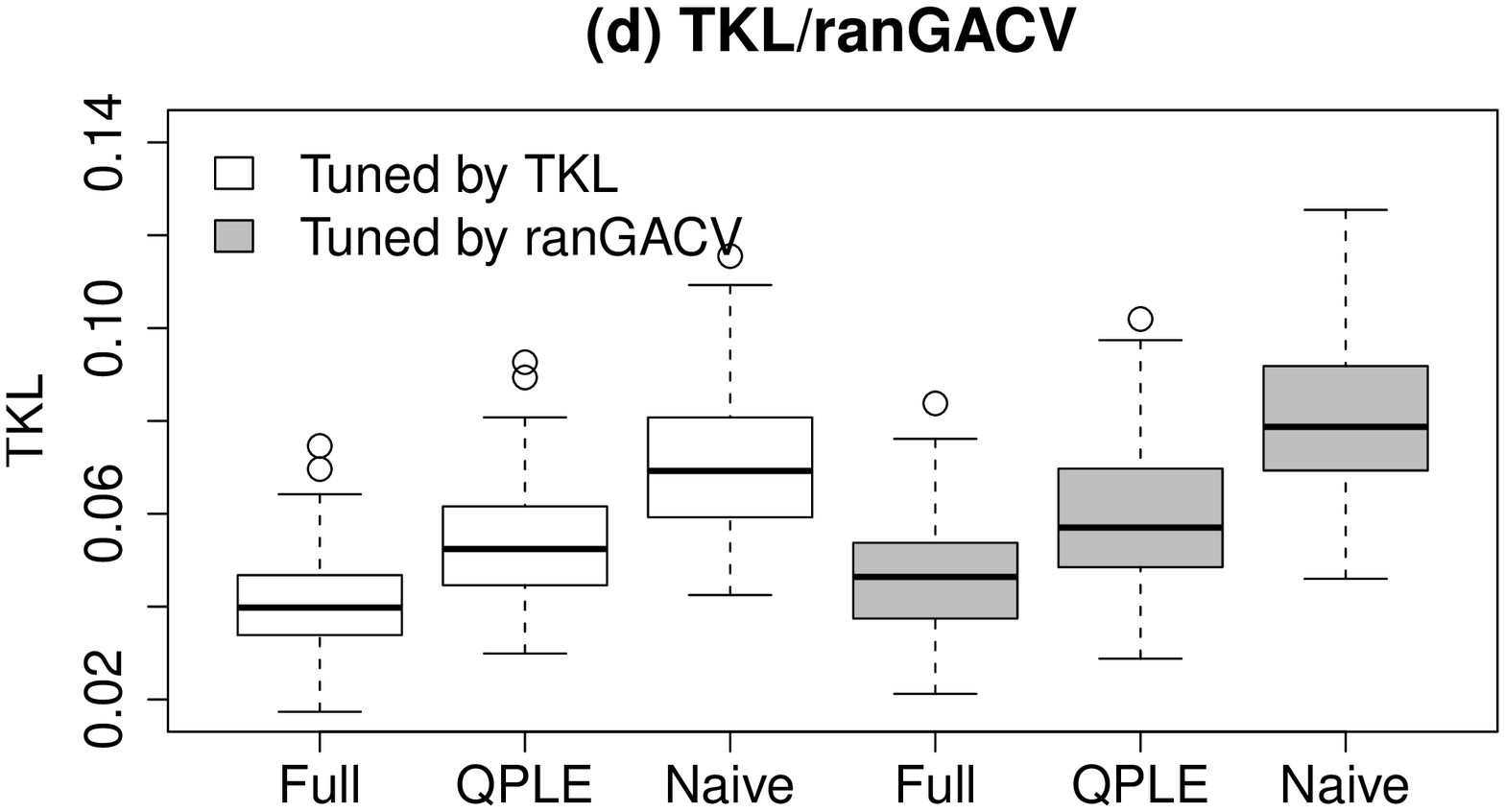}
}
 \caption{Estimated functions of $p(x_1,x_2)$
and TKL distances for case (i). (a) Full data estimate.
(b) QPLE estimate. (c) Naive estimate. The $\lambda$'s in
(a), (b) and (c) are  tuned by ranGACV.
(d) Box plots of TKL distances when tuned by TKL and by ranGACV.}\label{fig4}
\end{figure}
\begin{figure}[htbp]
\centerline{
\includegraphics[width=2.8in,height=1.8in]{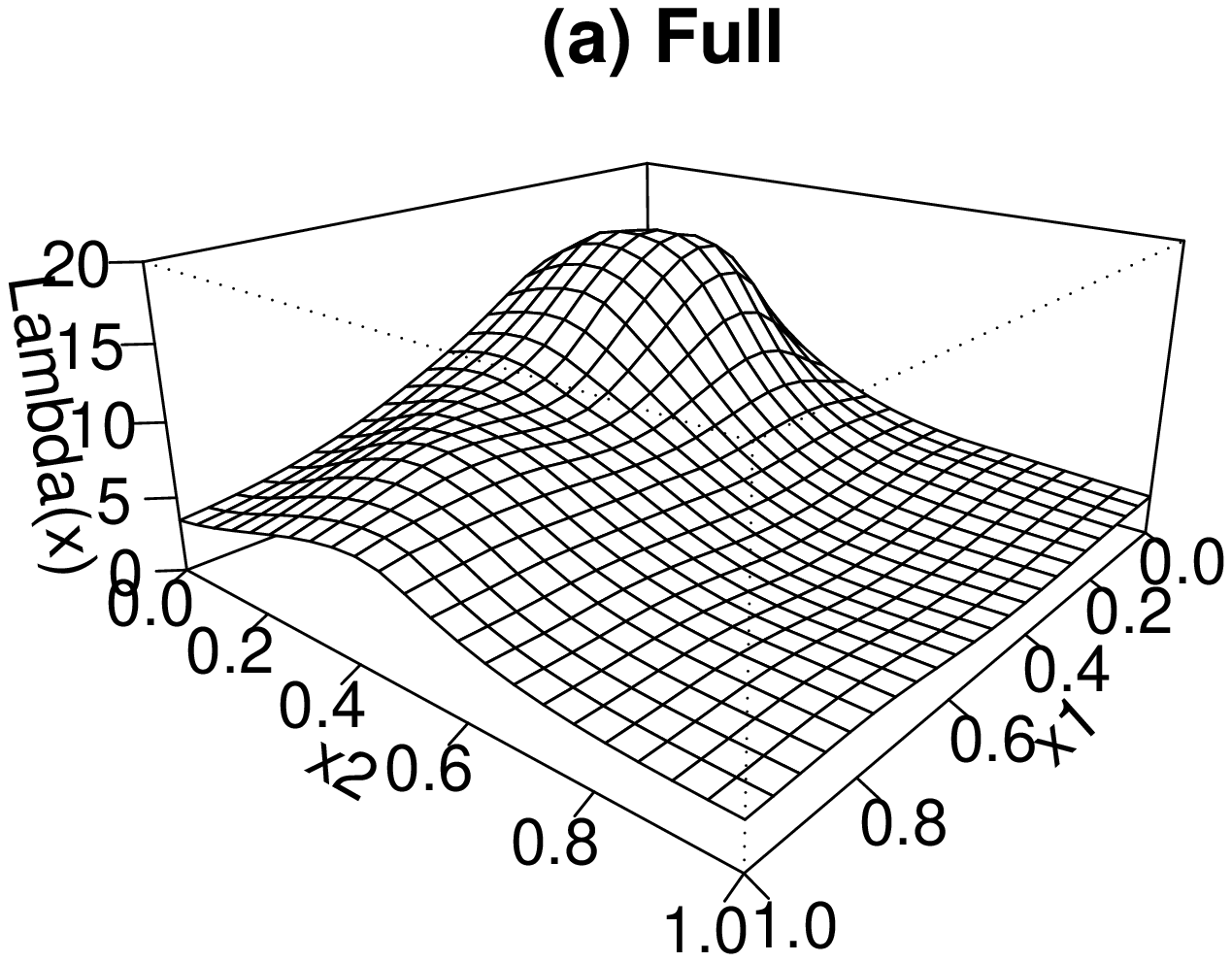}\includegraphics[width=2.8in,height=1.8in]{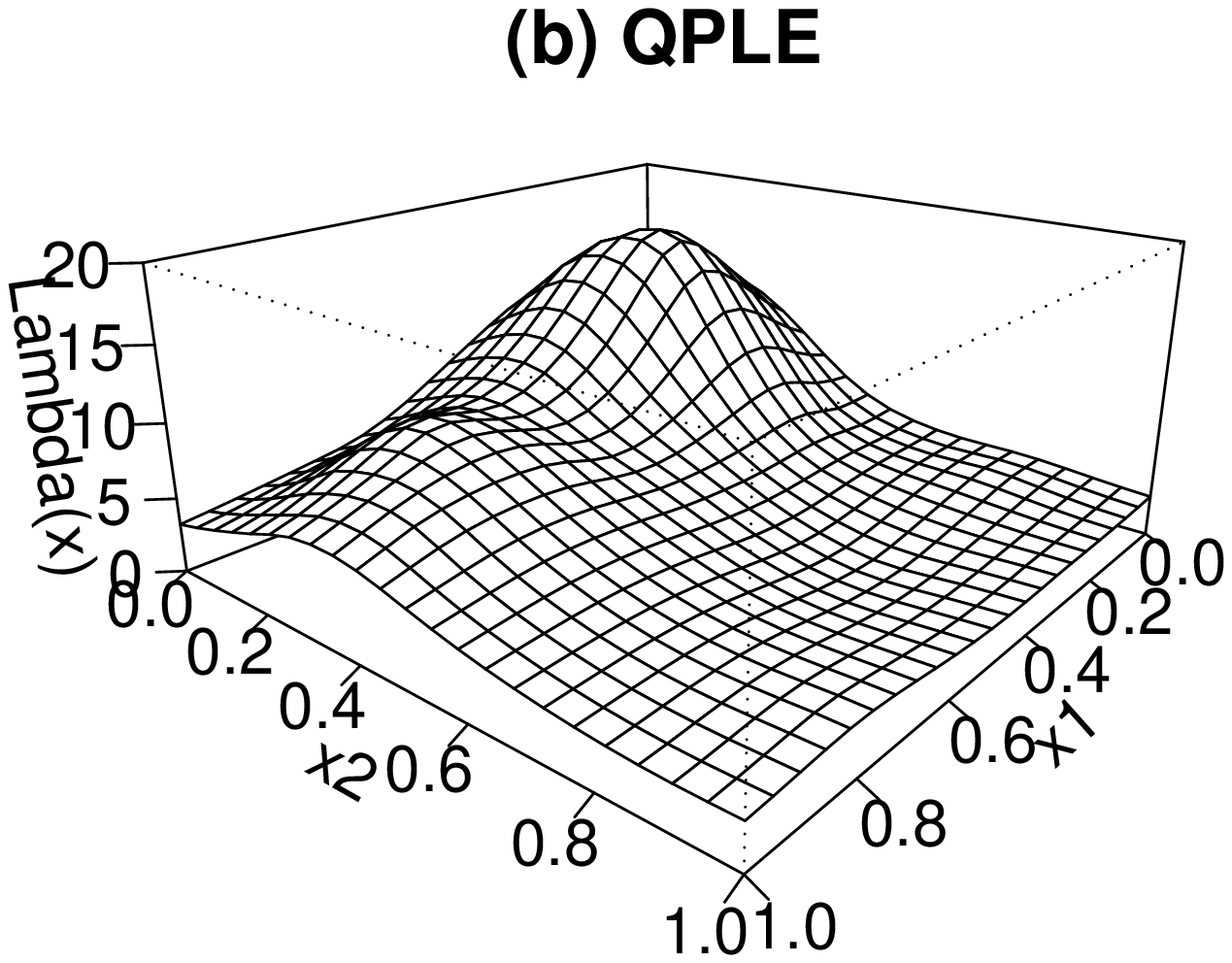}
}
\centerline{
\includegraphics[width=2.8in,height=1.8in]{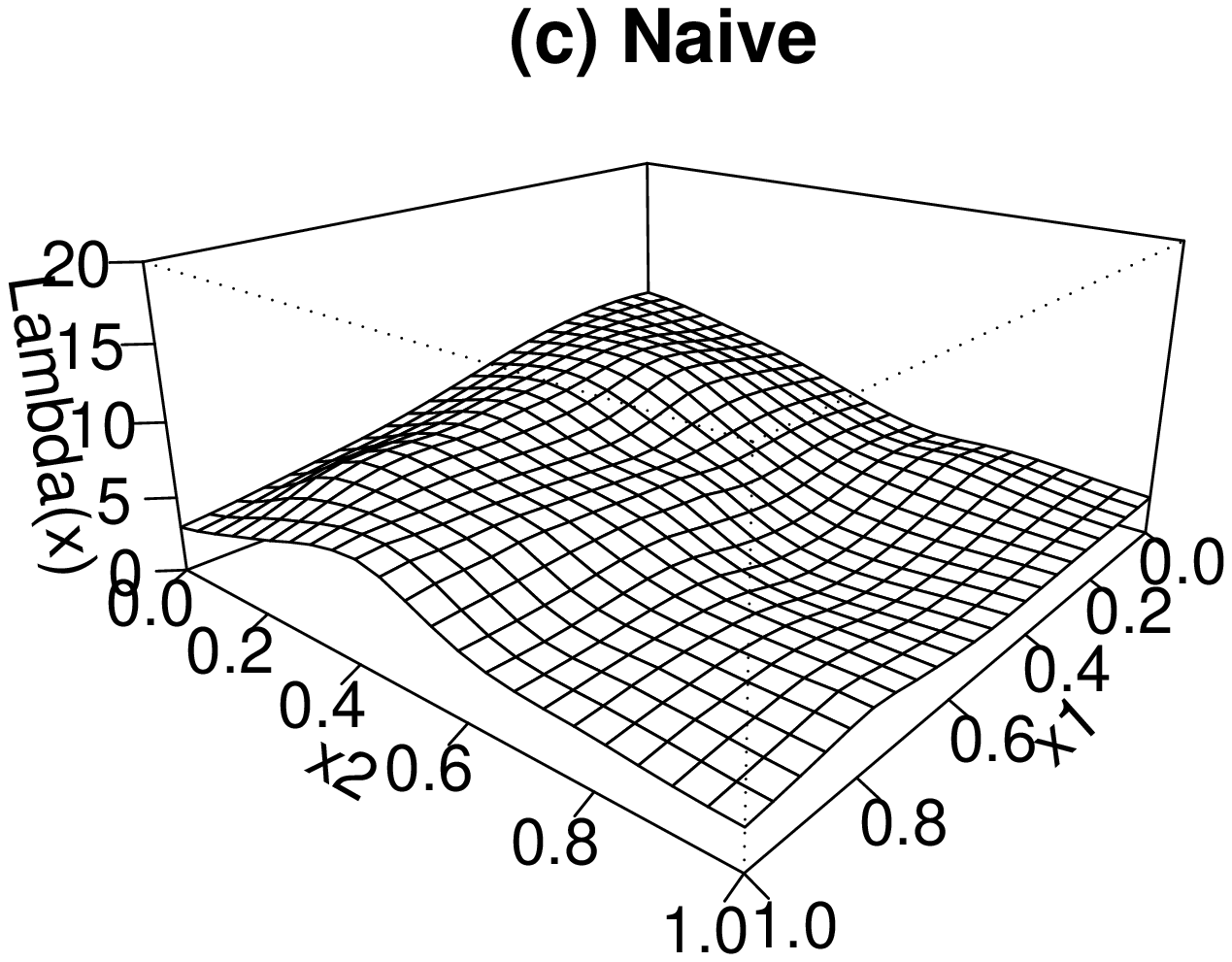}\includegraphics[width=2.8in,height=1.8in]{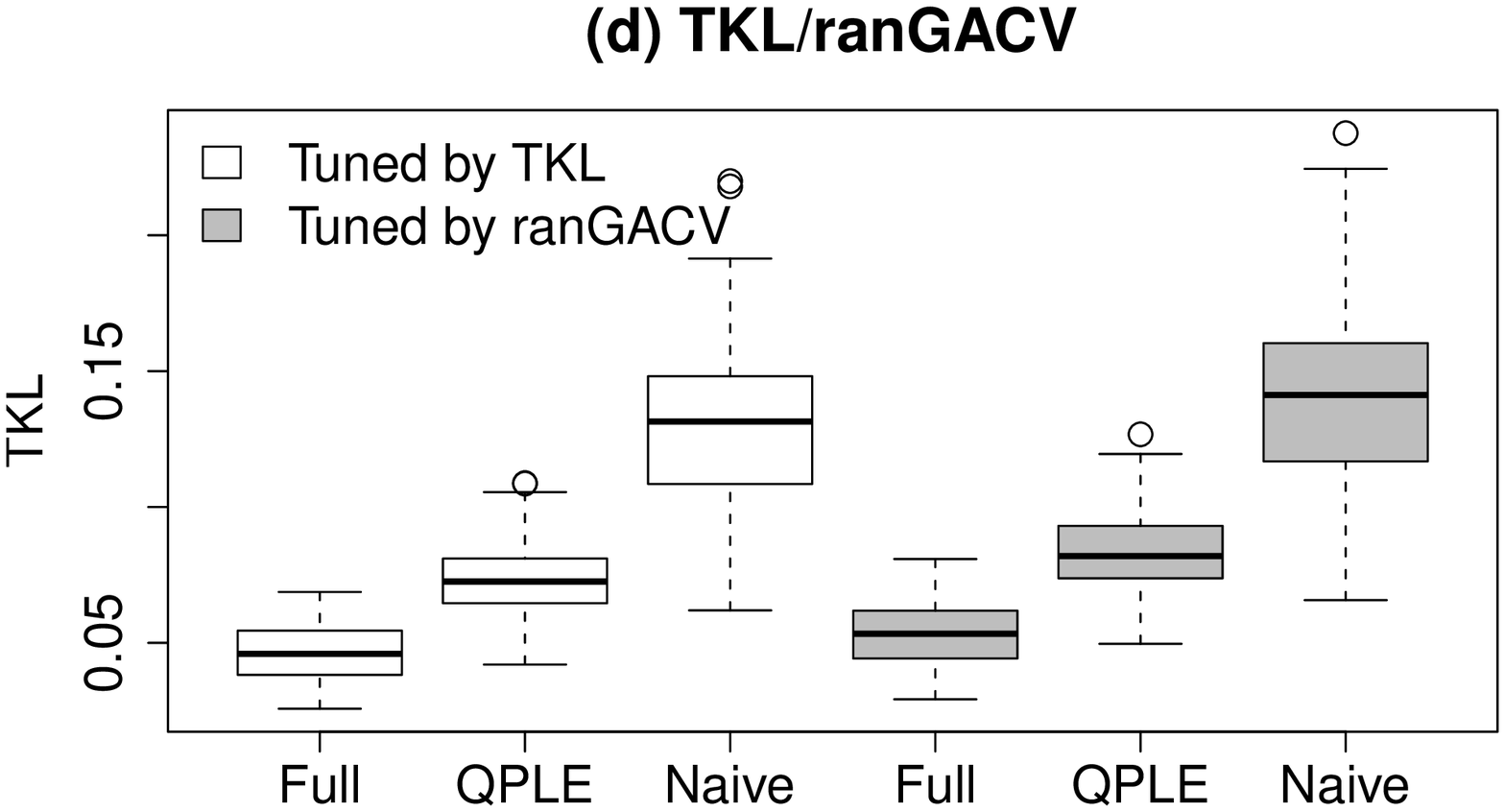}
}
 \caption{Estimated functions of $\Lambda(x_1, x_2)$ and TKL distances for case (ii).  (a) Full data estimate.
(b) QPLE estimate. (c) Naive estimate. The $\lambda$'s in
(a), (b) and (c) are  tuned by ranGACV.
(d) Box plots of TKL distances when tuned by TKL and by ranGACV. }\label{fig5}
\end{figure}

We will test our method by thin plate spline regression.  In order to implement QPLE, we specify for $x$ a bivariate normal distribution $N(\mu, \Sigma)$, where $\mu= (\mu_1, \mu_2)^T$ and $\Sigma = \{\sigma_{ij}\}_{2\times2}$ (an arbitrary covariance matrix) are to be estimated. At each EM iteration,  we construct for each incomplete $x_i$ a Gaussian quadrature rule, where $11$ nodes are created for each missing component.  Simulation results are summarized at Figure \ref{fig4} and \ref{fig5}.

Figure \ref{fig4} and \ref{fig5} show the estimated functions where the smoothing parameter is tuned by ranGACV.  The bottom right panel reports the TKL distances based on 100 simulations, when $\lambda$ is selected by TKL and ranGACV.  The performance of QPLE is also impressive in the case of missing covariate data.   Note that most incomplete observations appeared near the `peak' of the test function.  In this case, if these incomplete observations are left out, we will miss the information about the peak, as indicated by the naive estimator.  On the other hand, by incorporating most information in the data including the observations with paritally missing covariates, QPLE provides encouraging results, even though we actually specified a wrong covariate distribution.

\subsection{Case study}
In this section, we illustrate our method on an observational
data set that has been previously analyzed, by deleting some covariates,
and then comparing our method with the original analysis and the
naive method of dropping files with missing covariates.

The Beaver Dam Eye Study is an ongoing population-based study of age-related
ocular disorders. Subjects were a group of 4926 people aged
43-86 years at the start of the study who lived in
Beaver Dam, WI and were examined at baseline, between 1988 and 1990.
A description of the population and details
of the study at baseline may be found in Klein, Klein,
Linton and Demets (1991)\cite{Klein1991}. Pigmentary abnormalities
are one of the ocular disorders of interest in that study.
Pigmentary abnormalities are an early sign
of age-related macular degeneration and are defined by the
presence of retinal depigmentation and increased retinal
pigmentation.

Lin, Wahba, Xiang, Gao, Klein and Klein
(2000)\cite{Lin2000} and Gao, Wahba, Klein and Klein(2001)\cite{Gao2001}
considered only the $n=2545$ womem members of this
cohort.   $11.88\%$ of them showed evidence of pigmentary
abnormalities.
They examined the association of pigmentary abnormalities with
six other attributes at baseline, by fitting a
Smoothing Spline ANOVA (SS-ANOVA)  model.
The six attributes are are listed in Table \ref{tab1}.

Let $p(x)$ be  the probability
that a subject with attribute vector $x$ at baseline will
be found to have a pigmentary abnormality in at least one
eye, at baseline.

\begin{table}
\begin{tabular}{llll}
\hline Attributes    &     unit   &  range & code  \\
\hline
systolic blood pressure  &  $mmHg$ & 71-221 & \emph{sys}\\
serum cholesterol  &  $mg/dL$ & 102-503 &  \emph{chol}\\
age at baseline&  $years$  & 43-86 & \emph{age}\\
body mass index &  $kg/m^2$ & 15-64.8 & \emph{bmi}\\
undergoing hormone replacement therapy  &  yes/no & yes,no & \emph{horm}\\
history of heavy drinking & yes/no & yes,no & \emph{drin} \\
\hline
\end{tabular}
\caption{Covariates for Pigmentary Abnormalities}\label{tab1}
\end{table}

The model fitted was of the form
\begin{eqnarray}
&&f(x)=\text{constant} + f_1(sys) + f_2(chol) + f_{12}(sys, chol) \\
  &&~~~~~~~~~+ d_{age}\cdot
age + d_{bmi} \cdot bmi + d_{horm} \cdot I_2( horm) + d_{drin} \cdot I_2(drin).
\nonumber
\end{eqnarray}
Here $x$ denotes the vector of covariates listed in Table \ref{tab1} and $f(x)$ is
the logit
form of the probability: $f(x) = \log \frac{p(x)}{1-p(x)} $.

\begin{figure}[htbp]
\begin{center}
\includegraphics[width=1.8in,height=1.8in]{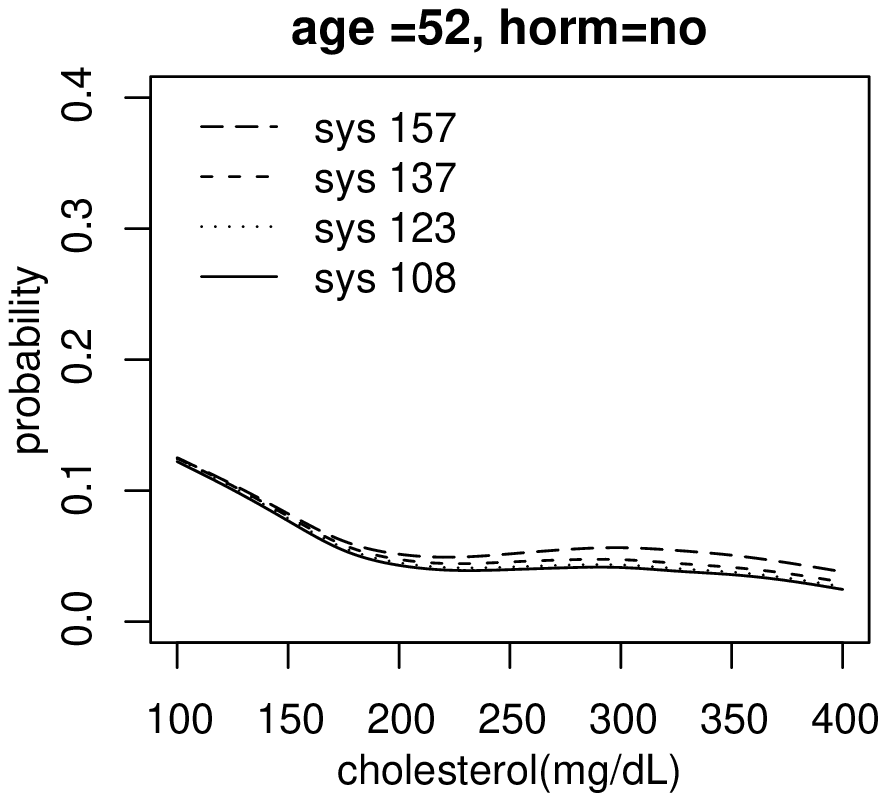}~\includegraphics[width=1.8in,height=1.8in]{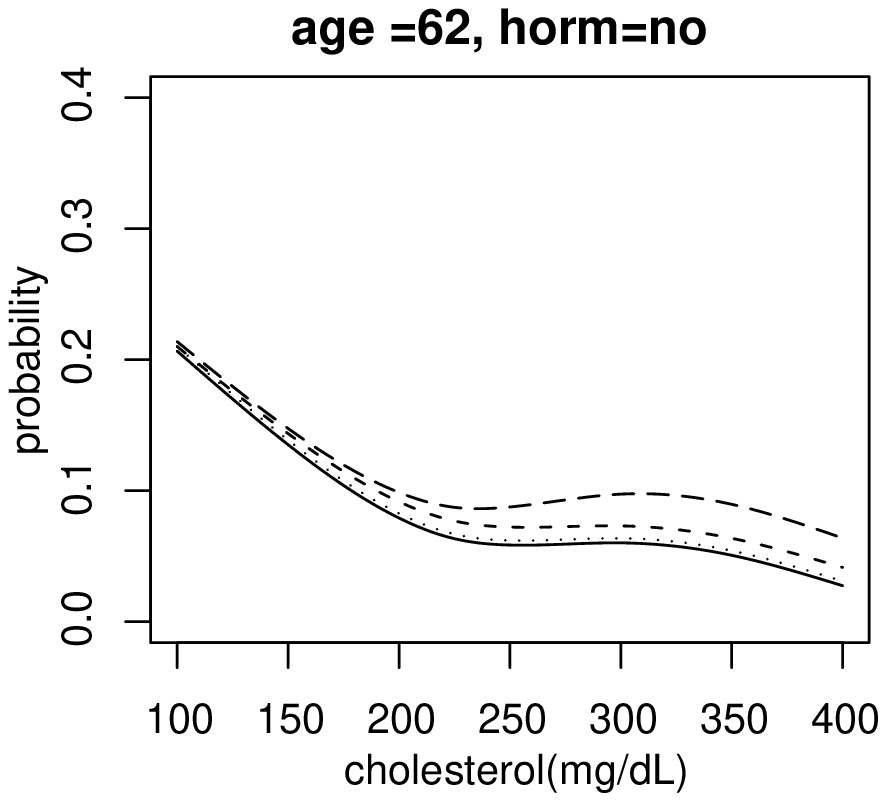}~\includegraphics[width=1.8in,height=1.8in]{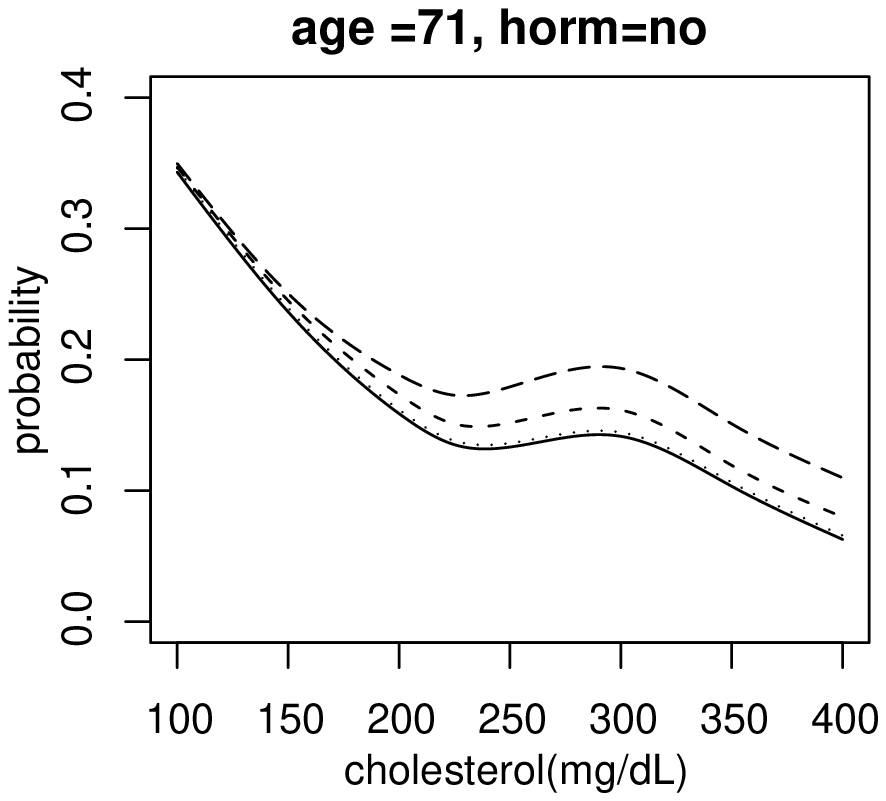}
\includegraphics[width=1.8in,height=1.8in]{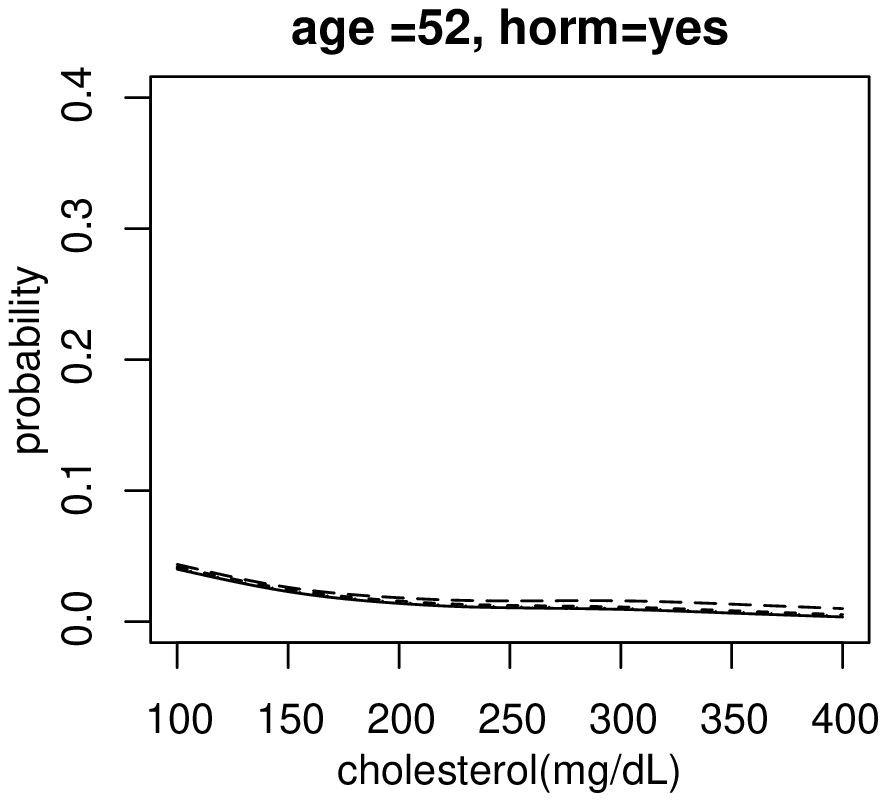}~\includegraphics[width=1.8in,height=1.8in]{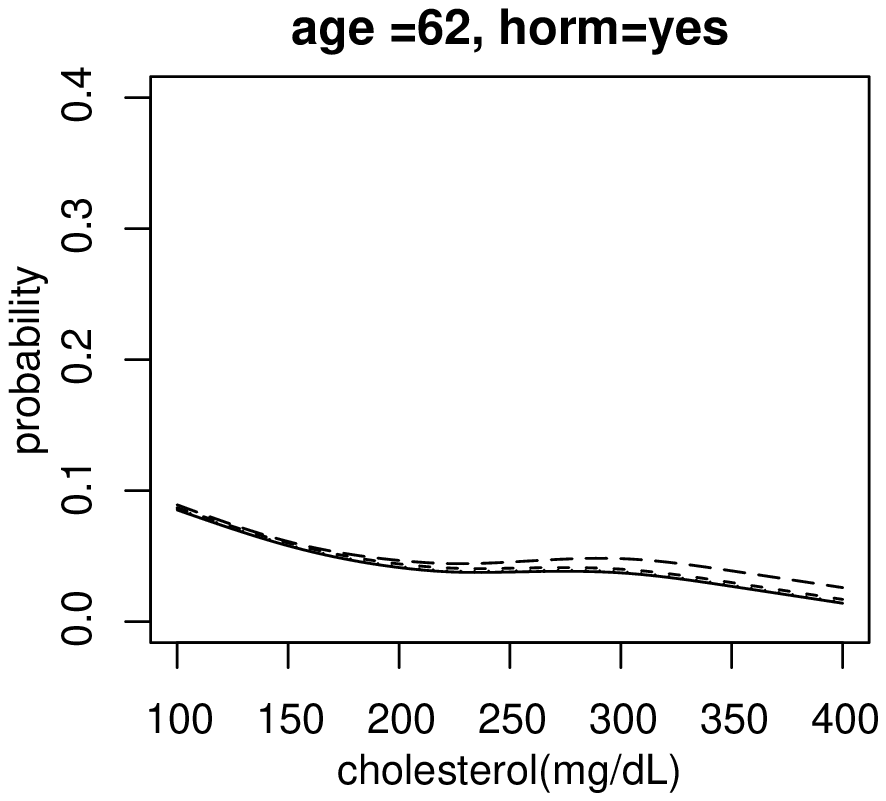}~\includegraphics[width=1.8in,height=1.8in]{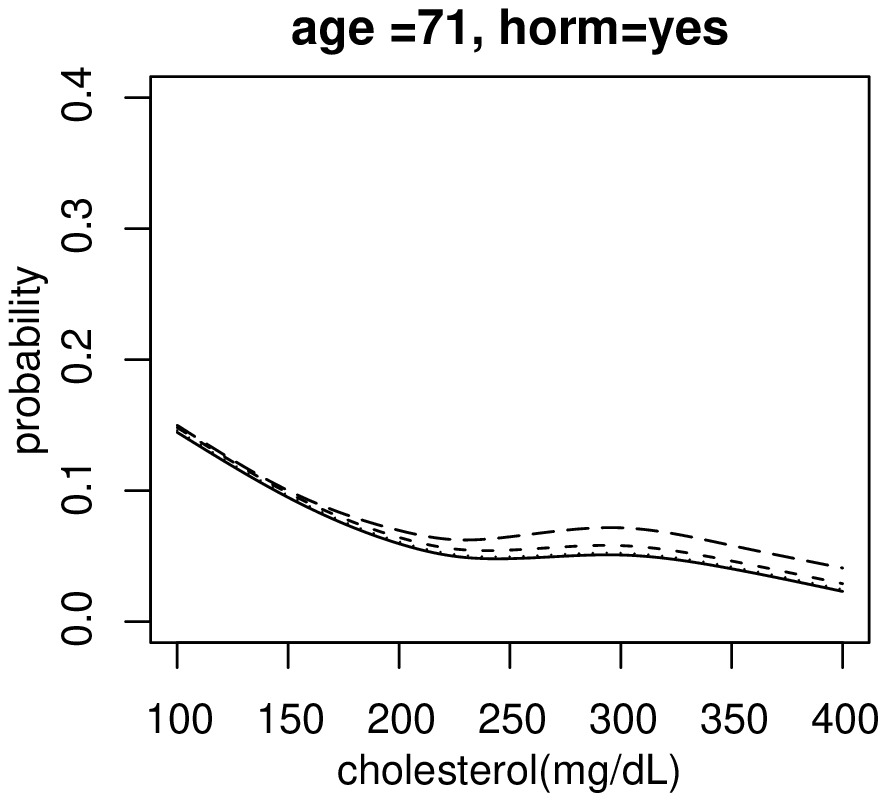}
\end{center}
 \caption{Probability curves estimated from the full data analysis.
This figure is adapted from Figures 9 and 10 from Lin, Wahba, Xiang, Gao, Klein
and Klein (2000)\cite{Lin2000}.  Each panel plots the estimated
probability of pigmentary abnormalities as a function of cholesterol,
for four different values of $sys$.  The six panels correspond
to different values of $age$ and $horm$, when \emph{drin}=no and
\emph{bmi}=27.5 are fixed.}\label{fig6}
\end{figure}

The data analysis is summarized in  Figure \ref{fig6},
which is adapted from Lin, Wahba, Xiang, Gao, Klein and Klein (2000)\cite{Lin2000}.
On each panel, we plot the estimated
probability of pigmentary abnormalities  as a function of \emph{chol},
for various values of $sys, age$ and $horm$.  Note that we only
plot for \emph{bmi} = 27.5 and \emph{drin} = no, because \emph{bmi}
has relatively small effect in the fitted model while only 152 out
of 2585 subjects have \emph{drin} = 1.  Hence Figure \ref{fig6}
is adequate to demonstrate the estimated association patterns.\\

Generally speaking, higher \emph{chol} was associated with a protective effect.
However, when \emph{chol} goes
from 250 to 350, a ``bump" appears on the estimated curves.
This phenomenon
provides us a good opportunity to test our method. In order to `hide'
the bump, we create a  data set with missing covariates
by deleting some attribute values
for those subjects whose cholesterol is between 250 and 350.
Consequently, 517 incomplete subjects are created with values of \emph{sys},
\emph{bmi} and \emph{horm} randomly removed.  More exactly,
30 subjects missed \emph{sys}, \emph{bmi} and \emph{horm}, 109
subjects missed both \emph{sys} and \emph{bmi}, 118 subjects missed
both \emph{sys} and \emph{horm} and 260 subjects missed only one
attribute value.

We shall first claim that the methodology in this paper can be
extended to SS-ANOVA models  without any extra effort, as illustrated in
Appendix C.   In this case, QPLE can be conducted following Ibrahim, Lipsitz and Chen
(1999)\cite{Ibrahim1999}.   We first model the joint covariate distribution
via a sequence of one-dimensional conditional distributions.
Note that (\emph{age}, \emph{chol}, \emph{drin}) are always
observed and hence we do not need to model them.
Also, very few subjects have $drin=1$, hence $drin$ will  be ignored  in the modeling.
Given (\emph{age}, \emph{chol}), we adopt a bivariate normal distribution
$\text{(\emph{sys}, \emph{bmi})}\sim N(\mu, \Sigma)$, where $\mu =
(\mu_1,~\mu_2)$ with $\mu_k = a_{k0} + a_{k1} sys + a_{k2}
bmi, ~k=1,2$ and $\Sigma=\{\sigma_{ij}\}_{2\times2}$ is an arbitrary
covariance matrix,
and the $a$'s and $\Sigma$ are to be estimated.
Now conditionally on other
attributes, $horm$ is modeled via a logistic regression model
\begin{eqnarray}\nonumber
p(horm=1)=\frac{\exp\{a_{30}+a_{31}
age+a_{32}chol+a_{33} sys+a_{34}bmi\}}
{1+\exp\{a_{30}+a_{31} age+a_{32}chol+a_{33}
sys+a_{34} bmi\}}.
\end{eqnarray}

\begin{figure}[htbp]
\begin{center}
\includegraphics[width=1.8in,height=1.8in]{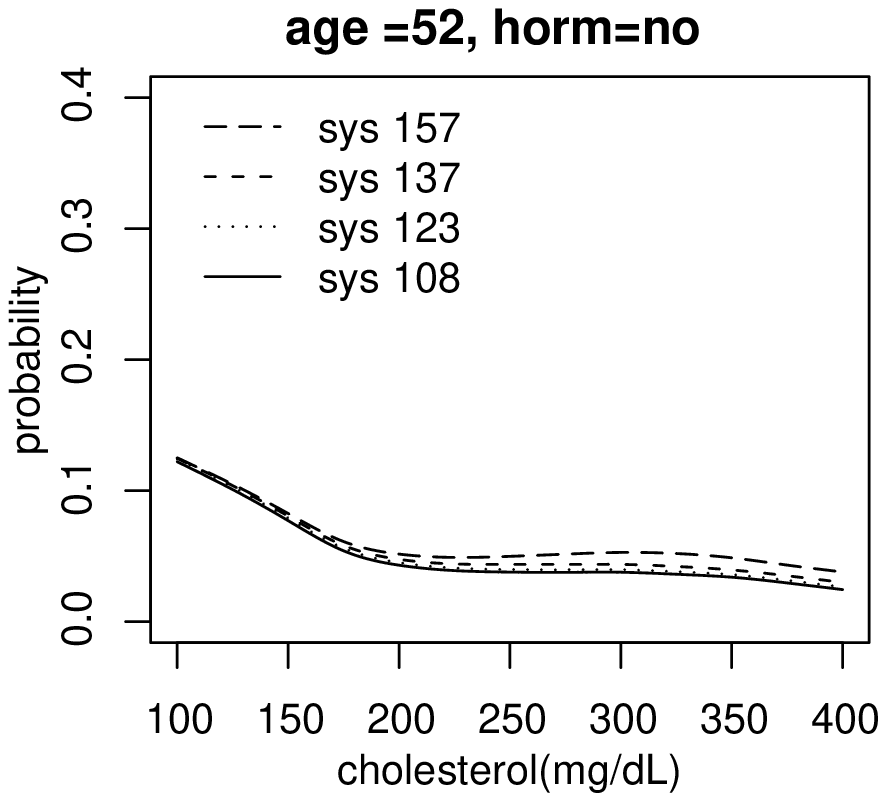}~\includegraphics[width=1.8in,height=1.8in]{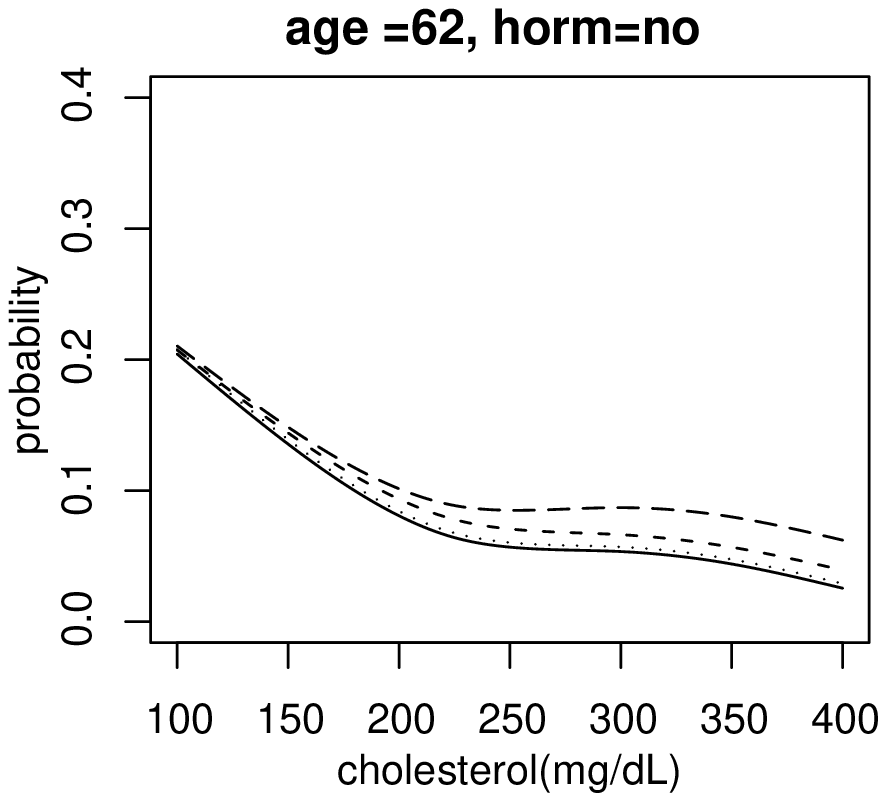}~\includegraphics[width=1.8in,height=1.8in]{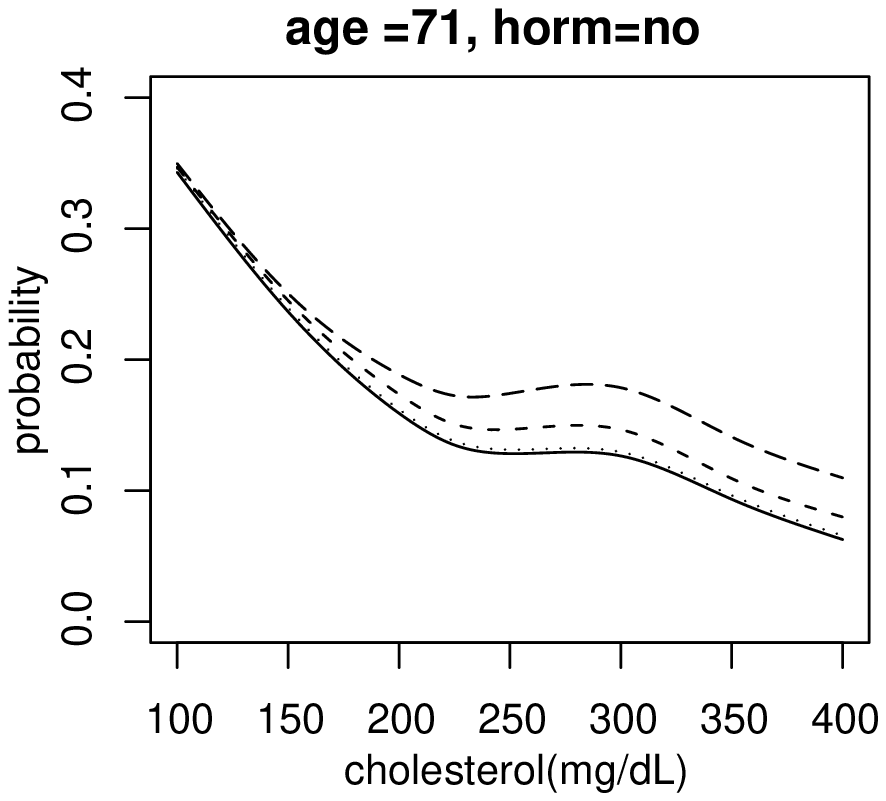}
\includegraphics[width=1.8in,height=1.8in]{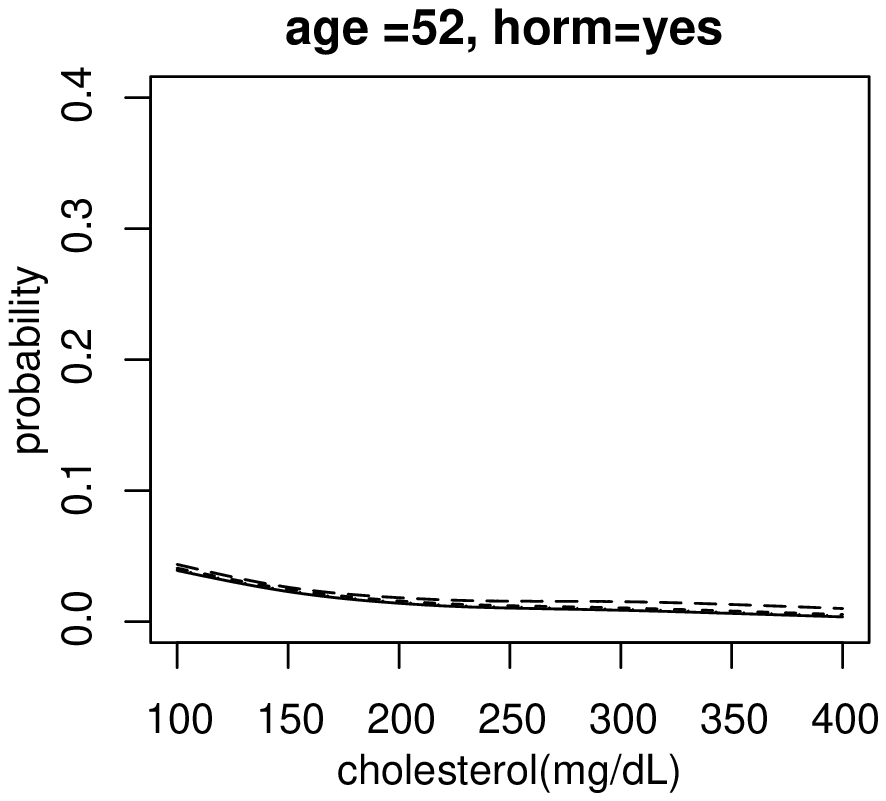}~\includegraphics[width=1.8in,height=1.8in]{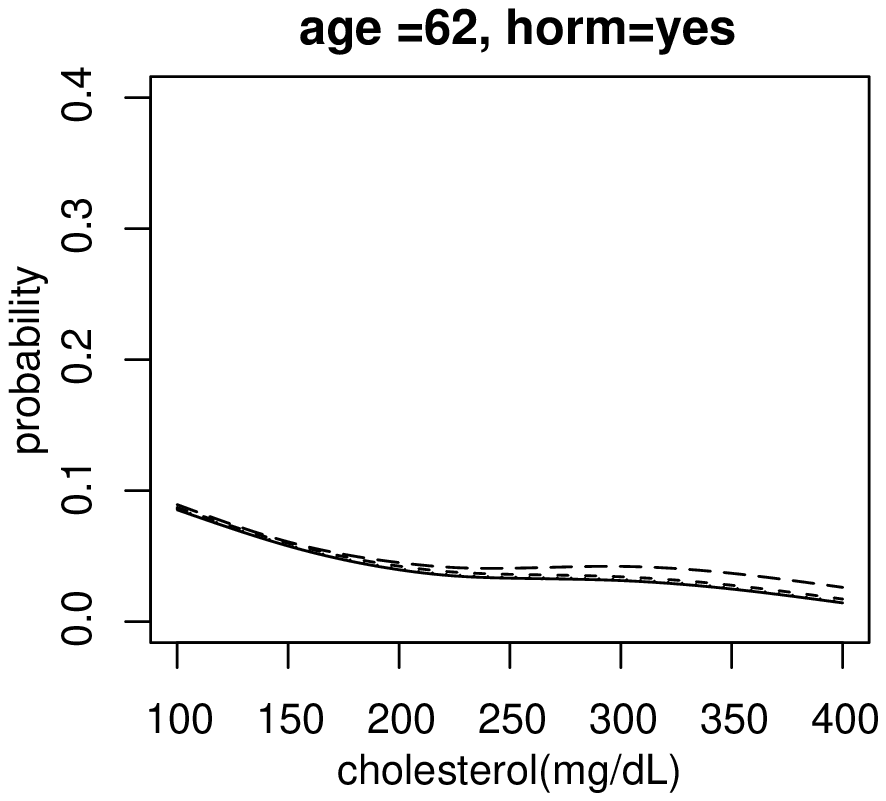}~\includegraphics[width=1.8in,height=1.8in]{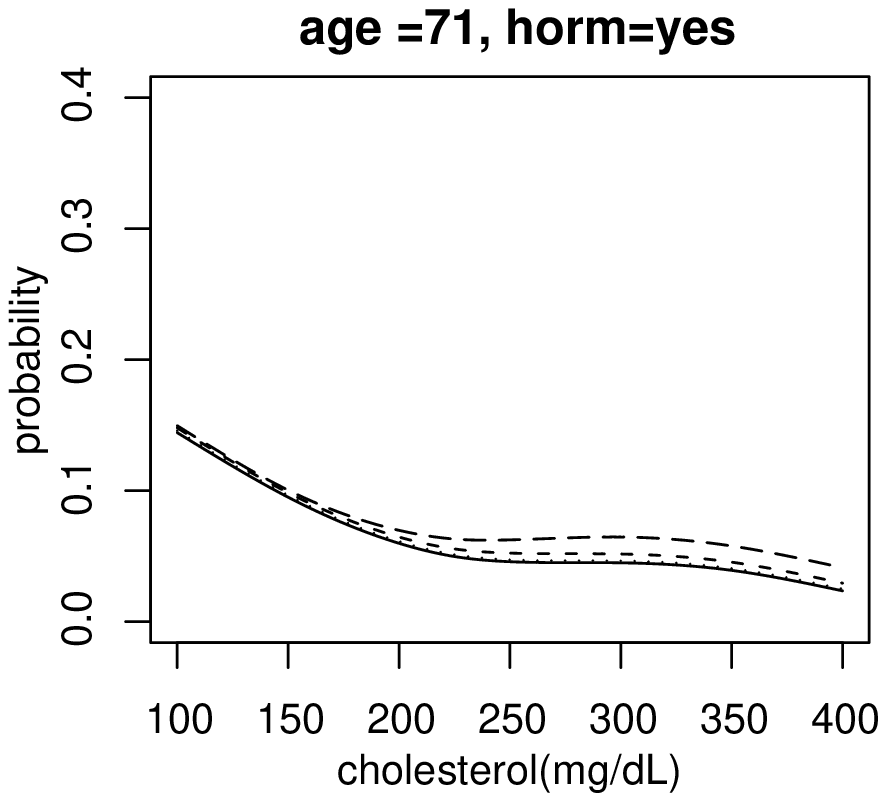}
\end{center}
 \caption{Probability curves obtained from QPLE.  Each panel plots the estimated
probability of pigmentary abnormalities  as a function of cholesterol, for four
different values of $sys$.  The six panels correspond to different values of $age$
and $horm$, when \emph{drin}=no and
\emph{bmi}=27.5 are fixed.}\label{fig7}
\end{figure}

\begin{figure}[htbp]
\begin{center}
\includegraphics[width=1.8in,height=1.8in]{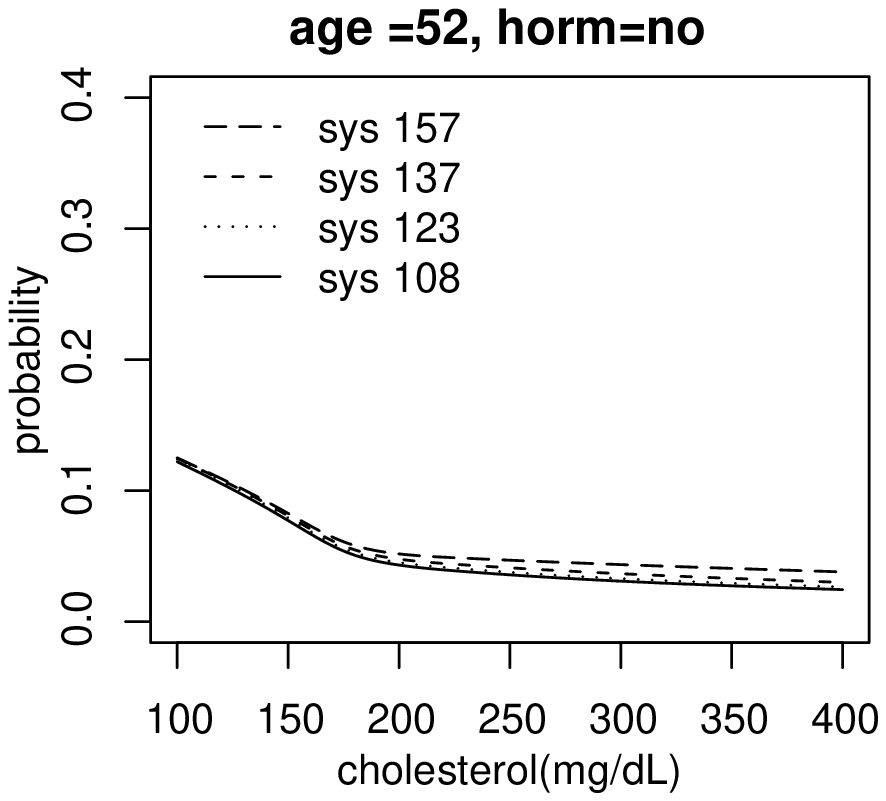}~\includegraphics[width=1.8in,height=1.8in]{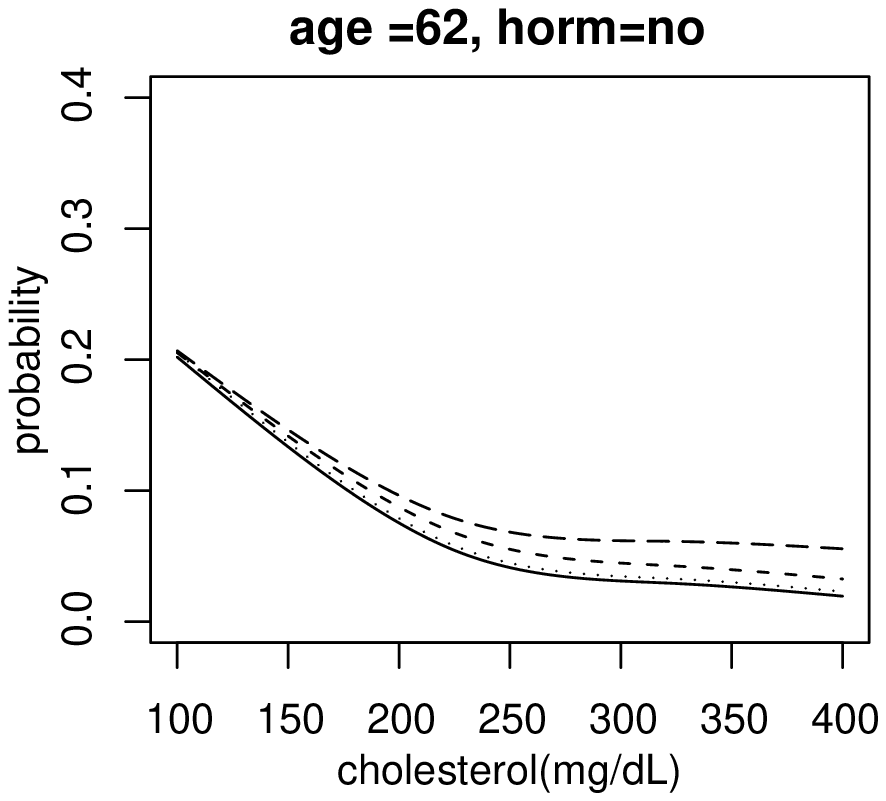}~\includegraphics[width=1.8in,height=1.8in]{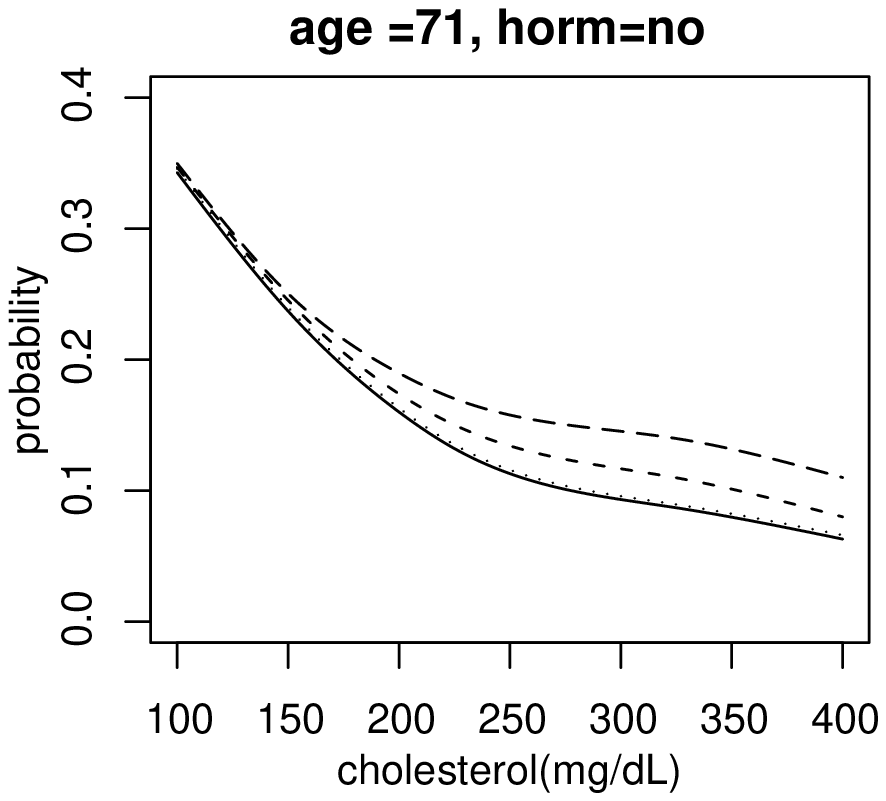}
\includegraphics[width=1.8in,height=1.8in]{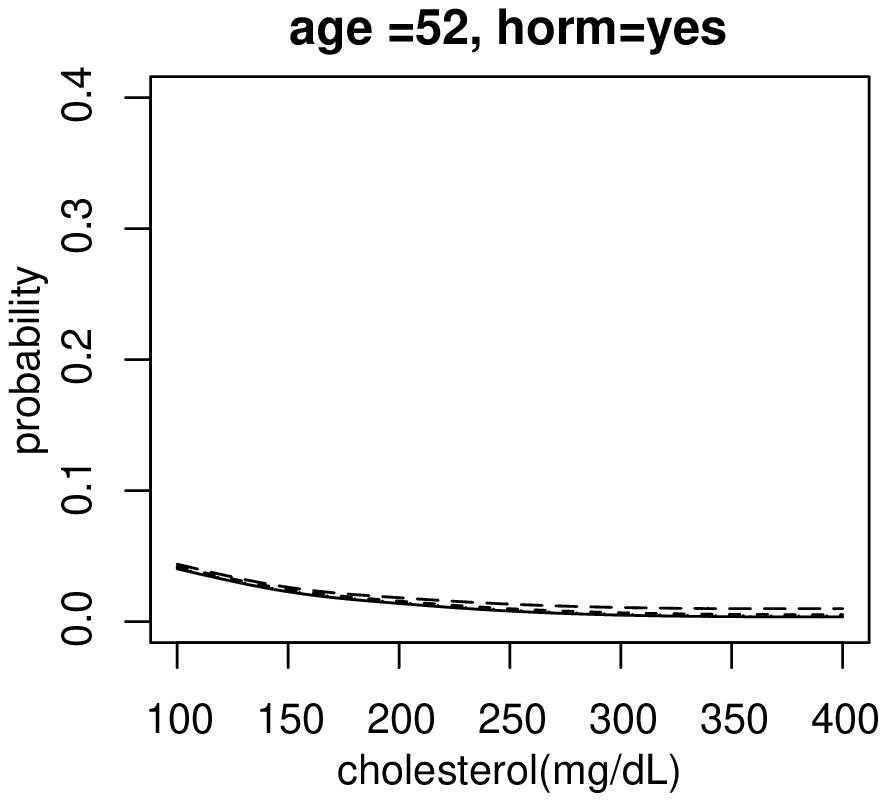}~\includegraphics[width=1.8in,height=1.8in]{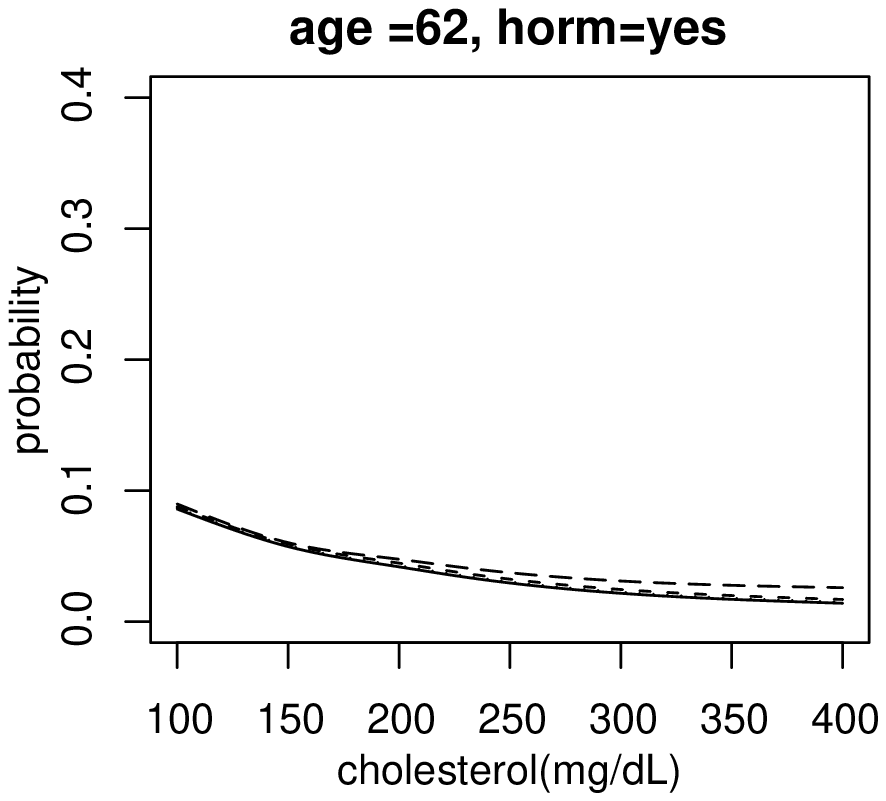}~\includegraphics[width=1.8in,height=1.8in]{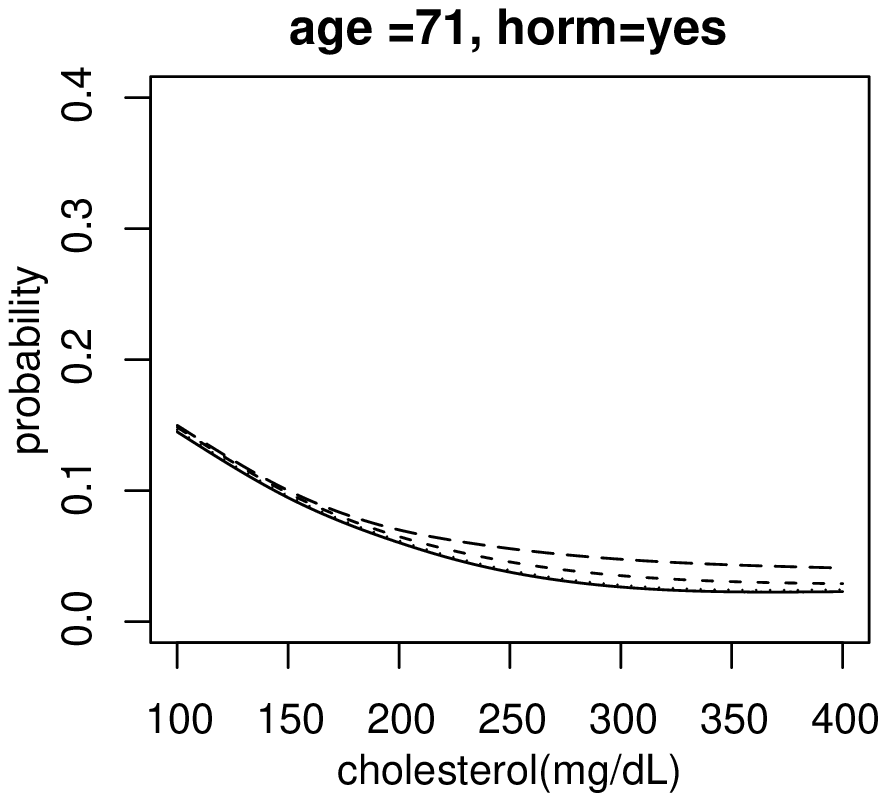}
\end{center}
 \caption{Probability curves obtained from the naive method.  Each panel plots the
estimated probability of pigmentary abnormalities as a function of cholesterol, for
four different values of $sys$.  The six panels correspond to different values of
$age$ and $horm$, when \emph{drin}=no and
\emph{bmi}=27.5 are fixed.}\label{fig8}
\end{figure}

Following this construction of covariate distributions
and using the method described in Section 3.1, a quadrature rule can be obtained
recursively at each EM iteration. In the computation, the numbers of nodes generated
for
\emph{sys}, \emph{bmi} and \emph{horm} are 10, 10 and 2 respectively.
Results of QPLE are given at Figure \ref{fig7}.
Figure \ref{fig8} shows the naive estimator computed
over the 2068 subjects without missing covariates.

Note that only the incomplete subjects contain
information about the bumps. Consequently, the naive estimator omitted these bumps,
leading to monotone decreasing probability curves.
In words, high cholesterol appears to generally lower the risk of pigmentary
abnormalities especially in the older, $horm=no$ group, aside from the ``bump",
from the full data analysis shown at Figure \ref{fig6}.
However, the naive estimator appears to make this risk decrease
substantially more rapidly due to missing the ``bump" completely,
while the QPLE did an excellent job of recovering the
original analysis--
the QPLE estimated curves are very close to those of the full data analysis.
This can be
understood from the fact that most of the incomplete subjects missed only one or two
(out of six) covariates.  Hence most information is still retained in the missing data.

\section{Concluding remarks}
We have presented a direct extension of penalized likelihood regression to the situation when the observed covariates are probability spaces.  The regression function is estimated by minimizing a penalized likelihood that incorporates distributional information of the covariates.  Numerically, we compute a finite dimensional estimator after approximating the integrals in the likelihood function by quadrature rules.  Using the same approximation, GACV and its randomized version have been derived to select the smoothing parameters.  Our method is computationally efficient, as it only require a small number of quadrature nodes to obtain a good estimate.  A direct implementation of our method is to handle incomplete covariate data such as covariate measurement error and partially missing covariates.  In the examples we have investigated, the resulting estimator substantially outperformed the naive estimator and appeared to be close to the full data analysis.

Our methods can also be extend to other regularization settings, for example, the LASSO and support vector machine with hinge loss function and $L_2$ penalty.  In these cases, it might be more complicated to develop a likelihood-based frequentist approach.  We would like to investigate these extensions in the future research.

\appendix

\section{Technical proofs}

\textbf{Proof of Proposition 2.1.}  Any linear combination of measurable
functions is still measurable.  Therefore it suffices to prove that
$\mathcal{H}_B$ is complete. Let $f_1,f_2,...$ be a Cauchy sequence
in $\mathcal{H}_B$ and $f^*$ be its limit in $\mathcal{H}$. Then
$f_1,f_2,...$ converge pointwise to $f^*$. Note that the pointwise
limit of measurable functions is still a measurable function. Therefore $f^*\in \mathcal{H}_B$.  $\Box$\\

Now to simply the notation in the proofs of Lemma 2.4-2.6, let's define
\begin{equation}
l_i(t) =  y_i\cdot t -b(t) + c(y_i)
\end{equation}
the log-density as a function of the natural parameter.  Then
$l_i(t)$ is strictly concave and bounded from above.  Therefore
there are three possible cases of the limit of $l_i(t)$:
\begin{eqnarray}
\label{type1}&&\text{(1)~~~}\lim_{t\rightarrow -\infty} l_i(t) =
\overline{l}_i
~\text{and}~\lim_{t\rightarrow +\infty} l_i(t) =  -\infty; \\
\label{type2}&&\text{(2)~~~}\lim_{t\rightarrow -\infty} l_i(t) =
-\infty
~\text{and}~\lim_{t\rightarrow +\infty} l_i(t) = \overline{l}_i; \\
\label{type3}&&\text{(3)~~~}\lim_{t\rightarrow -\infty} l_i(t) =
-\infty ~\text{and}~ \lim_{t\rightarrow +\infty} l_i(t) = -\infty
\end{eqnarray}
where $\overline{l}_i =  \sup_t l_i(t)
<\infty$.\\

\textbf{Proof of Lemma 2.4.}  Without loss of generality, we suppose that A.1 is satisfied with the
first $m$ cases (hence they are completely observed).  In order to show Lemma 2.4, we first prove that under A.1, $L(f) = \sum_{i=1}^m \log
p(y_i|x_i, f)$ is positively coercive over $\mathcal{H}_0$. Suppose to the contrary that this is not true.  Then there exists a constant $U>0$ and a
sequence $\{g_k\}_{k\in \mathbb{N}} \subseteq \mathcal{H}_0$ with $
||g_k||_\mathcal{H} = 1$ such that
\begin{equation}
\label{proof1} - \sum_{i=1}^{m}l_i(k \cdot g_k(x_i))\le U,~~~k\in
\mathbb{N}.
\end{equation}

Since the unit sphere $ \{g \in \mathcal{H}_0~:~||g||_\mathcal{H} =
1\} $ is sequence compact, there exists a subsequence
$\{g_{k_j}\}_{j \in \mathbb{N}}$ converging to some $g^*$ with
$||g^*||_\mathcal{H}=1$. We claim that
\begin{equation}
\label{case} g^*(x_i)~\left\{
                       \begin{array}{ll}
                         \le 0, & \text{ if } i \text{ belongs to Case 1 as (\ref{type1})} \\
                         \ge 0, &  \text{ if } i \text{ belongs to Case 2 as (\ref{type2})}\\
                         =0, &  \text{ if } i \text{ belongs to Case 3 as (\ref{type3})}.
                       \end{array}
                     \right.
\end{equation}
Suppose to the contrary that (\ref{case}) is not true.  If $i$
belong to case (1), then $g^*(x_i) = a >0$.  Since $\{g_{k_j}\}_{j \in
\mathbb{N}}$ converges to $g^*$, there exists $N>0$ such that
\begin{equation}
g_{k_j}(x_i) \ge a/2, ~~\text{ for all } j>N.
\end{equation}
From (\ref{proof1}), we have
\begin{equation}
l_i(k_j\cdot  g_{k_j}(x_i)) \ge U - \sum_{s \neq i} \overline{l}_s <
\infty, ~~~ j \in \mathbb{N}.
\end{equation}
This is a contradiction of (\ref{type1}) since when $j > N$
\begin{equation}
k_j \cdot g_{k_j}(x_i) \ge k_j \cdot a/2 \rightarrow +\infty.
\end{equation}
Similar contradiction can be observed when $i$ belongs to case (2) or
case (3).  Therefore the claim in Equation (\ref{case}) follows.

Now let $g_0$ be the unique minimizer of $-\sum_{i=1}^m l_i(g(x_i))$
in $\mathcal{H}_0$.  Consider $ g_0 + rg^*$ with $r>0$.  Combining
(\ref{type1})--(\ref{type3}) and (\ref{case}), we can see that
\begin{equation}
-\sum_{i=1}^m l_i(g_0(x_i)+r g^*(x_i) ) \le -\sum_{i=1}^m
l_i(g_0(x_i)),  ~~\forall r>0.
\end{equation}
But this is a contradiction.  Hence $L(f)$ is positively coercive over $\mathcal{H}_0$, which means that
\begin{equation}
\label{proof2}||g||_\mathcal{H}\rightarrow \infty  \Rightarrow
-\sum_{i=1}^m l_i(g(x_i)) \rightarrow +\infty,~~~g\in \mathcal{H}_0.
\end{equation}

Consider the orthogonal decomposition $f=g+h$ where $g \in
\mathcal{H}_0\bigcap \mathcal{H}_B$ and $h \in \mathcal{H}_1 \bigcap
\mathcal{H}_B$. The Lemma can be proved in steps.

(i) $||h||_\mathcal{H}\rightarrow +\infty$.  In this case
\begin{equation}
I^R_\lambda(f)  \ge  -\frac{1}{n}\sum_{i=1}^n \overline{l}_i +
\frac{1}{2}\lambda||h||_\mathcal{H}  \rightarrow +\infty.
\end{equation}

(ii) $||h||_\mathcal{H} \le U$ for some $U>0$ but $||g||_\mathcal{H}
\rightarrow +\infty$.  In this case
\begin{equation}\nonumber
|h(x_i)| = |\langle h, K(\cdot, x_i)\rangle | \le ||h||_\mathcal{H}
K^{1/2}(x_i,x_i) \le U \cdot K^{1/2}(x_i,x_i), ~~i=1,2,...m
\end{equation}
which implies that
\begin{eqnarray}\nonumber
f(x_i) = g(x_i) + h(x_i) = g(x_i) +
O(1),~~i=1,...,m,~||h||_\mathcal{H} \le U.
\end{eqnarray}
Let $||g||_\mathcal {H} \rightarrow \infty$, we have
\begin{eqnarray}
I^R_\lambda(f) &\ge& -\frac{1}{n}\sum_{i=1}^n \log \int_{\mathcal
{X}_i} p(y_i|x_i, f) d P_{i}{} \nonumber \\
&\ge& -\frac{1}{n}\sum_{i=1}^m l_i(g(x_i)+h(x_i)) - \frac{1}{n}\sum_{j=m+1}^n \overline{l}_j \nonumber \\
&=& -\frac{1}{n}\sum_{i=1}^m l_i(g(x_i)+O(1)) - \frac{1}{n}\sum_{j=m+1}^n \overline{l}_j \nonumber \\
\label{limit}&\rightarrow & +\infty
\end{eqnarray}
where (\ref{limit}) follows from the claim in Equation
(\ref{proof2}).

The Lemma is now proved by combining (i) and (ii). $~\Box$\\

\textbf{Proof of Lemma 2.5.}  Let $\{f_k\}_{k\in \mathbb{N}}$ be a sequence
in $\mathcal {H}_B$ which converges weakly to $f^*$.  Since
pointwise limit of measurable functions is still a measurable
function, $f^*\in \mathcal{H}_B$. From the continuity of $l_i(t)$,
$\{e^{l_i(f_k(x_i))}\}_{k\in \mathbb{N}}$ pointwise converges to
$e^{l_i(f^*(x_i))}$ over $\mathcal{X}_i$.
Note that $e^{l_i(f_k(x_i))}\le e^{\overline{l}_i}$ and every
constant is integrable with respect to $(\mathcal{X}_i,
\mathcal{F}_i, P_{i})$.  By the Dominated Convergence Theorem,
we have that
\begin{eqnarray}
\lim_{k\rightarrow \infty} \int_{\mathcal{X}_i}e^{l_i(f_k({x_i}))}
dP_{i}{} = \int_{\mathcal{X}_i}e^{l_i(f^*({x_i}))}
dP_{i}{}.
\end{eqnarray}
The Lemma now follows since $\log(\cdot)$ is continuous.  $~\Box$\\

\textbf{Proof of Lemma 2.6.}  Let $\{f_k\}_{k\in \mathbb{N}}$ be a sequence
in $\mathcal {H}_B$ which weakly converges to $f^*$. Consider the
orthogonal decomposition of each $f_k$ by $f_k = g_k+h_k$ with
$g_k\in \mathcal {H}_0\bigcap \mathcal {H}_B$ and $h_k\in
\mathcal{H}_1\bigcap \mathcal{H}_B$.  It is straightforward to see
that $\{h_k\}_{k\in \mathbb{N}}$ weakly converges to $h^*$, the
smooth part of $f^*$. Therefore we can write
\begin{equation}
\label{inequal}0\le ||h_k-h^*||_\mathcal{H}^2 =
||h_k||_\mathcal{H}^2 +||h^*||_\mathcal{H}^2 - 2 \langle h_k,
h^*\rangle.
\end{equation}
Let $k\rightarrow \infty$, we observe that
\begin{eqnarray}
0\le \liminf_{k}||h_k||_\mathcal{H}^2 - ||h^*||_\mathcal{H}^2
\end{eqnarray}
and the Lemma is proved by definition.  $~\Box$ \\
~\\

\textbf{Proof of Theorem 4.1.} For any fixed $\theta\in \Theta$, by Theorem 2.2, $I_\lambda^E(f,\theta)$ is minimizable in
$\mathcal{H}$. Let
\begin{equation}
T(\theta) \triangleq \min_{f\in \mathcal {H}} I_\lambda^E(f,\theta)
\end{equation}
denote the minimum penalized likelihood given $\theta$.  We claim
that $T(\theta)$ is continuous.

For any sequence $\{\theta_k\}_{k\in \mathbb{N}} \in \Theta$ that converges
to $\theta^*$, let $P_{\theta_k}$ and $P_{\theta^*}$ denote the probability measures on $\mathbb{R}^d$ with density functions $p({u}|\theta_k)$ and $p({u}|\theta^*)$.  Since $F({u}|\theta_k) \rightarrow F({u}|\theta^*)$ for any ${u} \in \mathbb{R}^d$, $P_{\theta_k}$ weakly converges to $P_{\theta^*}$.  Note that, for any fixed $f\in \mathcal{H}$, $G({u}) \triangleq p(y_i|x^{err}_i-{u}, f)$ is a real-valued, continuous and bounded function on $\mathbb{R}^d$.  Thus $\int G({u}) d P_{\theta_k} \rightarrow \int G({u}) d P_{\theta^*}$.  Equivalently, that is
\begin{equation}
\int_{\mathbb {R}^d} p(y_i|x^{err}_i-{u}_i, f)p({u}_i|\theta_k)d{u}_i \rightarrow \int_{\mathbb {R}^d} p(y_i|x^{err}_i-{u}_i, f)p({u}_i|\theta^*)d{u}_i
\end{equation}
which implies that $I_\lambda^E(f,\theta)$ is continuous in $\theta$ for any fixed $f$.  This is sufficient to prove the continuity of $T(\theta)$.  The theorem now follows from the compactness of $\Theta$. $~\Box$.\\

\textbf{Proof of Theorem 6.1.}  For any fixed $\theta\in \Theta$, by (\ref{plkm2})
and Theorem 2.2, $I_\lambda^M(f,\theta)$ is minimizable in $\mathcal{H}$.  Thus, we can define
\begin{equation}
T(\theta) \triangleq \min_{f\in \mathcal {H}} I_\lambda^M(f,\theta).
\end{equation}
We claim that $T(\theta)$ is continuous.

By Assumption M.1 and M.2, there exists $U>0$ such that $p(x_i|\theta)
< U$ for all $x_i^{mis}\in \mathcal{D}^{\theta}_i$, $\theta \in \Theta$ and $1\le i \le n$. Now for
any sequence $\{\theta_k\}_{k\in \mathbb{N}} \in \Theta$ that converges
to $\theta^*$, $p(y_i|x_i, f)p(x_i|\theta_k)$ pointwise converges to
$p(y_i|x_i, f)p(x_i|\theta^*)$.  Note that
$p(y_i|x_i, f)p(x_i|\theta_k) \le e^{\overline{l}_i} \cdot U$ and
any constant is integrable on the compact domain $\mathcal{D}^\theta_i$.
By Dominated Convergence Theorem,
we conclude that
\begin{equation}
\lim_{k \rightarrow \infty} \int_{\mathcal{D}^\theta_i}
p(y_i|x_i, f)p(x_i|\theta_k) d x_i^{mis} =   \int_{\mathcal{D}^\theta_i}
p(y_i|x_i, f)p(x_i|\theta^*) d x_i^{mis}
\end{equation}
which implies that $I_\lambda^M(f,\theta)$ is continuous in $\theta$
for any fixed $f$.  This is sufficient to prove the continuity of
$T(\theta)$.  The theorem now follows from the compactness of
$\Theta$. $~\Box$\\

\section{Derivation of GACV}

Our GACV is derived based on the cross validation function (\ref{CV2}).  Let us use the notations (\ref{vf1}) and (\ref{vf2}). It can be seen from (\ref{EW2}) that $\sum_{j=1}^{m_i} w^{[-i]}_{\lambda,ij}\hat{f}^{[-i]}_\lambda(z_{ij})$ can be treated as a function of $\vec{f}^{~[-i]}_{\lambda i}$. Note that $\vec{f}^{~[-i]}_{\lambda i}$ is expected to be close to $\vec{f}_{\lambda i}$, Thus using the first order Taylor expansion to expand $\sum_{j=1}^{m_i} w^{[-i]}_{\lambda,ij}\hat{f}^{[-i]}_\lambda(z_{ij})$ at $\vec{f}_{\lambda i}$, we have that
\begin{eqnarray}
\text{CV}(\lambda) &\approx& \widehat{\text{OBS}}(\lambda) + \frac{1}{n}\sum_{i=1}^{n} y_i (\vec{f}_{\lambda i} - \vec{f}^{~[-i]}_{\lambda i})^T \frac{\partial~ \sum_{j=1}^{m_i} w_{ij}(\tau)\tau_j }{\partial~ \tau} \Big|_{\vec{f}_{\lambda i}} \nonumber \\
&=& \widehat{\text{OBS}}(\lambda) + \frac{1}{n}\sum_{i=1}^{n} y_i (\vec{f}_{\lambda i} - \vec{f}^{~[-i]}_{\lambda i})^T \left(
                                                                                                                         \begin{array}{c}
                                                                                                                           d_{i1} \\
                                                                                                                           \vdots \\
                                                                                                                           d_{im_i} \\
                                                                                                                         \end{array}
                                                                                                                       \right)  \label{CVU}
\end{eqnarray}
where $w_{ij}(\tau)$ and $d_{ij}$ are defined by (\ref{Wfun}) and (\ref{WFDV}), respectively.  Thus, it remains to estimate $\vec{f}_{\lambda i} - \vec{f}^{~[-i]}_{\lambda i}$.  To do this, we first extend the leave-out-one lemma (Craven and Wahba,1979\cite{Craven1979}) to randomized covariate data.\\

LEMMA B.1  \emph{(leave-out-one-subject lemma) Let $l(y_i, t)  = y_i \cdot t- b(t)
+c(y)$ be the log-likelihood function and
$
I_\lambda^{{Z}, \Pi}(\vec{y}, f) = -\sum_{i=1}^n \log \sum_{j=1}^{m_i} \pi_{ij} \exp\{l(y_i, \\f(z_{ij})) \} + \frac{n\lambda}{2} J(f)
$,
where $\vec{y} = (\vec{y}^{~T}_1...,\vec{y}^{~T}_n)^T$ with $\vec{y}^{~T}_i = (y_i,...,y_i)^T$ being $m_i$ replicates of $y_i$. Suppose that  $\tau = (\tau_1,...,\tau_{m_i})^T$ is a $m_i\times 1$ vector and $h_\lambda(i, \tau, \cdot)$ is the minimizer in $\mathcal {H}$ of $I_\lambda^{{Z}, \Pi}( \vec{Y}, f)$, where $ \vec{Y} = (\vec{y}^{~T}_1,...,\vec{y}^{~T}_{i-1}, \tau^T, \vec{y}^{~T}_{i+1},\\...,\vec{y}^{~T}_n)^T$.  Then
\begin{equation}
h_\lambda(i, \vec{\mu}^{[-i]}_{\lambda i}, \cdot) = \hat{f}^{[-i]}_{\lambda}
\end{equation}
where $\hat{f}^{[-i]}_{\lambda}$ minimizes $-\sum_{k\neq i} \log \sum_{j=1}^{m_k} \pi_{kj} \exp\{l(y_i, f(z_{kj})) \} + \frac{n\lambda}{2} J(f)$, and $\vec{\mu}^{[-i]}_{\lambda i} = (b'(\hat{f}^{[-i]}_{\lambda}(z_{i1})),...,b'(\hat{f}^{[-i]}_{\lambda}(z_{im_i})))^T$ is the vector of means corresponding to $\hat{f}^{[-i]}_{\lambda}$.
}\\

\textbf{Proof of Lemma B.1.}  Firstly, we claim that
\begin{equation}\label{comparel}
l(b'(\hat{f}^{[-i]}_{\lambda}(z_{ij})),\hat{f}^{[-i]}_{\lambda}(z_{ij})) \ge l(b'(\hat{f}^{[-i]}_{\lambda}(z_{ij})),f(z_{ij})), ~1\le j \le m_i , ~\forall f \in \mathcal{H}.
\end{equation}
This follows since
\begin{equation}
\frac{\partial~ l(b'(\hat{f}^{[-i]}_{\lambda}(z_{ij})),t)}{\partial ~t}  = b'(\hat{f}^{[-i]}_{\lambda}(z_{ij})) -b'(t) \nonumber
\end{equation}
and using the fact that
$
\frac{\partial^2 l(y,t)}{\partial t^2} =  -b''(t) < 0
$.
Therefore $l(b'(\hat{f}^{[-i]}_{\lambda}(z_{ij})),t)$ achieves its unique maximum for $t = \hat{f}^{[-i]}_{\lambda}(z_{ij})$.

Define $\vec{y}^{~[-i]} = (\vec{y}^{~T}_1,...,\vec{y}^{~T}_{i-1},(\vec{\mu}^{[-i]}_{\lambda i})^T , \vec{y}^{~T}_{i+1},...,\vec{y}^{~T}_n)^T$.  Then for any  $f\in \mathcal {H}$,
\begin{eqnarray}
I_\lambda^{{Z}, \Pi}( \vec{y}^{~[-i]} , f) &=& -\log\sum_{j=1}^{m_i} \pi_{ij} \exp\{l(b'(\hat{f}^{[-i]}_{\lambda}(z_{ij})), f(z_{ij})) \} \nonumber \\
&&~~~~~-\sum_{k\neq i} \log \sum_{j=1}^{m_k} \pi_{kj} \exp\{l(y_k, f(z_{kj})) \} + \frac{n\lambda}{2} J(f) \nonumber\\
&\ge& -\log\sum_{j=1}^{m_i} \pi_{ij} \exp\{l(b'(\hat{f}^{[-i]}_{\lambda}(z_{ij})), \hat{f}^{[-i]}_{\lambda}(z_{ij})) \} \nonumber \\
&&~~~~~-\sum_{k\neq i} \log \sum_{j=1}^{m_k} \pi_{kj} \exp\{l(y_k, f(z_{kj})) \} + \frac{n\lambda}{2} J(f) \nonumber\\
&\ge& -\log\sum_{j=1}^{m_i} \pi_{ij} \exp\{l(b'(\hat{f}^{[-i]}_{\lambda}(z_{ij})), \hat{f}^{[-i]}_{\lambda}(z_{ij})) \} \nonumber \\
&&~~~~~-\sum_{k\neq i} \log \sum_{j=1}^{m_k} \pi_{kj} \exp\{l(y_k, \hat{f}^{[-i]}_{\lambda}(z_{kj})) \} + \frac{n\lambda}{2} J(\hat{f}^{[-i]}_{\lambda}) \nonumber.
\end{eqnarray}
The first inequality is due to (\ref{comparel}) and the second one is due to the fact that $\hat{f}^{[-i]}_{\lambda}$ minimizes $-\sum_{k\neq i} \log \sum_{j=1}^{m_k} \pi_{kj} \exp\{l(y_i, f(z_{kj})) \}  + \frac{n\lambda}{2} J(f)$. Thus we have $h_\lambda(i, \vec{\mu}^{[-i]}_{\lambda i}, \cdot) = \hat{f}^{[-i]}_{\lambda}$.$~\Box$\\

Consider the parametric form of the penalized likelihood in (\ref{class2}) and denote  $\vec{y}^{~[-i]} = (\vec{y}^{~T}_1,...,\vec{y}^{~T}_{i-1},(\vec{\mu}^{[-i]}_{\lambda i})^T , \vec{y}^{~T}_{i+1},...,\vec{y}^{~T}_n)^T$.  Then
Lemma B.1 says that $\vec{f}_\lambda^{~[-i]} = (\hat{f}^{[-i]}_\lambda(z_{11}),...,\hat{f}^{[-i]}_\lambda(z_{1m_1}),\hat{f}^{[-i]}_\lambda(z_{21}),...,\hat{f}^{[-i]}_\lambda(z_{nm_n}))^T$ minimizes \\ $I_\lambda^{{Z}, \Pi}( \vec{y}^{~[-i]} , \vec{f}~)$.   Note that $\vec{f}_\lambda = (\hat{f}_\lambda(z_{11}),...,\hat{f}_\lambda(z_{1m1}),\hat{f}_\lambda(z_{21}),...,\hat{f}_\lambda(z_{nm_n}))^T$ minimizes $I_\lambda^{{Z}, \Pi}( \vec{y} , \vec{f}~)$.  Thus,
\begin{equation}\label{dzero}
\frac{\partial ~I_\lambda^{{Z}, \Pi} }{\partial~ \vec{f}} (\vec{y} , \vec{f}_\lambda) =0, ~~\frac{\partial ~I_\lambda^{{Z}, \Pi} }{\partial~ \vec{f}} (\vec{y}^{~[-i]} , \vec{f}^{~[-i]}_\lambda) =0.
\end{equation}
Using first order Taylor expansion, we have that
\begin{eqnarray}
0 &=& \frac{\partial I_\lambda^{{Z}, \Pi}}{ \partial \vec{f}} (\vec{y}^{~[-i]} , \vec{f}^{~[-i]}_\lambda) \nonumber \\
&=& \frac{\partial I_\lambda^{{Z}, \Pi}}{ \partial \vec{f}} (\vec{y} , \vec{f}_\lambda) +  \frac{\partial^2 I_\lambda^{{Z}, \Pi}}{ \partial \vec{f}~\partial \vec{f}^T} (\vec{y}^{~*} , \vec{f}^{~*}_\lambda) (\vec{f}^{~[-i]}_\lambda - \vec{f}_\lambda) +  \frac{\partial^2 I_\lambda^{{Z}, \Pi}}{ \partial \vec{y}~\partial \vec{f}^T} (\vec{y}^{*} , \vec{f}^{*}_\lambda) (\vec{y}^{~[-i]} - \vec{y})  \nonumber \\
 &=& \frac{\partial^2 I_\lambda^{{Z}, \Pi}}{ \partial \vec{f}~\partial \vec{f}^T} (\vec{y}^{*} , \vec{f}^{*}_\lambda) (\vec{f}^{~[-i]}_\lambda - \vec{f}_\lambda) +  \frac{\partial^2 I_\lambda^{{Z}, \Pi}}{ \partial \vec{y}~\partial \vec{f}^T} (\vec{y}^{*} , \vec{f}^{*}_\lambda) (\vec{y}^{~[-i]} - \vec{y}) \nonumber \\
 \label{Iexp}
\end{eqnarray}
where $(\vec{y}^{*} , \vec{f}^{*}_\lambda)$ is a point between $(\vec{y} , \vec{f}_\lambda)$ and $(\vec{y}^{~[-i]} , \vec{f}^{~[-i]}_\lambda)$.

Consider any arbitrary vector $\vec{f}= (\vec{f}^T_1,...,\vec{f}^T_n)$ with $\vec{f}_i= (f_{i1},...,f_{im_i})^T$ being an $m_i\times 1$ vector. For $1\le i\le n$ and $1\le s, t \le m_i$, let's denote
\begin{eqnarray}
b^i_{st}(\vec{f}~) = \left\{
             \begin{array}{ll}
               -w_{is}(\vec{f}~)\left[1+(1-w_{is}(\vec{f}~))f_{is}(y_i - b'(f_{is}))\right], &\text{if } s=t \\
               w_{is}(\vec{f}~) w_{it}(\vec{f}~)f_{is}(y_i - b'(f_{it})), &\text{if } s\neq t
             \end{array}
           \right. \nonumber\\
d^i_{st}(\vec{f}~) = \left\{
             \begin{array}{ll}
               w_{is}(\vec{f}~)\left[b''(f_{is})-(1-w_{is}(\vec{f}~))(y_i -  b'(f_{is}))^2\right], &\text{if } s=t \\
               w_{is}(\vec{f}~)  w_{it}(\vec{f}~) (y_i - b'(f_{is}))(y_i - b'(f_{it})), &\text{if } s\neq t.
             \end{array}
           \right.  \nonumber
\end{eqnarray}
Define submatrices $B_i(\vec{f}~) = \left(b^i_{st}(\vec{f}~)\right)_{m_i\times m_i}$ and $D_i(\vec{f}~) = \left(d^i_{st}(\vec{f}~)\right)_{m_i\times m_i}$ and let
$B(\vec{f}~) = \text{diag}(B_1(\vec{f}~),...,B_n(\vec{f}~))$ and $D(\vec{f}~) = \text{diag}(D_1(\vec{f}~),...,D_n(\vec{f}~))$ be block diagonal matrices.
Then direct calculation yields
\begin{equation}
\frac{\partial^2 I_\lambda^{{Z}, \Pi}}{ \partial \vec{f}~\partial \vec{f}^T} (\vec{y}^{*} , \vec{f}^{*}_\lambda) = \frac{1}{n} D(\vec{f}^{*}_\lambda) + \Sigma_\lambda, ~~\frac{\partial^2 I_\lambda^{{Z}, \Pi}}{ \partial \vec{y}~\partial \vec{f}^T} (\vec{y}^{*} , \vec{f}^{*}_\lambda) = \frac{1}{n}B(\vec{f}^{*}_\lambda).
\end{equation}
Therefore, from (\ref{Iexp}), we have
\begin{equation}\label{influenceq}
 \vec{f}_\lambda - \vec{f}^{~[-i]}_\lambda = -(D(\vec{f}^{*}_\lambda) + n\Sigma_\lambda)^{-1}B(\vec{f}^{*}_\lambda) (\vec{y} - \vec{y}^{~[-i]}).
\end{equation}
Approximate $B(\vec{f}^{*}_\lambda)$ and $D(\vec{f}^{*}_\lambda)$ by $B(\vec{f}_\lambda)$ and $D(\vec{f}_\lambda)$.  Then denote $H = -(D(\vec{f}_\lambda) + n\Sigma_\lambda)^{-1}B(\vec{f}_\lambda)$ the influence matrix of $I_\lambda^{{Z}, \Pi}(\vec{y}, \vec{f}~)$ with respect to $\vec{f}$ evaluated at $\vec{f}_\lambda$.  From (\ref{influenceq}), we have
\begin{equation}\label{Approx}
\left(
  \begin{array}{c}
   \vec{f}_{\lambda 1} - \vec{f}^{~[-i]}_{\lambda 1} \\
    \vdots \\
    \vec{f}_{\lambda i} - \vec{f}^{~[-i]}_{\lambda i} \\
    \vdots \\
    \vec{f}_{\lambda n} - \vec{f}^{~[-i]}_{\lambda n} \\
  \end{array}
\right) \approx H \left(
                    \begin{array}{c}
                      0 \\
                      \vdots \\
                      \vec{y}_i - \vec{\mu}^{[-i]}_{\lambda i} \\
                      \vdots \\
                      0 \\
                    \end{array}
                  \right)_{\sum m_i\times 1}.
\end{equation}
Write
\begin{equation}\label{H}
H = \left(
      \begin{array}{cccc}
          H_{11} & * & * & * \\
          * & H_{22} & \cdots & * \\
          \vdots & \vdots & \ddots & \vdots \\
          * & * & \cdots & H_{nn} \\
      \end{array}
    \right)_{\sum m_i\times \sum m_i}
\end{equation}
where each $H_{ii}$ is a $m_i\times m_i$ submatrix matrix on the diagonal with respect to $(f_{i1},...,f_{im_i})^T$.  We observe from (\ref{Approx}) that
\begin{equation}\label{fact1}
\vec{f}_{\lambda i} - \vec{f}^{~[-i]}_{\lambda i} \approx H_{ii} (\vec{y}_i - \vec{\mu}^{[-i]}_{\lambda i}).
\end{equation}

Recall that $ \vec{\mu}^{[-i]}_{\lambda i} = (b'(\hat{f}^{[-i]}_\lambda(z_{i1})),...,b'(\hat{f}^{[-i]}_\lambda(z_{im_i})))^T$ is a vector of $b'(\cdot)$ evaluated at $\vec{f}^{~[-i]}_{\lambda i}$.  Hence, using a  first order Taylor expansion to expand $b'(\cdot)$ at $\vec{f}_{\lambda i}$, we have
\begin{equation}\label{fact2}
 \vec{\mu}^{[-i]}_{\lambda i}-\vec{\mu}_{\lambda i}   \approx {W}_i ( \vec{f}^{~[-i]}_{\lambda i} - \vec{f}_{\lambda i}  )
\end{equation}
where ${W}_i = \text{diag}(b''(\hat{f}_\lambda(z_{i1})),...,b''(\hat{f}_\lambda(z_{im_i})))$ is a diagonal matrix of variances.

Combining (\ref{fact1}) and (\ref{fact2}), we can show that
\begin{eqnarray}
\vec{f}_{\lambda i} - \vec{f}^{~[-i]}_{\lambda i} &\approx& H_{ii} (\vec{y}_i - \vec{\mu}^{[-i]}_{\lambda i}) \nonumber \\
 &=&  H_{ii} (\vec{y}_i - \vec{\mu}_{\lambda i} +\vec{\mu}_{\lambda i} -\vec{\mu}^{[-i]}_{\lambda i}) \nonumber \\
&\approx & H_{ii} (\vec{y}_i - \vec{\mu}_{\lambda i} + {W}_i ( \vec{f}_{\lambda i} - \vec{f}^{~[-i]}_{\lambda i} )). \label{factall}
\end{eqnarray}
Now, an approximation of  $\vec{f}_{\lambda i} - \vec{f}^{~[-i]}_{\lambda i}$ can be obtained by solving (\ref{factall})
\begin{equation}\label{keyf}
\vec{f}_{\lambda i} - \vec{f}^{~[-i]}_{\lambda i} \approx (I_{m_i\times m_i} - H_{ii}{W}_i)^{-1} H_{ii} (\vec{y}_i - \vec{\mu}_{\lambda i}).
\end{equation}
Plug (\ref{keyf}) into the CV function (\ref{CVU}), we obtain the approximate cross validation (ACV) function
\begin{equation}\label{ACV}
\text{ACV}(\lambda) = \widehat{\text{OBS}}(\lambda) + \frac{1}{n}\sum_{i=1}^{n} y_i (d_{i1},...,d_{im_i})  (I_{m_i\times m_i} - H_{ii}{W}_i)^{-1} H_{ii} (\vec{y}_i - \vec{\mu}_{\lambda i})
\end{equation}
where $ \widehat{\text{OBS}}(\lambda)$ is given in (\ref{OBS2}).  Define $G_{ii}= I_{m_i\times m_i} - H_{ii}{W}_i$.  Then a generalized form of approximate cross validation (GACV) can be obtained by replacing each $H_{ii}$ and $G_{ii}$ with the generalized average of submatrices defined in (\ref{AMatrix}).  Let $\bar{H}_{ii}$ and $\bar{G}_{ii}$ denote the generalized average of $H_{ii}$ and $G_{ii}$.  Then the generalized approximate cross validation (GACV) can be defined
\begin{eqnarray}
\text{GACV}(\lambda) &=& \widehat{\text{OBS}}(\lambda) + \frac{1}{n}\sum_{i=1}^{n} y_i (d_{i1},...,d_{im_i}) \bar{G}_{ii}^{-1} \bar{H}_{ii}(\vec{y}_i - \vec{\mu}_{\lambda i}) \nonumber\\
                   &=&  -\frac{1}{n} \sum_{i=1}^{n} \log \sum_{j=1}^{m_i} \pi_{ij}\exp\left\{ y_i \hat{f}_\lambda(z_{ij}) - b(\hat{f}_\lambda(z_{ij}))\right\} \label{GACV} \\
&&~~~+ \frac{1}{n}\sum_{i=1}^{n} y_i (d_{i1},...,d_{im_i})\bar{G}_{ii}^{-1}\bar{H}_{ii}\left(
                                                                                             \begin{array}{c}
                                                                                                    y_i-\hat{\mu}_{\lambda} (z_{i1}) \\
                                                                                                    \vdots \\
                                                                                                    y_i- \hat{\mu}_{\lambda} (z_{im_i})  \\
                                                                                                    \end{array}
                                                                                             \right). \nonumber
\end{eqnarray}

We remark that if all the $x_i$'s are exactly observed, then the above GACV function will reduce to the original GACV formula in Xiang and Wahba (1996)\cite{Xiang1996}.

\section{Extension to SS-ANOVA model}

Smoothing spline analysis of variance (SS-ANOVA) provides a general
framework for multivariate nonparametric function estimation.  The
application is very broad.  To extend the methodologies of the
paper, it suffices to show that the penalized likelihood for
SS-ANOVA model can be formulated in the form of (\ref{plk}).  The
following arguments are derived from Wahba (1990)\cite{Wahba1990}.

The penalized likelihood of smoothing Spline ANOVA model
takes the form of
\begin{eqnarray}
I_\lambda(f) =
-\frac{1}{n}\sum_{i=1}^{n}\log p(y_i|x_i, f)+\sum_{\beta=1}^{b}\lambda_\beta||\mathbb{P}^\beta_1
f||^2_{\mathcal {H}_1^\beta}
\end{eqnarray}
where $\mathcal {H}_1^\beta$ are nonparametric subspaces (smooth
spaces) which are assumed to be RKHS with reproducing kernel
$K_1^\beta(\cdot,\cdot)$ and $\mathbb{P}^\beta_1$ projects $f$ onto $\mathcal
{H}_1^\beta$.  Now For $\lambda_\beta>0$, define $\mathcal {H}_1 =
\sum_{\beta=1}^{b} \oplus \mathcal {H}_1^\beta$ with norm
\begin{equation}
||\eta||_{\mathcal {H}_1}^2 = \sum_{\beta=1}^{b}\lambda_\beta
||\mathbb{P}^\beta_1\eta||^2_{\mathcal {H}_1^\beta}, \eta \in \mathcal
{H}_1.
\end{equation}
It can be shown that $\mathcal {H}_1$ is a RKHS equipped
with RK $\sum_{\beta=1}^{b} \frac{1}{\lambda_\beta} K_1^\beta(s,t)$.
Then we can write that
\begin{eqnarray}
I_\lambda(f) = -\frac{1}{n}\sum_{i=1}^{n}\log p(y_i|x_i, f) +
||\mathbb{P}_1f||^2_{\mathcal {H}_1}
\end{eqnarray}
where $\mathbb{P}_1$ projects $f \in \mathcal{H}$ onto $\mathcal {H}_1$.  Set $J(f) = ||\mathbb{P}_1f||^2_{\mathcal {H}_1}$.  Then
the above expression takes the form of (\ref{plk}).  Therefore our
discussion in this paper can be extended to SS-ANOVA model.

\section*{Acknowledgments}
This work was partially supported by
NIH Grant EY09946,
NSF Grant DMS-0604572,
NSF Grant DMS-0906818,
ONR Grant N0014-09-1-0655(X.M., B.D. and G.W.),
NIH Grant EY06594 (R.K., B.K. and K.L.) and
the Research to Prevent Blindness Senior Scientific Investigator Awards
(R.K. and B.K.).

\end{document}